\documentclass{article}
\usepackage{amsfonts}

 \def\rql{\mbox{\strut\smash{\raisebox{-1.4ex}{''}}\hspace{0.1em}}}
 \def\rqr{\mbox{\hspace{0em}``}}

\def\dslash{\mbox{/\hspace{-0.3em}/}}

\makeatletter
\renewcommand{\@oddfoot}{--- ~\today~ ---\hfil
\raisebox{0.3ex}{\tiny --- ~D.~A.~Arbatsky~ ``QED. VI.
Quantization.''~ ---}\hfil --- ~\thepage~ ---}
 \makeatother

\begin{document}

\newcounter{parnum}
\newlength{\picwidth}
\newcommand*{\myparagraph}[1]{\refstepcounter{parnum}%
\paragraph{\arabic{parnum}. #1}}


\title{On quantization of electromagnetic field.\\
       VI. Quantization.}
\author{D.~A.~Arbatsky\footnote{http:\dslash d-a-arbatsky.narod.ru/}}
\date{\today}
\maketitle

\begin{abstract}
Here we describe a general method of quantization of linear
fields. We introduce a conception of quantization invariant with
respect to action of some group. A space of quantum states of
relativistic fields is constructed in apparently
relativistic-invariant way. A connection with quantization of the
field oscillator is established. We substantiate the necessity of
using an indefinite scalar product for electromagnetic field. We
discuss additional condition for ``physically allowed'' states of
electromagnetic field. We discuss properties of the space of
states of electromagnetic field from the point of view of
functional analysis. We consider the question about origin of
anti-unitary transformations in quantum field theory.
\end{abstract}


\myparagraph{Other approaches to quantization.}
 Before we start description of construction of a quantized field,
let us give here a brief comparison of our construction with other
approaches to description of quantized fields.

\begin{itemize}

\item {\bf Method of formal manipulations.}
This method was used even in the earliest papers on quantum field
theory. Its essence is that we calculate Poisson brackets for
classical values and accept that for corresponding quantum values
commutators can be calculated by the formula:
\begin{equation}
 [\,\hat a\,,\,\hat b\,] = i\,\{\,a\,,\,b\,\}\widehat~  \ .
\label{CommRel}
\end{equation}
Of course, this formula can not be true for all values.
Nevertheless, if we use it cautiously enough, we can ascertain
many properties of fields, even without clarifying what they are.
In fact, this approach turned out to be the most effective from
the point of view of practical applications.

With regards to how Poisson brackets were calculated, it was
performed in such a way that it was very difficult to see
relativistic invariance even in the classical theory. Later
Peierls~\cite{Peierls1952} have suggested an alternative method of
obtaining commutation relations, which was apparently
relativistic-invariant. The corresponding classical bracket used
in the expression (\ref{CommRel}), was even named ``the Peierls
bracket''. A quite complicated ``physical'' grounding of the
relations (\ref{CommRel}), used for quantization in the Peierls
method, can be found in \cite{DeWitt1987}.

The approach described is most close to ours.

About classical description of relativistic fields we can say that
invariant Hamiltonian formalism allows to define the Poisson
bracket in apparently relativistic-invariant way. In this approach
the notion of the Poisson bracket changes in such a way that the
Peierls bracket turns out to be its very particular case. Methods
of practical calculations of Poisson brackets were described by me
in the paper~[I].

With regards to quantization of fields, in this paper in will be
performed {\it constructively}: on the base of the developed
in~[I] invariant Hamiltonian formalism, using usual algebraic
methods, we will construct in this paper quantum fields evidently.

\item {\bf Method of box.}
This method was also very popular in the field theory from its
birth. The essence of it is that we consider field not in the
infinite space, but in a large box with periodical boundary
conditions. So, the description of the field becomes similar to
the description of infinite number of oscillators.

The introduction of a box breaks the relativistic invariance in
essence, and this reason is sufficient to not consider this method
in detail.

Nevertheless, as a result of using this method a useful notion of
the field oscillator appeared. In this connection we should notice
that in our method we have an analogous notion (introduced in the
paper~[IV]). We introduce it in a somewhat more abstract way with
the use of the notion of an induced symplectic representation.
Only by this way it is possible to ascertain the correct
group-theory nature of the field oscillator and to connect
invariant quantizations of field with invariant quantizations of
oscillator.

\item {\bf Fock construction and Wigner-Mackey theory.}
In the paper~\cite{Fock1932} Fock suggested that the space of
states of a quantum field is a symmetrized tensor exponent of
one-particle subspace. Later Wigner and Mackey developed a method
of searching for all possible one-particle subspaces, based on the
theory of induced unitary representations of Poincare group (see,
for example, a review~\cite{Mackey1968}). A quite detailed account
of this approach is given in the book~\cite{Weinberg1996}.

The shortcomings of this approach are the following. First,
operators of creation and destruction of quanta are introduced
with quite complicated formulas from the very beginning. Second,
even after introducing of these operators quite complicated
reasoning is necessary for construction of local operators of {\it
field}. Third, in this approach we can not consider fields with
indefinite scalar product at all.

In our approach the initial object is not a one-particle quantum
space, but it is a classical field described by the language of
the invariant Hamiltonian formalism. So, the second shortcoming
disappears, because field operators are constructed during
quantization.

Operators of creation and destruction of quanta we can also
introduce. And formulas for them are just the consequence of our
algebraic construction of quantum field, and they are not
postulated from the beginning.

With regards to quantum fields with indefinite scalar product, in
our approach they appear as naturally as fields with
positive-definite scalar product. Our scheme is really more
general, and it can be seen even with the example of massive
scalar field. Due to this generality, the scheme naturally
includes electromagnetic field as a particular case.

Let us notice also that in our approach there is also an analog of
Wigner-Mackey theory. This is the theory of symplectic
representations of Poincare group~[IV]. The theory of induced
symplectic representations is richer than the theory of induced
unitary representations.

\item {\bf Approach connected with Stone-von Neumann theorem.}
In quantum mechanics when a quantum oscillator is studied the
Stone-von Neumann theorem turns out to be useful; it says that
canonical commutation relations for coordinates and momenta
uniquely define corresponding irreducible unitary representation.
Of course, people tried to bring this approach to quantum field
theory.

But the Stone-von Neumann theorem can not be directly changed for
the case of quantization of infinite dimensional systems (see, for
example, \cite{Schwarz1975}). Great researches were made to
overcome this difficulty. Despite of their undoubted importance, I
still think that they turned out to be too far from practical
applications.

It should be noticed here also that if we do not require
positive-definiteness of scalar product in the space of states,
the uniqueness of quantization is lost even in the case of
one-dimensional harmonic oscillator. Therefore, our approach
differs from this essentially: we refuse to require
positive-definiteness of scalar product, but instead we introduce
the conception of invariant quantization.

Let us mention here the following fact. When the Stone-von Neumann
theorem is formulated, it is necessary to write commutation
relations in the form of Weyl relations. But in our approach we
use commutation relations just for unbounded operators. And the
construction of quantization rigorously define in what sense they
must be understood.

\item {\bf Geometric quantization.}
The method of geometric quantization is that using geometry of the
classical phase space we construct the space of quantum states as
a some set of functions on classical phase space (see, for
example, \cite{Kirillov1985}).

From the point of view of the invariant Hamiltonian formalism it
is natural to apply this scheme to invariant phase space $Z$. But
let us notice that substantial difficulties are seen immediately
for this approach. First, it is not simple at all to make this
approach mathematically rigorous for systems with infinite number
of degrees of freedom. Second, the connection with Fock
construction in this case is given by quite non-evident formulas.
For these reasons this approach turns out to be not very
convenient for practical applications. Third, it is not clear how
we can describe  by this approach the case of quantization with
indefinite scalar product: even in the case of one-dimensional
harmonic oscillator great difficulties arise with definition of
indefinite scalar product.

\item {\bf Method of path integral.}
This method became very popular now because it is supposed to be
apparently relativistic-invariant. In fact, people work with path
integral as with a symbolic form of writing of some expressions.

On the other hand, it is possible to consider path integral as
rigorously defined mathematical object (see, for example,
\cite{Berez1980}). But such an approach meets so substantial
difficulties that we can not talk about any ``apparent''
relativistic invariance at all.

\end{itemize}


\myparagraph{Quantization.} \label{quantization}
 Let us describe now, how having classical field we construct
quantum field.

\begin{enumerate}

\item
Consider the set of (complex) functions that are constant on the
whole space $Z$. Let us denote this set as $\mathcal C$. Consider
also the direct sum $\mathcal C\oplus Z^*_{\mathbb C}$. The
Poisson bracket turns this sum into a complex Lie algebra.

\item
Disengaging now from the algebraic structure of the set
 $\mathcal C\oplus Z^*_{\mathbb C}$ and considering all its elements
as completely independent let us call this set an {\it alphabet}
$\mathcal A$ and its elements {\it letters}. Let us make an
agreement that an element of $\mathcal C\oplus Z^*_{\mathbb C}$
considered as a letter of the alphabet $\mathcal A$ will be
written additionally with a ``hat'' on top or on side, for
example: $\hat a = a\widehat{~}$.

\item
Furthermore, we can introduce a formal product of elements of the
alphabet. Namely, let us suppose that if we multiply a (finite)
set of letters we get a {\it word} consisting of these letters
(the order of letters is important here):
\[
 \hat a_1 \cdot \hat a_2 \cdot \dots \cdot \hat a_k =
   \hat a_1 \hat a_2 \dots \hat a_k \ ,
 \qquad
 \hat a_1,\hat a_2, \dots,\hat a_k \in \mathcal A \ .
\]
Words we will also denote by symbols with the hat, for example:
$\hat w = w\widehat{~}$. Let us denote the set of all possible
words by the symbol $\mathcal W$; it is convenient to include in
this set the word of zero length that we can denote as
$()\widehat~$. It is natural to accept that the operation of
multiplication works on the set of words $\mathcal W$ also, and it
``joins'' words. So, $\mathcal A$ is a {\it system of free
generators}, and $\mathcal W$ is a {\it free semi-group with
neutral element} (the role of neutral element is played by the
special element $()\widehat~$).

\item
Similarly, let us define the formal sum of (finite) set of words
multiplied by arbitrary complex factors. Namely, let us call a
{\it phrase} a formal expression like:
\[
 \lambda_1 \hat w_1 + \lambda_2 \hat w_2 +
    \dots + \lambda_m \hat w_m \ ,
 \qquad
 \lambda_1,\lambda_2,\dots,\lambda_m\in\mathbb C \ , \ \
 \hat w_1,\hat w_2,\dots,\hat w_m\in\mathcal W \ .
\]
Let us also suppose two phrases to be equivalent, if for every
word they have the same sum of factors for this word. The set of
all such equivalence classes we will denote with the symbol
$\mathcal P$.

On the set of phrases $\mathcal P$ we can now naturally introduce
operations of addition and multiplication by a scalar. Namely, the
sum of phrases is the phrase that arise if we join initial phrases
by the symbol ``$+$''. Multiplication of a phrase by a number is
defined as multiplication of all factors by this number. So,
$\mathcal P$ gets natural structure of complex linear space
($\mathcal P$ is a {\it free $\mathbb C$-module generated by
$\mathcal W$}).

On the set of phrases $\mathcal P$ we can also naturally introduce
the operation of multiplication of two elements. Namely, the
product of two phrases, each of which contains one word, we define
by equality:
\[
 (\lambda_1 \hat w_1)\cdot(\lambda_2 \hat w_2) =
 (\lambda_1\lambda_2)\,(\hat w_1 \hat w_2)  \ ,
 \qquad
 \lambda_1,\lambda_2\in\mathbb C \ , \ \
 \hat w_1,\hat w_2\in\mathcal W \ .
\]
And we spread this definition to arbitrary phrases by the
requirement of distributivity:
\[
 \hat p_1\cdot(\hat p_2+\hat p_3) =
     \hat p_1 \hat p_2 + \hat p_1 \hat p_3 \ ,
 \qquad
 (\hat p_1 + \hat p_2)\cdot \hat p_3 =
     \hat p_1 \hat p_3 + \hat p_2 \hat p_3 \ ,
 \qquad
 \hat p_1,\hat p_2,\hat p_3\in\mathcal P \ .
\]

\item
So, the set of phrases $\mathcal P$ becomes an (associative)
algebra ($\mathcal P$ is a {\it semi-group algebra of the
semi-group $\mathcal W$} or a {\it free algebra over semi-group
$\mathcal W$}).

\item
Up to now when we constructed the algebra $\mathcal P$ the
alphabet $\mathcal A$ was considered as a completely arbitrary
set. Now, using structure of Lie algebra in
 $\mathcal C\oplus Z^*_{\mathbb C}$ and the correspondence
between the subspace $\mathcal C$ and the field of scalars
$\mathbb C$ we can additionally bring to the algebra $\mathcal P$
some {\it defining relations}. It is performed by factorization of
the algebra $\mathcal P$ with respect to the appropriate ideal.

Namely, consider all phrases of the following types:
\[
 (\lambda\, a)\widehat~ - \lambda\,\hat a \ ,
\]
\[
 (a+b)\widehat~ - (\hat a + \hat b) \ ,
\]
\[
 \{\,a\,,\,b\,\}\widehat~ + i\,(\hat a\,\hat b - \hat b\,\hat a) \ ,
\]
\[
 \hat1 - ()\widehat~ \ .
\]
 Here $\lambda\in\mathbb C$,
 $a,b\in\mathcal C\oplus Z^*_{\mathbb C}$,
 $\hat a,\hat b\in\mathcal A$.
 $1$~--- is a notation for the function that is equal to $1$
on the whole invariant phase space~$Z$, i.~e.
 $1\in\mathcal C\oplus Z^*_{\mathbb C}$; $\hat1$~---
the element of the alphabet that corresponds to it,
$\hat1\in\mathcal A$.

Let us span now on these phrases the two-sided ideal, i.~e. we
join to them the phrases that can be made from the given by using
finite number of operations of multiplication by a factor,
addition, and multiplication (from the left and the right) by
arbitrary phrases. Factorizing the algebra $\mathcal P$ with
respect to the obtained ideal we get an algebra that we will call
an {\it algebra of operators} and will denote as $\mathcal O$. The
elements of this algebra are called {\it operators} and are
denoted as phrases corresponding to them.

The performed factorization leads to the result that in the
algebra of operators we have the following defining relations:
\[
 (\lambda\, a)\widehat~ = \lambda\,\hat a \ ,
\]
\[
 (a+b)\widehat~ = \hat a + \hat b \ ,
\]
\[
 \{\,a\,,\,b\,\}\widehat~ = -\,i\,[\,\hat a\,,\,\hat b\,] \ ,
\]
\[
 \hat1 = ()\widehat~ \ .
\]
In the third relation we have used the standard notation for {\it
commutator} of two operators:
 $[\,\hat a\,,\,\hat b\,]=\hat a\,\hat b - \hat b\,\hat a$.

The fourth relation, in particular, allows us to escape using the
cumbersome notation $()\widehat~$, when we talk about operator and
to write always $\hat1$.

\item
Let us introduce now the operation of {\it conjugation}. This
operation will be always denoted by the symbol ${}^*$, without
distinction of an object that it is applied to.

Under conjugation of a complex number we will imply the usual
complex conjugation.

Elements of the space $\mathcal C\oplus Z^*_{\mathbb C}$ are
functions on $Z$. Under conjugation applied to such a function we
will imply a transition to the function with complex-conjugate
values. Obviously, in the result we also get an element of the
space $\mathcal C\oplus Z^*_{\mathbb C}$. Furthermore, so far as
the Poisson bracket in $\mathcal C\oplus Z^*_{\mathbb C}$ is a
complexified real Poisson bracket, we have the equality:
\begin{equation}
 \{\,a^*\,,\,b^*\,\}=\{\,a\,,\,b\,\}^* \ .
\label{poissconj}
\end{equation}

The definition of conjugation can be naturally spread to the
elements of the alphabet $\mathcal A$:
\[
 (\hat a)^*=(a^*)\widehat~ \ .
\]

To words and to phrases the operation of conjugation is spread by
the rules:
\[
 (\hat a\,\hat b)^* = \hat b^*\,\hat a^*    \ ,
 \qquad
 (\lambda\,\hat a)^* = \lambda^*\,\hat a^*  \ ,
 \qquad
 (\hat a + \hat b)^* = \hat a^* + \hat b^*  \ .
\]
Here $\hat a$ and $\hat b$ can be letters, words, and phrases;
$\lambda$ is a complex number.

Obviously, the operation of conjugation is correctly spread from
phrases to operators.

\item
Consider now the space $Z^*_{\mathbb C}$. Let us split it into a
direct sum of two subspaces:
\[
 Z^*_{\mathbb C} = Cr\oplus De \ .
\]
Let us require $Cr$ and $De$, first, to come to each other under
conjugation:
\begin{equation}
 Cr^*=De \ .
\label{cranconj}
\end{equation}
Second, let us require that the Poisson bracket on these subspaces
be zero\footnote{Taking into account~(\ref{poissconj}) and
(\ref{cranconj}), it is enough to require that it is zero on one
of these subspaces.}:
\[
 \{\,a\,,\,b\,\} = 0 \ , \qquad  a,b\in Cr \ ;
\]
\[
 \{\,a\,,\,b\,\} = 0 \ , \qquad  a,b\in De \ .
\]

The subspace $Cr$ we will call {\it creating}, and $De$~--- {\it
destructing}. In accordance with the construction given above, for
subspaces $Cr$ and $De$ we have some corresponding subspaces in
the algebra of operators; let us denote these subspaces
$Cr\widehat~$ and $De\widehat~$, correspondingly.

Let us introduce also a notation $\mathcal C\widehat~$ for the
subspace of the algebra of operators that corresponds to the
subspace $\mathcal C$ of the Lie algebra $\mathcal C\oplus
Z^*_{\mathbb C}$.

\item
Let us span now the left ideal on the subspace $De\widehat~$,
i.~e. let us join to operators there operators that are obtained
from them by finite number of operations of multiplication by
scalar, addition, and multiplication from the left with arbitrary
elements of the algebra $\mathcal O$.

Let us factorize now the algebra $\mathcal O$ with respect to the
constructed left ideal. In result we get some linear space. It is
not an algebra, but it is a left module, i.~e. there we have a
naturally defined action of operators from the left. Elements of
this factor-space are called {\it ket-vectors}. Ket-vectors are
denoted with the symbols like $|\,x\,\rangle$, where $x$ is any
symbol describing the given vector. The space of ket-vectors is
denoted $\mathcal H$ and is called {\it the space of states} of
the quantized field.

\item
We can perform the same construction with conjugate objects, i.~e.
instead of the subspace $De\widehat~$ we can consider the subspace
$Cr\widehat~$, span on it the right ideal instead of left, and,
factorizing, get the right module instead of left. The elements of
this module are called {\it bra-vectors}. And this module we will
call, if it does not make a misunderstanding, also {\it the space
of states} of the quantized field, and we will denote it also as
$\mathcal H$.

Bra-vectors are denoted by symbols like $\langle\,x\,|$. The
operation of conjugation is naturally spread to ket- and
bra-vectors and defines the one-to-one correspondence between
them. It is said that the ket- and bra-vectors, corresponding to
each other, belong to the same state of the quantized field.

\item
When we factorize the algebra of operators $\mathcal O$ and come
to the spaces of ket- and bra-vectors, the operator $\hat1$ turns
into some ket-vector $|\,0\,\rangle$ and bra-vector
$\langle\,0\,|$, correspondingly. These ket- and bra-vectors
describe the state which is called {\it vacuum}.

\item
Let us introduce now the so called {\it scalar product} of bra-
and ket- vectors. For an arbitrary bra-vector  $\langle\,x\,|$ and
a ket-vector $|\,y\,\rangle$ this product is written as
$\langle\,x\,|\cdot|\,y\,\rangle$, or, for shortness,
$\langle\,x\,|\,y\,\rangle$. As a result we get a complex number,
i.~e. $\langle\,x\,|\,y\,\rangle \in \mathbb C$.

Let us require that the operation of scalar product would satisfy
the following properties:
\[
 \langle\,0\,|\,0\,\rangle = 1 \ ,
\]
\[
 \big(\,\lambda\,\langle\,x\,|\,\big)\cdot|\,y\,\rangle =
               \lambda\cdot\langle\,x\,|\,y\,\rangle   \ ,
 \qquad
 \big(\,\langle\,x_1\,|+\langle\,x_2\,|\,\big)\cdot|\,y\,\rangle =
  \langle\,x_1\,|\,y\,\rangle+\langle\,x_2\,|\,y\,\rangle   \ ,
\]
\[
 \langle\,x\,|\cdot\big(\,\lambda\,|\,y\,\rangle\,\big) =
               \lambda\cdot\langle\,x\,|\,y\,\rangle   \ ,
 \qquad
 \langle\,x\,|\cdot\big(\,|\,y_1\,\rangle+|\,y_2\,\rangle\,\big) =
 \langle\,x\,|\,y_1\,\rangle+\langle\,x\,|\,y_2\,\rangle  \ ,
\]
\[
 \big(\,\langle\,x\,|\,\hat o\,\big)\cdot|\,y\,\rangle =
 \langle\,x\,|\,\cdot\big(\,\hat o\,|\,y\,\rangle\,\big)   \ .
\]
In the latter equality $\hat o\in\mathcal O$; this property allows
to write in such cases simply:
 $\langle\,x\,|\,\hat o\,|\,y\,\rangle$.

The given properties for the scalar product are defining, i.~e.
there is not more than one scalar product satisfying the given
properties.

\item
The scalar product is a Hermitian form in $\mathcal H$, i.~e.
always $\langle\,x\,|\,y\,\rangle=(\langle\,y\,|\,x\,\rangle)^*$.
If the scalar product turns out to be positive-definite, it
defines in $\mathcal H$ some topology. If we complete $\mathcal H$
with respect to this topology, it turns into a usual Hilbert
space. And operators appear to be defined on the corresponding
dense linear subspace in $\mathcal H$.

But the scalar product can be not positive-definite at all: it
depends on the Poisson bracket in $\mathcal C\oplus Z^*_{\mathbb
C}$ and the choice of the subspaces $Cr$ and $De$. The question
about definition of topology in this case for example of
electromagnetic field will be discussed in the
section~\ref{TFIPSEM}.

\end{enumerate}

So, in this section we have described how, if we have the Lie
algebra of observables of a classical field
 $\mathcal C\oplus Z^*_{\mathbb C}$, we construct the
algebra $\mathcal O$ of operators of the quantum field and the
space of states $\mathcal H$. This construction is called {\it
quantization} of a classical field.


\myparagraph{Graduation of algebra $\mathcal O$ and of space
$\mathcal H$. Connection with Fock construction.}
 Consider now in the algebra of operators the subspaces
 $\mathcal C\widehat~$, $Cr\widehat~$ and $De\widehat~$.
Let us call them subspaces of the {\it grade} $0$, $+1$ and $-1$,
correspondingly. Making all possible products of these operators,
let us assign to each such product a grade that is equal to the
sum of grades of factors. If two products have the same grade, let
us assign the same grade to their sum. It is easy to see that the
algebra of operators, considered as a linear space, can be
represented as a sum of its linear subspaces belonging to
different grades:
\[
 \mathcal O = \dots \oplus
   \mathcal O_{-2} \oplus
   \mathcal O_{-1} \oplus
   \mathcal O_0    \oplus
   \mathcal O_{+1} \oplus
   \mathcal O_{+2} \oplus \dots
\]
And if $\hat a\in\mathcal O_i$ and $\hat b\in\mathcal O_j$, then
$\hat a\,\hat b\in\mathcal O_{i+j}$. Shortly speaking, the algebra
$\mathcal O$ is graded.

This graduation is naturally transferred to the space of states
$\mathcal H$. And negative grades in the space $\mathcal H$
correspond to trivial subspaces. In other words, the space of
states $\mathcal H$ is positive-graded:
\begin{equation}
 \mathcal H =
   \mathcal H_0 \oplus
   \mathcal H_1 \oplus
   \mathcal H_2 \oplus \dots
\label{gradH}
\end{equation}

If a ket-vector belongs to one of the subspaces $\mathcal H_n$, we
say that there are $n$ particles in this state.

It is obvious also, that all this construction is possible also
for conjugate objects, and decomposition~(\ref{gradH}) looks
exactly the same for bra-vectors.

It is not out of place to mention here that, according to the
introduced definition, the number of particles is always
non-negative, and it does not depend on whether or not the scalar
product in $\mathcal H$ is positive-definite.

It easy to see also that subspaces $\mathcal H_i$ and $\mathcal
H_j$ are orthogonal with respect to the scalar product. In other
words, if $\langle\,a\,|\in\mathcal H_i$ and
$|\,b\,\rangle\in\mathcal H_j$ and $i\neq j$, then
$\langle\,a\,|\,b\,\rangle=0$. Taking it into account, we can
write the decomposition~(\ref{gradH}) in the form:
\begin{equation}
 \mathcal H =
   \mathcal H_0 \dot+
   \mathcal H_1 \dot+
   \mathcal H_2 \dot+ \dots
\label{OrthSum}
\end{equation}
In the case when the scalar product is positive-definite, the
decomposition~(\ref{OrthSum}) allows to establish connection with
the usual Fock construction~\cite{Fock1932}. So far as this is
quite simple, we will not discuss this connection in more details.


\myparagraph{Invariant quantization.} \label{invquant}
 In the section~\ref{quantization} quantization of a classical
field was described in ``internal'' notions of the invariant
Hamiltonian formalism. But for the choice of the subspaces $Cr$
and $De$ we have not suggested any concrete prescription: we have
just given some necessary conditions for the choice of these
subspaces. In practice it leads to the situation that for one
classical field we can get too many quantizations.

Let us suppose now that in the space $Z^*_{\mathbb C}$ there is a
linear symplectic representation of some group. Let us define then
{\it invariant quantization} by the requirement that subspaces
$Cr$ and $De$ are invariant with respect to the action of this
group. In other words, $Cr$ and $De$ are reducing subspaces of
this representation.

When we consider relativistic fields, as a main group of
invariance we have Poincare group~$\mathcal P$. A quantization,
invariant with respect to the Poincare group we will call
$\mathcal P$-invariant, for shortness. In the paper~[IV] we have
already seen, how the Poincare group acts in the space
$Z^*_{\mathbb C}$, and how its representations are reduced.

Quantization of relativistic fields we will consider a little
later, but now we will consider quantization of the harmonic
oscillator. Its symmetry group is just the one-parameter group of
time shifts.


\myparagraph{Quantization of harmonic oscillator.} \label{harmosc}
 Consider harmonic oscillator. It is described by the Lagrangian:
\begin{equation}
 \textstyle
 L = \frac12 \dot\varphi^2 - \frac{m^2}2\varphi^2    \ .
\label{osclag}
\end{equation}
Here $\varphi(x)$ is a real function of one real argument (time).

The equation of the motion is:
\[
 (\partial^2 + m^2)\,\varphi = 0   \ .
\]

The invariant phase space $Z$ is a two-dimensional real space. The
symplectic structure on it is given  by the formula:
\[
 \omega = \dot\varphi(t) \wedge \varphi(t)    \ .
\]
Here forms $\dot\varphi$ and $\varphi$ are taken in the same
arbitrary moment of time $t$.

So far as the Lagrangian~(\ref{osclag}) is invariant with respect
to time shifts, in the space $Z$ we have the action of the
additive group $\mathbb R$. The field representation
 $Z^*_{\mathbb C}$ in this case, like in the case of
relativistic fields, decomposes into the direct sum of positive-
and negative-frequency subrepresentations:
\[
 Z^*_{\mathbb C} = Z^{*\,(+)}_{\mathbb C} \oplus
                   Z^{*\,(-)}_{\mathbb C} \ .
\]
As a reducing basis let us take the following two elements:
\[
 \textstyle
 a = \frac1{\sqrt{2m}}\,(m\varphi(0) + i\dot\varphi(0)) \ ,
 \qquad
 a^* = \frac1{\sqrt{2m}}\,(m\varphi(0) - i\dot\varphi(0)) \ .
\]
Really, so far as we have decomposition
 $\varphi(t) =
  \frac1{\sqrt{2m}}(a\,e^{-imt} + a^*e^{+imt})$,
we get that $a\in Z^{*\,(+)}_{\mathbb C}$,
 $a^*\in Z^{*\,(-)}_{\mathbb C}$.

Symplectic structure can be written by the forms $a$ and $a^*$ in
the following way:
\[
 \omega = i\,a^*\wedge a  \ .
\]
The Poisson bracket of the corresponding linear functions is:
\[
 \{\,a\,,\,a^*\,\} = -i \ .
\]

So, in accordance with what we have said in the
section~\ref{invquant}, we have exactly two invariant
quantizations: either $Cr=Z^{*\,(-)}_{\mathbb C}$ and
$De=Z^{*\,(+)}_{\mathbb C}$, or $Cr=Z^{*\,(+)}_{\mathbb C}$ and
$De=Z^{*\,(-)}_{\mathbb C}$. It is easy to see that the first
quantization leads to positive-definite scalar product. In the
case of second quantization scalar product turns out to be
indefinite. And in the decomposition~(\ref{OrthSum}) on the
subspaces with even number of particles the scalar product is
positive-definite, and on subspaces with odd~--- negative.


\myparagraph{Connection of invariant quantizations of field and of
field oscillator.} \label{QuantFOsc}
 A field oscillator, as a finite-dimensional system, is more
convenient for investigation by algebraic means. On the other
hand, there is a close connection between quantization of a field
and quantization of its oscillator.

Really, it is easy to see that the choice of
 $\mathbb R\times\mathcal L_{k^{(0)}}$-invariant creation and
destruction subspaces for oscillator and the choice of
 $\mathcal P$-invariant creation and destruction subspaces
for field are connected by the operation of induction. So, there
is one-to-one correspondence between
 $\mathbb R\times\mathcal L_{k^{(0)}}$-invariant quantizations
of oscillator and $\mathcal P$-invariant quantizations of field.

Furthermore, corresponding quantizations of oscillator and field
have in the spaces of states scalar products either
positive-definite or indefinite simultaneously. So, we get the
convenient criterion of sign-definiteness of scalar product.


\myparagraph{Quantization of scalar field.}
 In the paper [IV] we have shown that with respect to the action
of Poincare group the space of linear observables of scalar field
$Z^*_{\mathbb C}$ decomposes into the direct sum of the two
irreducible subspaces:
 $Z^*_{\mathbb C}=Z^{*\,(+)}_{\mathbb C}\oplus Z^{*\,(-)}_{\mathbb C}$.
So, there are exactly two possibilities: either we accept
$Cr=Z^{*\,(-)}_{\mathbb C}$ and $De=Z^{*\,(+)}_{\mathbb C}$, or we
accept $Cr=Z^{*\,(+)}_{\mathbb C}$ and
 $De=Z^{*\,(-)}_{\mathbb C}$. According to the formulas for
Poisson brackets of corresponding functions, obtained in the paper
[IV], both decompositions satisfy all requirements of the
section~\ref{quantization}.

So, the scalar field allows exactly two $\mathcal P$-invariant
quantizations. The field oscillator in this case is just the usual
one-dimensional real oscillator. Using results of the
section~\ref{harmosc} and the criterion from the
section~\ref{QuantFOsc} we get that for the first quantization the
scalar product in the space $\mathcal H$ turns out to be
positive-definite, and for the second quantization~- indefinite.

The first quantization is, in fact, conventional. Nevertheless, it
should not be thought that quantization with indefinite metric is
senseless, because it leads to ``negative probabilities''. In
principle, we can not reject a usefulness of such a theory where
scattering states of the field constitute a subspace where the
scalar product is positive-definite.


\myparagraph{Quantization of electromagnetic field.}
 Consider now quantization of electromagnetic field. Under
electromagnetic field we will imply here ``non-physical''
electromagnetic field [I].

As we have explained in the paper [IV], with respect to the action
of the Poincare group $\mathcal P$ the space $Z^*_{\mathbb C}$ in
this case, like in the case of the scalar field, decomposes into
the direct sum of the two indecomposable subspaces. Therefore,
like in the case of the scalar field, we have exactly two
possibilities: either $Cr=Z^{*\,(-)}_{\mathbb C}$ and
 $De=Z^{*\,(+)}_{\mathbb C}$, or $Cr=Z^{*\,(+)}_{\mathbb C}$ and
$De=Z^{*\,(-)}_{\mathbb C}$.

Using the criterion from the section~\ref{QuantFOsc}, we easily
see that both quantizations lead to indefinite metric. Later on we
will consider only the first of the two quantizations, because
this quantization has useful physical interpretation.

The indefiniteness of the scalar product leads, of course, to
difficulties with probability interpretation.

In the paper [I] we have pointed out that scattering states of the
classical electromagnetic field satisfy additional condition~---
the Lorentz condition. It is natural to suppose that in the
quantum theory there is some analog of this condition. It is
important to emphasize, that this condition must appear in the
quantum theory not as a new postulate: it must be derivable from
the dynamical laws, like we have done it for the classical field.
Unfortunately, we do not have a satisfactory theory of interacting
fields yet. Because of this reason, we will not give here any
derivation\footnote{A formal derivation, of course, was present
even in~\cite{Bleuler1950}. On the other hand, it is possible to
appeal to Feynman rules, but this rules need grounding from the
point of view of this paper.}. The desired condition is:
\begin{equation}
 -\,i\,k_\mu\,\hat a_\mu^{(+)}(k)\,|\,\mathrm{rad}\,\rangle = 0 \ .
\label{Guptacond}
\end{equation}
So, a vector of state of radiated field satisfies this equality
for any~$k$.

The given condition looks the same like one of versions of this
condition in the classical theory. Nevertheless, it should be
noticed that it can not be written as
 $-\,i\,k_\mu\,\hat a_\mu^{(-)}(k)\,|\,\mathrm{rad}\,\rangle = 0$
or as $-\,i\,k_\mu\,\hat a_\mu(k)\,|\,\mathrm{rad}\,\rangle = 0$,
because even the vacuum does not satisfy these
conditions\footnote{In coordinate representation the condition
 $-\,i\,k_\mu\,\hat a_\mu(k)\,|\,\mathrm{rad}\,\rangle = 0$
looks as
 $\partial_\mu\hat A_\mu(x)\,|\,\mathrm{rad}\,\rangle = 0$.
In literature even up to now there is no unity of opinions on
whether or not it is possible to formulate the additional
condition in this way. In this way, on the grounding of analogy
with the classical field, it was formulated even by Fermi. It was
pointed out in the papers \cite{Ma1949,Belinf1949} that such an
additional condition leads to difficulties with normalizability of
states in the quantum case. In the paper \cite{Gupta1950} it was
substituted with
 $\partial_\mu\,\hat A_\mu^{(+)}(x)\,|\,\mathrm{rad}\,\rangle = 0$.
Later in literature some authors used condition
 $\partial_\mu\,\hat A_\mu(x)\,|\,\mathrm{rad}\,\rangle = 0$
\cite{Thirr1964,Umezawa1958,Dirac1979}. Other \cite{BogShir1984}
pointed that such a condition can not be applied, because there
appears some contradiction with commutation relations. The other
authors \cite{Prokh1988} claimed that such a condition can be
used, but special care is needed, because allowed vectors of
states turn out to be not-normalizable. These differences of
opinion have quite deep reasons:

\begin{enumerate}
\item No exact meaning was put into the term ``quantization''.
In fact, quantized electromagnetic field was studied by formal
manipulations with algebraic symbols, but it was not given any
constructive definition of this object. In such a situation the
additional condition in any form is not free from criticism.
\item There is no satisfactory formulation of the theory
of interacting fields (even in the frame of perturbation theory).
And the Feynman rules do not have any clear connection with the
operatoral formalism, and operatoral formalism turns out to be
separated from practical calculations (from scattering theory,
first of all).
\end{enumerate}

With regards to the first remark, I believe, the relation
 $\partial_\mu\,\hat A_\mu(x)\,|\,\mathrm{rad}\,\rangle = 0$
does not contradict directly to the commutation relations. It is
possible to say only that the construction of quantization
presented in this paper is badly co-ordinated with such a
condition.

The second remark is, of course, more important. When we
considered the dynamics of the classical field, it was pointed out
that the additional condition for scattering states automatically
follows from dynamics, and it is not an arbitrary postulate. We
have also clearly shown the role of the ``non-physical'' degrees
of freedom. And though in this paper we still do not formulate a
theory of quantum interacting fields, nevertheless, arguments
based on analogy with the classical theory give us reasons to
think that construction of theory with condition
 $\partial_\mu\,A_\mu(x)\,|\,\mathrm{rad}\,\rangle = 0$ is not a
realistic task.}.

Now we will show that the condition~(\ref{Guptacond}) leads to
positivity of scalar product.


\myparagraph{Positivity of scalar product of electromagnetic
field.}\footnote{The contents of this section does not have any
deep connection neither with constructivity of quantization, nor
with algebraic, nor with topological questions discussed in the
present papers. The only reason why we discuss here this old (and
very simple) question is that in all sources that I know an
erroneous proof is given (for some reason people think that
positivity of a quadratic form on vectors of some basis
necessarily leads to positivity of the form in general).}
 Consider the field oscillator of electromagnetic field. Let us
write here, for shortness, $a_\mu$ instead of $\hat a_\mu(+1)$ and
$a^*_\mu$ instead of $\hat a_\mu(-1)$. These operators satisfy
relations:
\[
 [\,a^*_\mu\,,\,a_\nu\,]\,= g_{\mu\nu}, \qquad
 [\,a^*_\mu\,,\,a^*_\nu\,]=0, \qquad
 [\,a_\mu\,,\,a_\nu\,]=0 \ .
\]
For the quantization under consideration we also have:
\[
 a_\mu\,|\,0\,\rangle=0 \ .
\]
The additional condition takes the form:
\begin{equation}
 k^{(0)}_\mu a_\mu\,|\,\mathrm{rad}\,\rangle = 0 \ ,
\label{OscAddCond}
\end{equation}
here $k^{(0)}_\mu$ is a fixed vector on the light cone. The
subspace of the vectors of states satisfying this condition we
will denote $\mathcal H^\mathrm{rad}$.

So far as the operator $k^{(0)}_\mu a_\mu$ has a definite grade
(its grade is equal to $-1$), the graduation of the space of
states $\mathcal H$ is transferred to the space
 $\mathcal H^\mathrm{rad}$:
\begin{equation}
 \mathcal H^\mathrm{rad}=
   \mathcal H^\mathrm{rad}_0 \dot+
   \mathcal H^\mathrm{rad}_1 \dot+
   \mathcal H^\mathrm{rad}_2 \dot+ \dots
\label{HradSum}
\end{equation}
In other words, the subspace, defined by the additional condition
(\ref{OscAddCond}), also decomposes into the orthogonal sum of
states with definite number of particles.

So far as the sum (\ref{HradSum}) is orthogonal, it is enough to
prove the non-negativity of the scalar product for each of the
subspaces $\mathcal H^\mathrm{rad}_n$.

An arbitrary state vector $|\,n\,\rangle$ from the subspace
 $\mathcal H^\mathrm{rad}_n$ can be represented in the form:
\[
 |\,n\,\rangle = T_{\mu\nu\dots\rho} \,
                 a^*_\mu\,a^*_\nu\dots a^*_\rho \,
                 |\,0\,\rangle    \ ,
\]
here $T_{\mu\nu\dots\rho}$ is an $n$-valent tensor. This tensor
can be supposed to be symmetrical, without loss of generality.

The condition (\ref{OscAddCond}) for the vector $|\,n\,\rangle$
turns out to be an additional condition for the tensor
$T_{\mu\nu\dots\rho}$:
\begin{equation}
 k^{(0)}_\mu T_{\mu\nu\dots\rho} = 0 \ .
\label{TOscAddCond}
\end{equation}

The scalar product for the vector $|\,n\,\rangle$ multiplied by
itself is:
\begin{equation}
 \langle\,n\,|\,n\,\rangle =
      (-1)^n\,n!\,\cdot\,
      T^*_{\mu\nu\dots\rho}\,T_{\mu\nu\dots\rho} \ .
\label{Tsum}
\end{equation}

The value
$(-1)^n\,n!\,\cdot\,T^*_{\mu\nu\dots\rho}\,T_{\mu\nu\dots\rho}$ is
a sum of positive numbers, some of them are included in the sum
with the sign ``plus'', and some of them with the sign ``minus''.
This sign is defined by the relativistic summation rule by
repeated indexes and by the factor $(-1)^n$. Let us show that this
sum is non-negative under condition (\ref{TOscAddCond}).

So far as the construction of the space of states is invariant
with respect to the choice of the vector $k^{(0)}_\mu$, this
vector can be supposed to be equal:
\begin{equation}
 k^{(0)}_\mu =
 \left(\begin{array}{c|ccc}
       k_0 & 0 & 0 & k_0
       \end{array}\right)_\mu    \ .
\label{kChoice}
\end{equation}

Consider now  in the sum (\ref{Tsum}) all addends of the type
$(-1)^n\,n!\,\cdot\,T^*_{0\nu\dots\rho}\,T_{0\nu\dots\rho}$.
Because of (\ref{TOscAddCond}) and (\ref{kChoice}), they are
completely neutralized by the addends of the type
$(-1)^n\,n!\,\cdot\,T^*_{3\nu\dots\rho}\,T_{3\nu\dots\rho}$.

Among remaining addends let us consider addends of the type
$(-1)^n\,n!\,\cdot\,T^*_{\mu0\dots\rho}\,T_{\mu0\dots\rho}$. They
are completely neutralized with the remaining addends of the type
$(-1)^n\,n!\,\cdot\,T^*_{\mu3\dots\rho}\,T_{\mu3\dots\rho}$.

Et cetera. Finally, in the sum (\ref{Tsum}) there remain only
addends that have indexes $1$ or $2$. All these addends are
included in the sum (\ref{Tsum}) with the sign ``plus''.


\myparagraph{Topology and completeness of invariant phase space of
electromagnetic field.} \label{TFIPSEM}
 When we studied the invariant Hamiltonian formalism, at the beginning
we included into the invariant phase space $Z$ only solutions of
the equations of the motion that are smooth in coordinate
representation and finite in space direction. But the question
about what topology should be really introduced in the space $Z$
we have not discussed at all. In fact, we have not given even an
exact definition of the space $Z^*$; we have not clarified, how
exactly the functional spaces $Z$ and $Z^*$ turn out to be
isomorphic; an exact definition of the Poisson brackets on $Z^*$
was not given also. In this section we will show, that in the
space $Z$ we can introduce such a topology (and to complete it
with respect to this topology), that it becomes very similar to
the Hilbert space. And all necessary for invariant Hamiltonian
formalism properties of this topology will be satisfied.

A symplectic structure itself does not define a topology. But
between elements of the classical phase space $Z$ and one-particle
states of a quantized field there exist one-to-one correspondence:
\begin{equation}
 \underline c \ \longleftrightarrow\
 (I^{-1}\underline c)\widehat~ \ |\,0\,\rangle \ .
\label{ZHisom}
\end{equation}

Consider now the scalar field as an example. Among its invariant
quantizations there is a quantization with positive-definite
scalar product. The correspondence (\ref{ZHisom}), firstly,
introduces a complex structure in the space $Z$, and secondly,
transfers there the scalar product from the one-particle quantum
space $\mathcal H_1$. So, $Z$ has natural structure of complex
Hilbert space\footnote{The obtained topology is the most natural
for quantization. Bohr and Rosenfeld \cite{BohrRos1933} noticed
that for consideration of the quantized electromagnetic field it
is necessary to perform an averaging of the field in a small
space-time volume: the symbol $\hat A_\mu(x)$ itself is not a
physical value. Our consideration shows that this statement is in
the same way true for the classical field also, if its phase space
is provided with the ``proper'' topology.}.

The symplectic structure turns out to be continuous with respect
to the pair of its arguments with respect to the obtained
topology; the symplectic structure defines the isomorphism of the
spaces $Z$ and $Z^*$; the Poisson brackets are correctly defined
on the whole $Z^*$.

Let us represent now the scalar field in the Fourier
representation in the form:
\[
 \widetilde\varphi(k) = 2\pi\,\delta(k^2-m^2)\cdot a(k)  \ .
\]
The correspondence (\ref{ZHisom}) can be written more manifestly
as:
\[
 \underline c \ \longleftrightarrow\
 \int d\mu^+_m \cdot i\,a(k)^{\,\underline c} \cdot
               \hat a^*(k) \ |\,0\,\rangle \ ,
 \qquad \mbox{here} \ \
 d\mu^+_m = \frac{d^4\!k}{(2\pi)^4} \cdot
            2\pi\,\delta(k^2-m^2) \cdot
            \theta(k) \ .
\]
And the scalar product in the space $Z$ takes the form:
\begin{equation}
 \langle\,\underline c\,,\,\underline d\,\rangle =
 \int d\mu^+_m \cdot
 a^*(k)^{\,\underline c} \cdot
 a(k)^{\,\underline d} \ .
\label{ScalScalProd}
\end{equation}

The one-to-one transference of this scheme to the case of
electromagnetic field appears to be impossible, because, as we
have seen, among $\mathcal P$-invariant quantizations of
electromagnetic field there are  no quantizations with
positive-definite scalar product.

But here is another opportunity. Let us represent electromagnetic
field in the Fourier representation as
\[
\widetilde A_\mu(k) = 2\pi\,\delta(k^2) \cdot a_\mu(k)  \ .
\]
By analogy with the formula (\ref{ScalScalProd}) let us introduce
in the classical phase space $Z$ of electromagnetic field a scalar
product by the formula:
\begin{equation}
 \langle\,\underline c\,,\,\underline d\,\rangle =
 \int d\mu^+_m \cdot
 M_{\nu\rho} \cdot
 a^*_\nu(k)^{\,\underline c} \cdot
 a_\rho(k)^{\,\underline d} \ .
\label{EMScalProd}
\end{equation}
Here $M_{\nu\rho}$ is an arbitrary positive-definite Hermitian
matrix. In the capacity of such a matrix we can take, for example,
$M_{\nu\rho}=\mathrm{diag}\,(+1,+1,+1,+1)_{\nu\rho}$.

The scalar product (\ref{EMScalProd}) is positive-definite.
Therefore it defines some topology in the space $Z$. The important
is the fact that this topology does not depend on the concrete
choice of matrix $M_{\nu\rho}$.

From this it follows that the introduced topology is
relativistic-invariant.

With respect to the introduced topology the symplectic structure
turns out to be continuous with respect to the pair of its
arguments; the symplectic structure defines an isomorphism of the
spaces $Z$ and $Z^*$; the Poisson brackets turn out to be
correctly defined on the whole $Z^*$.

So, the invariant phase space of electromagnetic field has natural
structure of linear complex topological space, and the topology
there is defined by a set of equivalent (in the sense of topology)
scalar products. Such spaces we will call {\it spaces of Hilbert
type}.


\myparagraph{Tensor product of spaces of Hilbert type.}
 Now we will show that tensor product of spaces of Hilbert
type is a space of Hilbert type.

Consider two such spaces $X$ and $Y$. Their tensor product (more
exactly, algebraic tensor product) is defined as follows.

Consider all formal products of the type $x\diamond y$, where
$x\in X$ and $y\in Y$. We will call such products pairs. Let us
suppose that pairs can be formally multiplied by complex numbers
forming expressions of the type: $\lambda\cdot x\diamond y$. And
consider all formal sums of the type:
\[
 \lambda_1\cdot x_1\diamond y_1 +
 \lambda_2\cdot x_2\diamond y_2 +
 \dots +
 \lambda_n\cdot x_n\diamond y_n     \ .
\]
Let us consider two such sums to be equivalent, if they have the
same sums of factors for each pair. So, we come to the complex
vector space that we denote as  $X\diamond Y$ (this is a {\it free
$\mathbb C$-module generated by the Cartesian product
 $X\times Y$}).

In the constructed space consider elements of the type:
\[
 1\cdot (\lambda x)\diamond y - \lambda \cdot x\diamond y \ ,
\]
\[
 1\cdot x\diamond (\lambda y) - \lambda \cdot x\diamond y \ ,
\]
\[
 1\cdot (x_1+x_2)\diamond y -
   1\cdot x_1\diamond y - 1\cdot x_2\diamond y \ ,
\]
\[
 1\cdot x\diamond (y_1+y_2) -
   1\cdot x\diamond y_1 - 1\cdot x\diamond y_2  \ .
\]
The linear shell of these elements we denote as $X\circ Y$.

Factorizing $X\diamond Y$ with respect to $X\circ Y$ we come to
the tensor product:
\[
 X\otimes Y = (X\diamond Y)/(X\circ Y) \ .
\]

For shortness, let us introduce the notation $x\otimes y$. Namely,
let us suppose that under the defined factorization an element
$1\cdot x\diamond y$ of the space $X\diamond Y$ transfers into the
element $x\otimes y$ of the space $X\otimes Y$.

If $X$ and $Y$ are usual Hilbert spaces, then, as it is known,
their tensor product $X\otimes Y$ has natural structure of Hilbert
space. The scalar product in $X\otimes Y$ in this case at the
beginning is defined for pairs by the formula:
\begin{equation}
 \langle\, x_1\otimes y_1\,,
        \, x_2\otimes y_2\,\rangle_{X\otimes Y}=
 \langle\, x_1\,,\, x_2\,\rangle_X \cdot
 \langle\, y_1\,,\, y_2\,\rangle_Y            \ ,
\label{ScalProdInTensProd}
\end{equation}
and for other elements it is spread by the requirement of
linearity with respect to the second argument and anti-linearity
with respect to the first. Performing completion of $X\otimes Y$
with respect to the given scalar product we come to Hilbert space.

In the case of spaces of Hilbert type each of the spaces $X$ and
$Y$ has many equivalent scalar products. The space $X\otimes Y$,
in accordance with the given scheme, gets many scalar products.
Let us show that all these scalar products are equivalent.

So, suppose that in one of the two spaces, for example in $X$, we
change the scalar product $\langle\,\cdot\,,\,\cdot\,\rangle_X$ to
equivalent scalar product $(\,\cdot\,,\,\cdot\,)_X$. And the
scalar product $\langle\,\cdot\,,\,\cdot\,\rangle_{X\otimes Y}$ in
the space $X\otimes Y$ changes to
 $(\,\cdot\,,\,\cdot\,)_{X\otimes Y}$.

Consider some element of the space $X\otimes Y$:
\[
 z=x_1\otimes y_1 + x_2\otimes y_2 + \dots + x_n\otimes y_n \ .
\]
It is easy to see that this element can be represented in such a
form that in this sum all $y_1,y_2,\dots,y_n$ are orthogonal to
each other with respect to the scalar product in $Y$. Then the
scalar products of this element with itself take especially simple
form:
\begin{equation}
 \langle\,z\,,\,z\,\rangle_{X\otimes Y} =
 \langle\,x_1\,,\,x_1\,\rangle_X \cdot
         \langle\,y_1\,,\,y_1\,\rangle_Y +
 \langle\,x_2\,,\,x_2\,\rangle_X \cdot
         \langle\,y_2\,,\,y_2\,\rangle_Y + \dots +
 \langle\,x_n\,,\,x_n\,\rangle_X \cdot
         \langle\,y_n\,,\,y_n\,\rangle_Y  \ ,
\label{ZSqrAng}
\end{equation}
\begin{equation}
 (\,z\,,\,z\,)_{X\otimes Y} =
 (\,x_1\,,\,x_1\,)_X \cdot \langle\,y_1\,,\,y_1\,\rangle_Y +
 (\,x_2\,,\,x_2\,)_X \cdot \langle\,y_2\,,\,y_2\,\rangle_Y +\dots +
 (\,x_n\,,\,x_n\,)_X \cdot \langle\,y_n\,,\,y_n\,\rangle_Y  \ .
\label{ZSqrBra}
\end{equation}

A convenient necessary and sufficient criterion of equivalence of
scalar products $\langle\,\cdot\,,\,\cdot\,\rangle_{X\otimes Y}$
and $(\,\cdot\,,\,\cdot\,)_{X\otimes Y}$ is the following: there
exists such $\varepsilon\in\mathbb R$, $\varepsilon>0$, that for
any $z\in X\otimes Y$ we have the following conditions:
\[
 \varepsilon\cdot(\,z\,,\,z\,)_{X\otimes Y} <
 \langle\,z\,,\,z\,\rangle_{X\otimes Y} \ , \qquad
 \varepsilon\cdot\langle\,z\,,\,z\,\rangle_{X\otimes Y} <
 (\,z\,,\,z\,)_{X\otimes Y} \ .
\]

But if such $\varepsilon$ exists for the scalar products
$\langle\,\cdot\,,\,\cdot\,\rangle_X$ and
$(\,\cdot\,,\,\cdot\,)_X$, then in accordance with formulas
(\ref{ZSqrAng}) and (\ref{ZSqrBra}), it is appropriate also for
the scalar products
 $\langle\,\cdot\,,\,\cdot\,\rangle_{X\otimes Y}$ and
 $(\,\cdot\,,\,\cdot\,)_{X\otimes Y}$.

So, all scalar products in $X\otimes Y$ are equivalent. Performing
completion of $X\otimes Y$ with respect to the topology defined by
these scalar products we come to the space of Hilbert type.


\myparagraph{Topology of space of states of quantized
electromagnetic field.}
 As we have seen, the space of states of quantized electromagnetic
field $\mathcal H$ decomposes into orthogonal sum of subspaces
with definite number of particles:
\begin{equation}
 \mathcal H =
   \mathcal H_0 \dot+
   \mathcal H_1 \dot+
   \mathcal H_2 \dot+ \dots
\label{PartDecompEMF}
\end{equation}

The space $\mathcal H_0$ is one-dimensional. So, the question
about its topology does not arise.

The subspace $\mathcal H_1$, as it was shown in the
section~\ref{TFIPSEM}, can be in fact identified with the phase
space of the classical field $Z$. It is the space of Hilbert type.
So, the question about its topology is also solved.

Consider now the tensor product of the space $Z$ with itself:
$Z\otimes Z$. There we have naturally defined action of the group
of transpositions of two elements. The simplest elements are
transformed under action of this group as
 $\underline a\otimes\underline b\to
  \underline b\otimes\underline a$, and to other elements this
action is spread by linearity.

Let us choose among the scalar products in the space $Z\otimes Z$
only those invariant with respect to the action of the group under
consideration. In practice it is convenient just to restrict
ourself with the products for which in the right part of the
formula~(\ref{ScalProdInTensProd}) we have product of the same
scalar products.

Then the group of transpositions acts unitarily\footnote{Under
unitary transformation of a space of Hilbert type we imply an
automorphism of this space, i.~e. one-to-one transformation on
itself preserving linear structure and each of scalar products.}
in $Z\otimes Z$. And $Z\otimes Z$ decomposes into the orthogonal
sum of two invariant subspaces: symmetric and anti-symmetric.

Each of these two subspaces inherits topology from $Z\otimes Z$.

It is easy to see that the two-particle quantum space
 $\mathcal H_2$ can be naturally identified with the symmetric
subspace in $Z\otimes Z$ and therefore it gets corresponding
topology.

Et cetera. $n$-particle subspace $\mathcal H_n$ is identified with
fully symmetric subspace of $n$-th tensor power of $Z$. And
inherits topology from there.

So, each of the subspaces in the orthogonal sum
(\ref{PartDecompEMF}) is a space of Hilbert type. From practical
point of view it is very convenient, because for practical
applications it is much easier to define the described spaces
using bases (and not as factor-spaces of free modules).

Consider now the space $\mathcal H$ as a whole. Using topologies
introduced in its subspaces we can introduce different topologies
in the whole space. For example, convergence of a directed set to
zero can be understood as independent convergence to zero of all
projections of this set. Such a topology seems to be the most
natural in this case.

On the other hand, this requirement can be strengthened if we
require additionally, for example, that starting from some moment
only finite number of projections of the directed set differ from
zero.

So, there are many natural topologies in $\mathcal H$. And the
space $\mathcal H$ is not a space of Hilbert type.

In this connection I want to pay attention to the following. In
the paper of Gupta~\cite{Gupta1950} it was suggested to quantize
electromagnetic field in usual Hilbert space with
positive-definite metric. In fact, it is possible to do so
considering quantizations invariant with respect to a more narrow
group than the Poincare group (namely, with respect to the
subgroup of the Poincare group that leaves the direction of the
time axis unchanged). The subspaces with fixed number of particles
turn out to be the same, in fact, that we constructed above. But
the whole space $\mathcal H$ acquires such a topology that is not
relativistic-invariant. And it turns out that some states of the
field belong to the Hilbert space in one frame of reference and do
not belong in other. So, the old Gupta formalism is not
relativistic-invariant, even implicitly.

In his later papers Gupta tried to refuse from forced introduction
of a sign-definite metric. But his last paper on this
topic~\cite{Gupta1959} clearly showed that he still had no
apparently relativistic-invariant construction of quantized
electromagnetic field (to say nothing of the problems with
functional analysis).


\myparagraph{About origin of anti-unitary transformations.}
 In accordance with the introduced procedure of quantization,
linear transformations of observables of a classical field that
preserve Poisson brackets and the subspaces $Cr$ and $De$ generate
unitary transformations of states of the quantized field. But it
is known that in quantum theory an important role is played also
by anti-unitary symmetry transformations. Consider their origin on
the example of the operation of time reversal~--- $T$.

Suppose that the Lagrangian of a classical field is $T$-invariant
(as examples we can use scalar and electromagnetic fields). Then
the action turns out to be $T$-invariant also. Therefore, field
functions satisfying the stationary action principle under the
operation of time reversal transfer into functions that also
satisfy the stationary action principle.

So, from the point of view of invariant Hamiltonian formalism, the
operation of time reversal is a one-to-one transformation of the
phase space $Z$ on itself. And how does the symplectic structure
change? It is obvious from the variational definition of
symplectic structure that it changes sign. It seems to be natural
to call such transformations, that change sign of symplectic
structure, {\it anti-symplectic}.

Consider now conjugate action of the operation of time reversal on
elements of the conjugate space $Z^*_{\mathbb C}$:
\[
 a \stackrel T{\longrightarrow} a_T \ , \qquad
 a,a_T \in Z^*_{\mathbb C} \ .
\]
The Poisson bracket changes sign under such a transformation:
\[
 \{\,a_T\,,\,b_T\,\} = - \,\{\,a\,,\,b\,\} \ .
\]
So, the operation of time reversal generates the automorphism of
the Lie algebra $\mathcal C \oplus Z^*_{\mathbb C}$.

In accordance with the procedure of quantization described in this
paper, it is natural to wish to construct corresponding
automorphism of the algebra of operators $\mathcal O$ for the
quantized field. But it is impossible to do it. The reason is that
for construction of the algebra of operators we used not only the
structure of the Lie algebra of the set
 $\mathcal C \oplus Z^*_{\mathbb C}$ but also the special
correspondence between the subspace $\mathcal C$ and the set of
scalars $\mathbb C$ (the relation $\hat1 = ()\widehat~$). It means
that for bringing of the operation of time reversal to the quantum
case we need some additional conventions, besides operation of
quantization that we have already introduced.

Let us agree that words (i.~e. elements of the semi-group
$\mathcal W$) under time reversal are transformed in accordance
with the following formula:
\[
 \hat a\,\hat b \dots \hat c \stackrel T{\longrightarrow}
 \hat c_T \dots \hat b_T\,\hat a_T \ , \qquad
 \hat a,\hat b,\dots,\hat c,\hat a_T,\hat b_T,\dots,\hat c_T
 \in\mathcal A \ ,
\]
i.~e. every letter is substituted by the corresponding, and after
that letters are written in the reverse order. It is obvious that
time reversal acts on a product of two words in the following way:
\[
 \hat p\,\hat q \stackrel T{\longrightarrow} \hat q_T\,\hat p_T \ ,
 \qquad \hat p,\hat q,\hat p_T,\hat q_T \in \mathcal W \ .
\]
A one-to-one transformation of a semi-group on itself that
satisfies such a property can be called an {\it
anti-auto\-mor\-phism}.

The constructed anti-automorphism of the semi-group of words
$\mathcal W$ by linearity spreads to anti-automorphism of the
algebra of phrases $\mathcal P$. Furthermore, it is easy to check
that this anti-automorphism of the algebra of phrases generates an
anti-automorphism of the algebra of operators $\mathcal O$.

Suppose now that under time reversal creation subspace $Cr$ and
destruction $De$ transfer to one another. This condition is
usually satisfied because these subspaces are negative- and
positive-frequency, correspondingly. The left ideal spanned on
$De\widehat~$ under time reversal transfers into the right ideal
spanned on $Cr\widehat~$. The constructed by factorization left
module transfers into the corresponding right module.

So, the operation of time reversal defines the one-to-one linear
correspondence of ket- and bra-vectors\footnote{Schwinger
suggested~\cite{Schwin1992} to interpret ket-vectors as symbols
denoting creation of system in the past; bra-vectors as symbols
denoting destruction of system in future; and operators as action
of instruments during experiment. Such an interpretation turns out
to be quite natural here.}.

As it can be easily seen, the scalar product is preserved under
time reversal:
\begin{equation}
 \langle\,x\,|\,y\,\rangle=\langle\,y_T\,|\,x_T\,\rangle \ .
\label{TScale}
\end{equation}

So far as there is a one-to-one anti-linear correspondence between
ket- and bra-vectors, we could make all our discussion inside one
space, for example, the space of ket-vectors. Then the operation
of time reversal could be considered as one-to-one anti-linear
transformation of this space. The relation~(\ref{TScale}) shows
that under this transformation the scalar product changes to
complex-conjugate. A transformation satisfying such properties is
called {\it anti-unitary}.


\bigskip
At the end I want to thank V.~M.~Shabaev, L.~D.~Faddeev,
V.~A.~Franke, V.~D.~Lyakhovsky, L.~V.~Prokhorov, V.~V.~Vereschagin
and Yu.~M.~Pis'mak for fruitful discussions.



\end{document}


\newcounter{parnum}
\newlength{\picwidth}
\newcommand*{\myparagraph}[1]{\refstepcounter{parnum}%
\paragraph{\arabic{parnum}. #1}}


\title{On quantization of electromagnetic field.\\
       I. Classical electrodynamics.}
\author{D.~A.~Arbatsky\footnote{http:\dslash d-a-arbatsky.narod.ru/}}
\date{\today}
\maketitle

\begin{abstract}
A technique for investigation of classical fields is developed on
the base of invariant Hamiltonian formalism. Electromagnetic and
scalar fields are considered as particular examples of using the
general method. Poisson brackets for these fields are calculated.
The necessity of introduction of ``non-physical'' degrees of
freedom for electromagnetic field is explained.
\end{abstract}


\myparagraph{Introducing remarks.}
 This paper is the first in the set of six joined by the same
title. References to these papers will be given here in Roman
numbers: [I],~\dots,~[VI].

One of the causes for this research was the attempt to answer the
question: What is the structure of the space of states of
quantized electromagnetic field, considered from functional
analysis?

\begin{enumerate}

\item
When they describe Gupta-Bleuler quantization scheme modern
textbooks silently imply that the question about the topology of
quantum space of states of quantized electromagnetic field is
solved in full analogy with scalar field. And electromagnetic
field is quantized either in usual Hilbert space (such a
construction is not relativistic invariant, even in non-apparent
way [VI]) or in ``Hilbert space with indefinite metric'' (in such
an approach the question about the topology has not been
considered yet with all necessary mathematical strictness; in
fact, it turns out that the space of states has not been defined
constructively at all).

Anyway, founded on different analogies, it is usually supposed
that the space of states, from {\it topological} point of view,
{\it must} be a Hilbert space. From our research it will be clear,
that this point of view is erroneous. And this result is quite
general: it forces us to make a new look at quantization of
fields, even those fields that were quantized quite well up to now
(for example, the scalar field).

\item
It will be shown that the root of the problem is not in functional
analysis, but in algebra. The process of quantization in Fock
space was not satisfactorily described from {\it algebraic} point
of view yet. This is because, following Fock~\cite{Fock1932}, for
construction of quantized electromagnetic field people take as a
starting object the one-particle quantum space of states. But such
an algebraic structure is too poor: there is no algebraic process
that allows to construct quantized field, based on this structure
(Namely, using only structure of one-particle state, it is
impossible to introduce local field operators (see, for example,
\cite{Weinberg1996}, chapter~5)).

It will be clear from this research that an adequate algebraic
structure for construction of a quantized field is a classical
field, described with the use of the invariant Hamiltonian
formalism.

\item
After solving algebraic questions (i.~e. formulating satisfactory
quantization scheme) the solution of the question about topology
appears to be a quite ``working'' (non-fundamental) problem. It
appears to be important to realize that it is necessary for
constructing the space of states of electromagnetic field to
abandon to use Hilbert space. In the paper [VI] for solving this
problem we will introduce new types of functional spaces.

\item
There is no reason to discuss the question about relativistic
invariance of the suggested construction of quantized field. The
suggested construction is not only invariant: the requirement of
relativistic invariance is the part of the construction.

If we use the quantization scheme, given here, for quantizing the
harmonic oscillator, the requirement of invariance with respect to
some group substitutes (in some sense) the requirement of using
Hilbert space as a space of states. And we get two quantizations:
one in Hilbert space and one~--- in the space with indefinite
scalar product.

So, we do not even try to generalize Stone-von~Neumann theorem for
the case of relativistic fields: instead, we suggest completely
different conception of quantization, which works {\it
differently} even in the case of the harmonic oscillator.

\item
So far as our approach to quantization is consistently oriented to
explicit control of a symmetry (in the case of relativistic
fields~--- with respect to the Poincare group), it becomes
possible to see, how the action of a group of symmetry is
transferred from classical space of states to quantum. It will not
be an exaggeration to say that these are our papers where it is
for the first time explained (at the rigorous level) how a quantum
system gets integrals of motion of the same type that
corresponding classical system has (and these integrals are
connected with the group of symmetry).

\item
It is well known from the classical paper of Bohr and Rosenfeld
\cite{BohrRos1933} that some problems arise when we consider field
operators in a fixed point of space-time. In modern books
pretending to mathematical strictness it is said that we have in
this case an ``operator distribution''. But as far as I know no
pithy mathematical theory of such distributions (like theory of
generalized functions) was created. So, the term ``operator
distribution'' still only expresses the fact that symbols like
$\hat A_\mu(x)$ have {\it no} rigorous sense at all.

If we use the invariant Hamiltonian formalism as a base for
construction of quantum fields, we can see that this problem can
be investigated even in the classical mechanics, and much more
clearly.

From the formal point of view, it is possible to investigate this
question even in this paper. Especially as rigorous approach to
calculation of Poisson brackets requires discussion of the
topology of invariant phase space. Nevertheless, I prefer to
postpone this question to the last paper. This is because the
choice of topology for invariant phase space can be clearly and
simply {\it motivated} only after consideration relativistic
fields from the point of view of group theory [IV] and discussion
of quantization [VI].

\item
It is known that it is possible to add some divergence to the
Lagrangian of a field without changing the equations of the
motion. It is shown in the paper [II] that invariant Hamiltonian
description of a field is not changed in this case. So far as the
suggested scheme of quantization is totally based on the structure
of the invariant Hamiltonian formalism, obtained quantization does
not depend on this substitution of Lagrangian also.

\item
So far as for the base for construction of quantized fields we use
classical field, not one-particle quantum space, Wigner-Mackey
theory (about unitary representations of the Poincare group) does
not play a fundamental role for us. The analogy of this theory is
the theory of symplectic representations of Poincare group (in the
paper [IV] I tried to describe some foundations).

The theory of symplectic representations of Poincare group is
similar to the theory of unitary representations. But the example
of electromagnetic field shows that there is not full analogy
here.

\item
In our account we will orientate ourself mainly to linear real
Bose fields. The requirement of linearity here is of principle: we
cannot quantize non-linear fields. As to requirements of reality
and being Bose-field, these restriction are accepted here only to
simplify notations and formulations. ``Complexity'' of a field
just means the existence of additional complex structure. As to
Fermi fields, for their description we need to use Grassmanian
variables (see, for example,~\cite{OgiMez1975}). It is not
important for our consideration, whether we use usual or
Grassmanian numbers.

\end{enumerate}

In this paper we give classical description of electromagnetic
field. The main goal is to give an account of well-known facts in
a new language. However this approach allows us to look at some
questions completely differently: for example, the question about
energy of electrostatic field was treated incorrectly till now.


\myparagraph{Notations.}
 In tensor notations we will write vectors of Minkowski space
as $a_\mu$, $b_\nu$ etc. And we will use only {\it
contravariant\/} components of tensors; indexes numbering
components will be written always {\it down}. The frame of
reference will be always supposed to be orthonormal with respect
to the scalar product $g(\,\cdot\,,\,\cdot\,)$, so that metric
tensor has standard appearance:
 $g_{\mu\nu} = \mathrm{diag}\,(+1,-1,-1,-1)_{\mu\nu}$.
Repeating indexes always imply summation with due regard for
signs. For example, the scalar product will be written as
\[
 a\,b = a_\mu\, b_\mu = g_{\mu\nu}\, a_\mu\, b_\nu =
 a_0\, b_0 - a_1\, b_1 - a_2\, b_2 - a_3\, b_3 \ .
\]

The derivatives with respect to space-time coordinates will be
marked by the symbol $\partial_\mu$:
\[
 \partial_\mu =
   \left(\begin{array}{c|ccc}
          \frac\partial{\partial x_0} &
         -\frac\partial{\partial x_1} &
         -\frac\partial{\partial x_2} &
         -\frac\partial{\partial x_3}
   \end{array}\right)_\mu                   \ ,
\]
so, the index of a derivative, like all indexes, is contravariant.

For reducing of notations the symbol of antisymmetrization is
introduced: ${}^{[\mu\nu]}$. This symbol means that we make
antisymmetrization with respect to the nearest indexes~$\mu$
and~$\nu$ following after it. For example, the tensor of
electromagnetic field is written as
 $F_{\mu\nu}=\partial_\mu A_\nu - \partial_\nu A_\mu =
 {}^{[\mu\nu]}\,2\,\partial_\mu A_\nu$.

D'Alembert operator is given by formula:
$\partial^2=\partial_\mu\partial_\mu$.

The Fourier transformation everywhere looks like:
\begin{equation}
 \widetilde\varphi(k) = \int d^4\!x\,e^{+ikx}\,\varphi(x)  \ ,
 \qquad
 \varphi(x) = \int \frac{d^4\!k}{(2\pi)^4}\,e^{-ikx}\,
              \widetilde\varphi(k) \ .
\label{Fourier}
\end{equation}

The system of units is so that $\hbar=c=1$.


\myparagraph{Invariant Hamiltonian formalism.}
\label{InvHamForm}
 It was shown in papers~\cite{Witten1986,Crnkovic1987,Zucker1987}
that basic notions of the Hamiltonian formalism (namely, phase
space and symplectic structure on it) can be treated in a
relativistic invariant sense. In those papers formulas for
symplectic structure of the most important systems were obtained
(in coordinate representation). So, there was given a base of a
scheme that can be called an invariant Hamiltonian formalism. A
detailed account of this formalism see in original
reviews~\cite{CrWitt1987,Crnkovic1988}. The question of
equivalence of this approach and usual Hamiltonian mechanics for
the case of systems with constraints was discussed also
in~\cite{BHS1991}.

Nevertheless I will remind here some basic notions. This is
necessary in order to introduce our own notations.

Let field $\varphi_i(x)$ is described by the Lagrangian
$L(\varphi_i,\partial_\mu\varphi_i)$. Here we will suppose that
Lagrangian depends only on field values $\varphi_i$ and their
derivatives\footnote{It will be sufficient here to restrict
ourself with Lagrangians depending only on derivatives of the
first order. In general case we can consider also Lagrangians with
higher derivatives~\cite{BHS1991}.} and does not depend on the
point~$x$. Furthermore, we will consider only linear fields and
Lagrangian will be supposed to be a quadratic function.

The least-action principle leads us to Euler-Lagrange equations:
\begin{equation}
 \frac{\partial L}{\partial\,\varphi_i} -
 \partial_\mu
    \frac{\partial L}{\partial\,(\partial_\mu\varphi_i)} = 0 \ .
\label{EuLag}
\end{equation}

Invariant phase space~$Z$ is defined as the set of all
functions\footnote{More precisely, we will suppose for the present
that we consider only smooth functions. We will also suppose that
these functions differ from zero only on such a set, that
intersection of this set with any space-like plane is finite. In
paper~[VI] we will discuss the question about natural topology of
the space~$Z$.} $\varphi_i$ that satisfy equations~(\ref{EuLag}).
So long as we suppose the equation~(\ref{EuLag}) to be linear, the
space~$Z$ has natural linear structure.

The elements of the space~$Z$ we will denote by underlined sets of
symbols; for example, $\underline c\in Z$. Functions on this space
we will write as $f^{\,\underline c}$. For example, the value of
the function $\varphi$ corresponding to an element $\underline c$
in point $x$ is written as $\varphi(x)^{\,\underline c}$.

We will usually identify elements of tangent bundle~$TZ$ with
elements of the space~$Z$. If it is necessary to point out that
tangent vector $\underline c$ comes from a point $\underline b\in
Z$, we can write it as~$\underline{\smash{c\,[b]}}$. Elements of
cotangent bundle~$T^*\!Z$ will be similarly identified with the
elements of adjoint space~$Z^*$. At the same time, differentials
of linear functions will be identified with functions, i.~e.
instead of writing $df$ we write simply: $f$.

On the invariant phase space~$Z$, like on usual phase space, there
is a symplectic structure~$\omega$:
\begin{equation}
 \omega = \int\limits_\Sigma d\sigma_\mu\,\delta j_\mu(x) \ ,
 \qquad \mbox{here} \ \
 \delta j_\mu(x) =
       \frac{\partial L}{\partial\,(\partial_\mu\varphi_i)}
            \wedge \varphi_i  \ .
\label{symplstr}
\end{equation}
The integration is assumed over any space-like
hyper-surface~$\Sigma$, which behaves well enough in infinity.
2-form~$\delta j_\mu$ is called a symplectic current. So long as
the symplectic current is conserved, the result does not depend on
the choice of the surface~$\Sigma$.

The freedom in the choice of Lagrangian and non-uniqueness of the
symplectic structure will be considered in the paper~[II]. In
reasonable sense we can say that non-uniqueness in the choice of
Lagrangian does not influence the symplectic structure.

The symplectic structure~$\omega$, as usually~\cite{Arnold1989},
defines an isomorphism $I:T^*\!Z\to TZ$ of cotangent $T^*\!Z$ and
tangent $TZ$ bundles\footnote{Rigorously speaking, we can talk
about such an isomorphism only after discussing the topology of
the invariant phase space. Without such a discussion even the
notion of cotangent bundle does not have a clear sense. For the
reasons explained in ``Introducing remarks'' discussion of this
question for the case of relativistic fields we postpone to paper
[VI]. \par It should be also kept in mind that values like
$\varphi(x)$ that we use here appear to be not belonging to
cotangent bundle after we introduce the proper topology. This
shortcoming can be easily eliminated, but prefer just to ignore it
till discussion of the topology.}. It transfers any 1-form
$l^{\,\underline\cdot}\in T^*\!Z$ to tangent vector
$\underline{Il}\in TZ$, so that for any tangent vector $\underline
c\in TZ$ we have the equality:
\begin{equation}
 \omega^{\,\underline c\,;\,\underline{Il}} = l^{\,\underline c} \ .
\label{cotanisom}
\end{equation}

If a form $l^{\,\underline\cdot}$ is a differential of some
function $g^{\,\underline\cdot}$, i.~e.
$l^{\,\underline\cdot}=dg^{\,\underline\cdot}$, we say that
$g^{\,\underline\cdot}$ is a generator of the vector field
$\underline{Il}=\underline{\smash{I\,dg}}$.

The Poisson bracket of two functions $f^{\,\underline\cdot}$ and
$g^{\,\underline\cdot}$ is defined by equality:
\begin{equation}
 \{\,f\,,\,g\,\} = df^{\,\underline{\smash{I\,dg}}} \ .
\label{Poisdef}
\end{equation}
If functions $f^{\,\underline\cdot}$ and $g^{\,\underline\cdot}$
are linear this definition can be written just as:
\begin{equation}
 \{\,f\,,\,g\,\} = f^{\,\underline{\smash{Ig}}} \ .
\label{Poislin}
\end{equation}

It will be enough here to restrict ourself with real fields, so
the space~$Z$ will be a real linear space. Nevertheless, it is
useful to consider complex functions on the phase space. Their
differentials belong to complexified adjoint space. This is why
instead of using the space~$Z^*$ we will always use its
complexification~$Z^*_{\mathbb C}$. Furthermore, when we define
the Poisson bracket by formula~(\ref{Poisdef}), it is necessary to
consider complex vectors $\underline{\smash{I\,dg}}$. It could be
avoided if we define the Poisson bracket in a more abstract
way\footnote{We prefer to avoid complexification of $Z$, because
this space can be considered as the set of {\it physical\/} states
existing in nature. The space~$Z^*_{\mathbb C}$ can be considered
as the set of {\it mathematical\/} values that we use to describe
physical states.}.


\myparagraph{Scalar field.} \label{ScalF}
 Consider the scalar field. Its Lagrangian is:
\[
\textstyle
 L = \frac12 \, \partial_\nu \varphi \, \partial_\nu \varphi -
     \frac{m^2}2 \, \varphi^2  \ .
\]

The equation of the motion~(\ref{EuLag}) takes form
(Klein-Fock-Gordon equation):
\begin{equation}
 \left(\partial^2\! + m^2\right)\varphi = 0        \ .
\label{KlFock}
\end{equation}

The symplectic structure~(\ref{symplstr}) on the invariant phase
space~$Z^\mathrm{scal}$ in coordinate representation takes the
form (given in~\cite{Witten1986}):
\begin{equation}
 \omega =
 \int\limits_\Sigma d\sigma_\nu \,\partial_\nu\varphi\wedge\varphi \ .
\label{symplKF}
\end{equation}

Let us make a Fourier transformation of the function $\varphi(x)$
by the formulas~(\ref{Fourier}). So far as the function
$\varphi(x)$ satisfies the equation~(\ref{KlFock}), its
Fourier-transform $\widetilde\varphi(k)$ can be represented as:
\begin{equation}
 \widetilde\varphi(k) = 2\pi\,\delta(k^2-m^2) \cdot a(k)  \ ,
\label{defa}
\end{equation}
where $a(k)$ is a usual function defined on the mass surface
$k^2=m^2$.

Performing Fourier transformation of the symplectic
structure~(\ref{symplKF}) we have:
\begin{equation}
 \omega^{\,\underline\cdot\,;\,\underline\cdot} =
 \int d\mu_m \cdot i\,\varepsilon(k) \cdot
            a(-k)^{\,\underline\cdot} \, a(k)^{\,\underline\cdot} \ ,
 \qquad \mbox{here} \ \
 d\mu_m = \frac{d^4\!k}{(2\pi)^4} \cdot 2\pi\,\delta(k^2-m^2) \ .
\label{sKFmom}
\end{equation}
Here integrated function is a usual product of numbers\footnote{If
we denote $d\mu_m^+=d\mu_m \cdot \theta(k)$, the
formula~(\ref{sKFmom}) can be written with external product of
forms:
\[
 \omega = \int d\mu_m^+ \cdot i\, a^*(k) \wedge a(k) \ .
\]}.
It is implied that arguments in the right part follow in the same
order as in the left.

Now in order to calculate the Poisson bracket of two field values
$\{\,\widetilde\varphi(k)\,,\,\widetilde\varphi(k')\,\}$,
according to the formula~(\ref{Poislin}), let us consider the
vector $\underline{\smash{I\widetilde\varphi(k')}}$. From the
formulas~(\ref{cotanisom}) and~(\ref{sKFmom}) we have that for any
vector~$\underline c$:
\[
 \int d\mu_m \cdot i\,\varepsilon(k)\cdot a(-k)^{\,\underline c\,}
      a(k)^{\,\underline{\smash{I\widetilde\varphi(k')}}} =
 \widetilde\varphi(k')^{\,\underline c}   \ .
\]
Using linear independence of the form~$a(k)^{\,\underline\cdot}$
with different values of~$k$, $k^2=m^2$, and the definition of
this form~(\ref{defa}), we get that the
vector~$\underline{\smash{I\widetilde\varphi(k')}}$ satisfies the
following equation:
\[
 a(k)^{\,\underline{\smash{I\widetilde\varphi(k')}}} =
 -\,i\,\varepsilon(k) \cdot (2\pi)^4\delta(k+k') \ .
\]

Therefore, using the formula~(\ref{Poislin}), for the Poisson
bracket we have:
\begin{equation}
 \{\,\widetilde\varphi(k)\,,\,\widetilde\varphi(k')\,\} =
 -\, \Big[ \,i\,\varepsilon(k) \cdot 2\pi\,\delta(k^2-m^2) \,\Big]
     \cdot (2\pi)^4\delta(k+k')   \ .
\label{PoisKF}
\end{equation}

Sometimes it is useful to have an analogous relation in coordinate
representation. In order to get it we make the Fourier
transformation of the formula~(\ref{PoisKF}) with respect to the
arguments~$k$ and~$k'$:
\[
 \{\,\varphi(x)\,,\,\varphi(x')\,\} = - D_m(x-x') \ .
\]
Here function~$D_m(y)$ is:
\begin{equation}
 D_m(y)=\int \frac{d^4\!k}{(2\pi)^4} \, e^{-iky} \cdot
        \Big[\, i\,\varepsilon(k) \cdot
                2\pi\, \delta(k^2-m^2) \,\Big]  \ .
\label{Dfunc}
\end{equation}


\myparagraph{``Physical'' electromagnetic field.} \label{emfield}
 Let us choose the Lagrangian of electromagnetic field
gauge-invariant:
\begin{equation}
\textstyle
 L = -\, \frac14\, F_{\mu\nu}\, F_{\mu\nu} =
     -\,{}^{[\mu\nu]}\,\partial_\mu A_\nu \, \partial_\mu A_\nu \ ,
 \qquad \mbox{here} \ \
 F_{\mu\nu}={}^{[\mu\nu]}\,2\,\partial_\mu A_\nu \ .
\label{emlag}
\end{equation}

The equations of the motion~(\ref{EuLag}) take the form (the
second Maxwell equation):
\begin{equation}
 \partial_\mu F_{\mu\nu}=0 \ .
\label{Maxwell}
\end{equation}

In this case it is convenient to define the invariant phase
space~$Z^\mathrm{phys}$ as the set of functions $F_{\mu\nu}(x)$
that satisfy the equation~(\ref{Maxwell}). The value of vector
potential~$A_\mu$ in a fixed point~$x$ is not a well-defined
function on~$Z^\mathrm{phys}$: an expression
like~$A_\mu(x)^{\,\underline\cdot}$ has no sense if we do not say
which gauge we use. For this reason, it is impossible to give any
one-valued formula for the Poisson bracket of vector potential.

The symplectic structure~(\ref{symplstr}) in this case takes the
form (shown in~\cite{Crnkovic1987,CrWitt1987}):
\begin{equation}
 \omega =
 - \int\limits_\Sigma d\sigma_\mu\, F_{\mu\nu} \wedge A_\nu  \ .
\label{symplem}
\end{equation}

It is appropriate to mention that, though the form~$\omega$ is
written with using gauge-dependent value~$A_\nu$, it is still
gauge-independent, of course.

Let us now choose the vector potential~$A_\mu$ so that it
satisfies the Lorentz condition $\partial_\nu A_\nu=0$. According
to the equation~(\ref{Maxwell}), it will also satisfy the
D'Alembert equation: $\partial^2\! A_\mu=0$. Let us split the
expression~(\ref{symplem}) into two terms:
\[
 \omega =
 - \int\limits_\Sigma d\sigma_\mu\, \partial_\mu A_\nu \wedge A_\nu +
   \int\limits_\Sigma d\sigma_\mu\, \partial_\nu A_\mu \wedge A_\nu   \ .
\]
The second integral is equal to zero. We can check it in the
following way. First, from the Lorentz condition and antisymmetry
of external product we have that the integrated function is a
conserved current:
 $\partial_\mu(\partial_\nu A_\mu \wedge A_\nu)=0$.
Therefore in the second integral, independently from the first, we
can integrate on the surface $t=0$ instead of integration
on~$\Sigma$. So, we get the integral:
\[
 \int d^3\!\vect x\, \partial_\nu A_0 \wedge A_\nu \ .
\]
Using the Lorentz condition we can write it as:
\[
 \int d^3\!\vect x\, \partial_\nu (A_0 \wedge A_\nu) \ .
\]
But $A_0\wedge A_0=0$. Therefore the summation by index~$\nu$ can
be performed just from $1$ to $3$. We get:
\[
 \int d^3\!\vect x\, \partial_n (A_0 \wedge A_n) \ .
\]
This integral is an integral of the three-dimensional vector $A_0
\wedge A_n$. By Ostrogradsky-Gauss divergence theorem, it is equal
to the flux of this vector through a far two-dimensional closed
surface. This flux is equal to zero because the vector potential
is implied to be chosen so that it is equal to zero there.

So, under the Lorentz condition $\partial_\nu A_\nu=0$, the
symplectic structure~(\ref{symplem}) can be written in the
following way:
\begin{equation}
 \omega =
 - \int\limits_\Sigma d\sigma_\mu\, \partial_\mu A_\nu \wedge A_\nu  \ .
\label{symplemlor}
\end{equation}

Let us make the Fourier transformation~(\ref{Fourier}) of
functions $A_\mu(x)$ and $F_{\mu\nu}(x)$. Similarly to the case of
the scalar field we can write these Fourier transforms as:
\[
 \widetilde A_\mu(k) = 2\pi\,\delta(k^2) \cdot a_\mu(k)  \ , \qquad
 \widetilde F_{\mu\nu}(k) = 2\pi\,\delta(k^2) \cdot f_{\mu\nu}(k)  \ .
\]
Here $a_\mu(k)$ and $f_{\mu\nu}(k)$ are usual functions defined on
the light cone $k^2=0$.

The symplectic structure~(\ref{symplemlor}) in the Fourier
representation takes the form\footnote{Using an external product
we can write it as:
\[
 \omega = - \int d\mu_m^+ \,i\, a_\nu^*(k) \wedge a_\nu(k) \ .
\]}:
\begin{equation}
 \omega^{\,\underline\cdot\,;\,\underline\cdot} =
 - \int d\mu_m \cdot i\,\varepsilon(k)\cdot
       a_\nu(-k)^{\,\underline\cdot} \, a_\nu(k)^{\,\underline\cdot} \ .
\label{semmom}
\end{equation}

The formulas~(\ref{symplemlor}) and~(\ref{semmom}) look very
similar to the corresponding formulas~(\ref{symplKF})
and~(\ref{sKFmom}) for the scalar field. More rigorously, they
look like formulas for the symplectic structure of four scalar
fields. But it is important to understand that the invariant phase
space~$Z^\mathrm{phys}$ introduced in this section is not a direct
sum of phase spaces of four scalar fields. For this reason, as we
mentioned above, it is impossible to calculate the Poisson bracket
for vector potential.

Consider now the structure of the space~$Z^\mathrm{phys}$ in
Fourier representation. This space can be considered as the set of
functions $a_\mu(k)$ on the light cone $k^2=0$ satisfying the
Lorentz condition $-\,i\,k_\mu\,a_\mu(k)=0$ and considered
accurate up to the gauge transformation:
\[
 a_\mu(k) \ \sim\ a_\mu(k) - i\,k_\mu\,\lambda(k)  \ .
\]
From here and from the formula~(\ref{semmom}) we see that the
symplectic structure on the space~$Z^\mathrm{phys}$ is not
degenerate.

Now it is clear that, though it is impossible to calculate the
Poisson brackets for components of vector potential, it is
possible to calculate the brackets
 $\{\, \widetilde F_{\mu\rho}(k) \, , \,
       \widetilde F_{\nu\sigma}(k')\,\}$ and
$\{\,F_{\mu\rho}(x)\,,\,F_{\nu\sigma}(x')\,\}$. Nevertheless, we
postpone this problem till the section~\ref{Poisspp}.


\myparagraph{Heisenberg-Pauli term.} \label{HeisPau}
 After the paper of Heisenberg and Pauli~\cite{HeisPau1929} when
quantum theory of electromagnetic field is discussed it is
customary to change the Lagrangian adding to it an additional term
$\frac\varepsilon2\,(\partial_\mu A_\mu)^2$. For such a step in
modern textbooks there are many different ``explanations''.
Generally they can be divided into two types.

The first type are ``explanations'' appealing to formal
difficulties of using the Hamiltonian formalism and watching
relativistic invariance. As we have shown in the
section~\ref{emfield}, even if there were any difficulties, they
disappear, if we use the invariant Hamiltonian formalism.

The second type are ``explanations'' appealing to quantum theory
in some way. Now our first goal is to explain that the reasons for
changing the Lagrangian exist even in the classical theory.

At first sight, the term
 $\frac\varepsilon2\,(\partial_\mu A_\mu)^2$
only fixes the gauge: if the Lorentz condition was satisfied in
the past, it will be satisfied in the future. But actually there
is a more deep consequence: when we develop scattering theory and
perturbation theory we get the opportunity to get rid of such a
fiction as separation of field to ``own'' field of particles and
``radiation''.

Indeed, it is impossible in practice to solve a dynamic problem
exactly. So, we calculate first some first approximation and then
we calculate radiative corrections. In order to take into account
radiative recoil we have to divide field into ``own'' and
``radiative'' in some way. But it is impossible to do it
rigorously even in simplest situations. It is not surprising,
because in the Lagrangian formalism there is no hint of such a
separation.

But we can approach problems of scattering differently. Consider,
for example, scattering of two particles interacting with each
other by electromagnetic field. Let us use some naive mechanical
analogy: for example, two heavy balls sliding on a smooth elastic
membrane. Membrane corresponds to electromagnetic field.

If we do not divide the field into own and radiation we have to
accept that all field is ``created'' by particles. In order to
formalize such an approach we have to introduce turning
interaction on and off.

The process of scattering of particles and of electromagnetic
field then can be described in the following way:
\begin{enumerate}
\item
In the far past the particles are free and they do not interact
with the field. There exists also the field and it is far from the
particles.
\item
Then, while particles are far from each other, we turn on
interaction adiabatically. The particles put on their field at
this stage.
\item
After interaction is fully turned on particles (and field) run to
each other near, interact and then run away from each other.
\item
When they are far enough we turn interaction off adiabatically.
\item
Then there remain only bare particles, and there is no field near
them, and free field which is not connected with the particles.
\end{enumerate}

It is obvious for balls sliding on an elastic membrane how to turn
on and off interaction: it is necessary to turn on and off the
gravitation. But in the case of electromagnetic field we can not
do anything so simple. Consider, for example, interaction of
electromagnetic field $A_\mu(x)$ with a given current $J_\mu(x)$.
The Lagrangian in this case takes the form:
\begin{equation}
\textstyle
 L = -\, \frac14\, F_{\mu\nu}\, F_{\mu\nu} - \alpha\,J_\mu\,A_\mu \ .
\label{gaugeinvlag}
\end{equation}
Here $\alpha(x)$ is a function that turns on and off interaction.
In order to write formulas shorter we will use a notation:
\[
 J^\alpha_\mu(x) = \alpha(x)\,J_\mu(x)    \ .
\]

From the Lagrangian~(\ref{gaugeinvlag}) we get the equation of the
motion:
\begin{equation}
 \partial_\mu F_{\mu\nu}=J^\alpha_\nu \ .
\label{emeq}
\end{equation}
Applying to both sides the operation~$\partial_\nu$ we see that,
if $\partial_\nu J^\alpha_\nu\neq0$, the equation~(\ref{emeq})
does not have any solutions. So, the interaction can not be turned
on and off.

But if we add to the Lagrangian the term
$\frac\varepsilon2\,(\partial_\mu A_\mu)^2$, i.~e. use the
Lagrangian
\[
\textstyle
 L = -\, \frac14\, F_{\mu\nu}\, F_{\mu\nu} +
     \frac\varepsilon2\,(\partial_\mu A_\mu)^2 -
     J^\alpha_\mu\,A_\mu \ ,
\]
then the equation of the motion takes the form:
\begin{equation}
 \partial_\mu F_{\mu\nu} -
 \varepsilon\,\partial_\nu\partial_\mu A_\mu =
 J^\alpha_\nu \ ,
\label{modeq}
\end{equation}
and solution~$A_\mu(x)$ can be easily found for any
function~$\alpha(x)$.

Later we will always imply that $\varepsilon=-1$. In such a case
the equation~(\ref{modeq}) takes especially simple form:
\begin{equation}
 \partial^2\! A_\nu = J^\alpha_\nu \ .
\label{DAlabA}
\end{equation}
Its solution is the Li\'enard-Wiechert potential. For this reason
this ``gauge'' should be called the Li\'enard-Wiechert gauge.
Nevertheless, in quantum electrodynamics it is called the Feynman
gauge.

If we calculate a four-dimensional divergence of both parts of
equation (\ref{DAlabA}), we get:
\[
 \partial^2(\partial_\nu A_\nu) = \partial_\nu J^\alpha_\nu    \ .
\]
It can be shown from here that, if interaction is turned on and
off adiabatically and the Lorentz condition $\partial_\nu A_\nu=0$
took place in the far past, then it will take place everywhere.

From this we also see that the additional term
$\frac12\,(\partial_\mu A_\mu)^2$ in the Lagrangian will make a
zero contribution in the variation of action. Therefore in the
region, where interaction is completely turned on, the system of
particles and fields will satisfy the same equations of motion as
the system with no additional term in the Lagrangian.

So, introduction of the additional term in the Lagrangian does not
substantially change the scattering. But it allows to avoid
separation of ``own'' field of particles and allows to construct a
local perturbation theory.

It is difficult to say now, how far we can go in constructing of
perturbation theory for classical electrodynamics. But it is
apparent that without introduction of the additional term in
Lagrangian situation would be completely hopeless.


\myparagraph{``Non-physical'' electromagnetic field.}
\label{nonphys}
 Now consider again the free electromagnetic field, but include
already in Lagrangian the term $-\,\frac12\,(\partial_\mu
A_\mu)^2$. Then the Lagrangian can be written in the following
way:
\[
\textstyle
 L = -\, \frac14\, F_{\mu\nu}\, F_{\mu\nu} -
     \frac12\,(\partial_\mu A_\mu)^2 =
     -\, \frac12\,\partial_\mu A_\nu\,\partial_\mu A_\nu +
     \frac12\,\partial_\mu(A_\nu\,\partial_\nu A_\mu -
                           A_\mu\,\partial_\nu A_\nu)          \ .
\]

As we will show in the paper~[II] the term that is a full
divergence can be thrown off. After that the Lagrangian takes
especially simple form:
\[ \textstyle
 L = -\, \frac12\,\partial_\mu A_\nu\,\partial_\mu A_\nu   \ ,
\]
i.~e. it is in fact the Lagrangian of four scalar
fields\footnote{The term corresponding to the field with index
$\nu=0$ is included with negative sign.} with $m=0$ considered in
the section~\ref{ScalF}. The equation of the motion has the form:
\[
 \partial^2\! A_\mu = 0 \ .
\]

Now the invariant phase space is a direct sum of four phase spaces
of scalar fields. Let us denote it as~$Z^4$. The symplectic
structure in coordinate and Fourier representations is given by
formulas:
\begin{equation}
\omega =
 - \int\limits_\Sigma d\sigma_\mu\, \partial_\mu A_\nu \wedge
                                                 A_\nu         \ .
\label{symcoor2}
\end{equation}
\begin{equation}
 \omega^{\,\underline\cdot\,;\,\underline\cdot} =
 - \int d\mu_m \cdot i\,\varepsilon(k)\cdot
          a_\nu(-k)^{\,\underline\cdot} \,
          a_\nu(k)^{\,\underline\cdot} \ .
\label{symmom2}
\end{equation}

The formulas~(\ref{symcoor2}) and~(\ref{symmom2}) look the same as
corresponding formulas from the section~\ref{emfield}. But the
difference in their meaning is important: now we can calculate the
Poisson bracket for the vector potential. In Fourier
representation it has the form:
\begin{equation}
 \{\,\widetilde A_\mu(k)\,,\,\widetilde A_\nu(k')\,\} =
 g_{\mu\nu}
 \cdot \Big[\, i\,\varepsilon(k)
               \cdot 2\pi\,\delta(k^2) \,\Big]
 \cdot (2\pi)^4\delta(k+k')       \ .
\label{PoissAmom}
\end{equation}
In coordinate representation:
\begin{equation}
 \{\,A_\mu(x)\,,\,A_\nu(x')\,\} = g_{\mu\nu}\,D_0(x-x')       \ .
\label{PoissAcoord}
\end{equation}
Here $D_0(y)$ is the function~(\ref{Dfunc}) with $m=0$. In this
case it can be written just as:
\[
 D_0(y)=\frac1{2\pi}\,\varepsilon(y)\,\delta(y^2)  \ .
\]

Let us apply to both sides of the formula~(\ref{PoissAmom}) symbol
 ${}^{[\rho\mu]}\,2\,(-\,i\,k_\rho) \,
  {}^{[\sigma\nu]}\,2\,(-\,i\,k'_\sigma)$.
We get the formula for the Poisson bracket of the tensor of
electromagnetic field in the Fourier representation:
\begin{equation}
 \{\,\widetilde F_{\mu\rho}(k)\,,\,\widetilde F_{\nu\sigma}(k')\,\} =
 {}^{[\mu\rho]}\, {}^{[\nu\sigma]}\,
     4\, g_{\mu\nu}\, k_\rho\, k_\sigma \cdot
 \Big[\, i\,\varepsilon(k) \cdot 2\pi\,\delta(k^2) \,\Big]
 \cdot (2\pi)^4\delta(k+k')       \ .
\label{PoissFmom}
\end{equation}

In the coordinate representation this relation can be written in
the following way:
\begin{equation}
 \{\,F_{\mu\rho}(x)\,,\,F_{\nu\sigma}(x')\,\} =
 - \,{}^{[\mu\rho]}\, {}^{[\nu\sigma]}\, 4\, g_{\mu\nu}\,
     \partial_\rho\, \partial_\sigma D_0(x-x')     \ .
\label{PoissFcoord}
\end{equation}
Here both derivatives are taken with respect to the argument
without dash.

It should be mentioned already that formulas~(\ref{PoissFmom}) and
(\ref{PoissFcoord}) are the same as formulas for the Poisson
brackets of the ``physical'' electromagnetic field described in
the section~\ref{emfield}. It will be proven in the
section~\ref{Poisspp}.


\myparagraph{Energy of electrostatic field.}
 In the paper~[III] we will get a formula that can be used in
the invariant Hamiltonian formalism for obtaining generators of
the Poincare group. With its help we easily get the formula for
the momentum of the ``non-physical'' electromagnetic field:
\begin{equation}
 P_\nu = - \int d\mu^+_m \cdot k_\nu\, a^*_\rho(k)\,a_\rho(k) \ .
\label{emmom}
\end{equation}

From this formula we see that the energy of the ``non-physical''
field is not positive-definite. And what is more, it follows from
this formula that the energy stored in electrostatic field created
by a system of fixed charges is always {\it negative}.

Here we will discuss this paradox with some details. This is
necessary because it was usually incorrectly treated in the frame
of {\it quantum\/} theory (see, for example,
\cite{Gupta1950,Gupta1957,Nagy1969,Haag1961}).

The problem of negativity of energy of ``time'' photons in quantum
electrodynamics was noted a long time ago. It was supposed that
quantization in the space with indefinite scalar product allows
change the sign of this energy. And it was considered as an
argument for introduction of indefinite scalar product. It seemed
so because, if we use indefinite scalar product in quantum case,
it is not quite clear what we should call the energy of a state:
the average value of the Hamiltonian or its eigenvalue. This
vagueness, of course, appeared because, first, there was no clear
conception of quantization, and second, classical theory was not
formulated in the appropriate way.

But from our consideration it is clear that the discussed energy
{\it must\/} be negative: only such value looks natural from the
point of view of relativistic theory. And indefiniteness of metric
in quantum case has no relation to this question.

It should be also noted here that the notion of vacuum of a field
we do not connect in any way with minimum of energy. The vacuum of
a classical field is just the zero vector of the space~$Z$. A
notion of vacuum of a quantum field we will introduce in
description of quantization in the paper~[VI].

But what can we say now about the positiveness of electrostatic
energy which is well-known in the usual electrostatics? The fact
is that this notion implies usually a different meaning: it
usually imply the work that the system can make, if charges will
be split in infinite small parts and they will be carried far away
from each other. But the formula~(\ref{emmom}) gives the energy
that remains in the field, if its interaction with the charges is
instantly turned off.

We will not discuss this question in more details here. But let us
notice the following. In electrostatics there is the formula for
energy:
\[
 E^\mathrm{\,full}={\textstyle\frac12}\int\varphi\,dQ   \ .
\]
The multiplier $\frac12$ is usually explained so that, if we use
the formula for the energy in external field, we count the mutual
energy of two charges two times.

But in the frame of the theory that we discuss here the field is
always considered as external. Therefore, we can say that the
energy of charges is always given by the formula:
\[
 E^\mathrm{\,charges} = \int\varphi\,dQ \ .
\]
We have to add to it the energy of the field:
\[
 E^\mathrm{\,field} = \,- \,{\textstyle\frac12}\int\varphi\,dQ  \ .
\]
The sum of $E^\mathrm{\,charges}$ and $E^\mathrm{\,field}$ is
exactly $E^\mathrm{\,full}$.


\myparagraph{Scattered states of ``non-physical'' field.}
\label{ScatStat}
 Consider now scattering of the ``non-physical'' field on a given
current. As it was explained in the section~\ref{HeisPau}, we will
suppose that the interaction with electromagnetic field is turned
on and off adiabatically and the particles constituting the
current are accelerated only in the region of space-time where the
interaction is fully turned on, i.~e. $\alpha(y)=1$.

In an arbitrary point of space-time the field will consist of two
components: the field that existed in the far past and the field
created by the particles. This additional field can be calculated
with the Li\'enard-Wiechert formula:
\begin{equation}
 A_\mu^\mathrm{ret}(x) = \int d^4\!x'\,
      D_0^\mathrm{c}(x-x')\,J^\alpha_\mu(x')  \ ,
 \qquad \mbox{here} \ \
 D_0^\mathrm{c}(y)=\theta(y)\,D_0(y)  \ .
\label{LienWiech}
\end{equation}

In the invariant Hamiltonian formalism a state of a field implies
a solution of the corresponding homogeneous equation. In our case
this is the equation $\partial^2\!A_\mu=0$. Therefore when we talk
about a state of the field in the far past or in the far future it
is natural to continue the function $A_\mu(x)$ by the homogeneous
equation to the whole space. So, we come to the definition of the
in- and out- fields:
 $A_\mu(x)^{\,\underline{\mathrm{in}}}$ and
 $A_\mu(x)^{\,\underline{\mathrm{out}}}$. Here
 $\underline{\mathrm{in}}$ and $\underline{\mathrm{out}}$ are
vectors of the invariant phase space~$Z^4$. It is natural to call
these vectors in- and out- states. The difference of these two
vectors corresponds to the radiated field:
\[
 \underline{\mathrm{out}} - \underline{\mathrm{in}} =
 \underline{\mathrm{rad}} \ .
\]

In order to find the field $A_\mu(x)^{\,\underline{\mathrm{rad}}}$
we have just to make a minor correction in the Li\'enard-Wiechert
formula~(\ref{LienWiech}):
\[
 A_\mu(x)^{\,\underline{\mathrm{rad}}} =
 \int d^4\!x'\, D_0(x-x')\,J^\alpha_\mu(x')  \ .
\]
In the Fourier-representation this formula can be written in an
especially simple way:
\[
 \widetilde A_\mu(k)^{\,\underline{\mathrm{rad}}} =
 i\,\varepsilon(k)\cdot 2\pi\,\delta(k^2)\cdot
    \widetilde J^\alpha_\mu(k) \ ,
\]
or, with using non-singular function $a_\mu(k)$ on the light cone,
even shorter:
\begin{equation}
 a_\mu(k)^{\,\underline{\mathrm{rad}}} =
 i\,\varepsilon(k)\,\widetilde J^\alpha_\mu(k) \ .
\label{myLienWiech}
\end{equation}

We see that in the given formulas we can approach to the adiabatic
limit. The set of all possible vectors $\underline{\mathrm{rad}}$
forms a linear subspace of the invariant phase space~$Z^4$. Let us
denote this space~$Z^\perp$. As it was explained in the
section~\ref{HeisPau}, if the interaction is turned on and off
adiabatically, the Lorentz condition is not broken. Therefore the
space~$Z^\perp$ does not coincide with the whole~$Z^4$. On the
other hand, it follows from the formula~(\ref{myLienWiech}) that
there is no other condition that restricts the space~$Z^\perp$.
So, $Z^\perp$ is the subspace of the space~$Z^4$ defined by the
Lorentz condition:
\begin{equation}
 \underline{\mathrm{rad}} \in Z^\perp
 \ \ \Longleftrightarrow \ \
   -\,i\,k_\mu\, a_\mu(k)^{\,\underline{\mathrm{rad}}} = 0 \ .
\label{Lorcond0}
\end{equation}
This explains the notation~$Z^\perp$.

It is natural to believe that all possible in-states were created
from the vacuum by interaction with currents. Then vectors of in-
and out- fields also lie in the subspace~$Z^\perp$. So far as all
possible radiation fields fill the whole~$Z^\perp$, all possible
in- and out- fields also fill the whole~$Z^\perp$.

Using the condition~(\ref{Lorcond0}) and the formula~(\ref{emmom})
for the energy-momentum vector we see that the states from the
space~$Z^\perp$ have non-negative energy.


\myparagraph{Remainders of gauge invariance.}
 Consider now such a vector
 $\underline{\smash{\mathrm{gauge}}}\in Z$
that the function
 $a_\mu(k)^{\,\underline{\smash{\mathrm{gauge}}}}$
can be written as:
\[
 a_\mu(k)^{\,\underline{\smash{\mathrm{gauge}}}} =
 -\,i\,k_\mu\,\lambda(k) \ ,
\]
where $\lambda(k)$ is some function on the light cone. The set of
all vectors of this type form a linear subspace in~$Z^4$. Let us
denote this space~$Z^\parallel$. It is obvious that $Z^\parallel$
lies in~$Z^\perp$.

Consider again the scattering of particles and field when the
motion of particles is not given and there is, generally, non-zero
in-field. This scattering is described by the Lagrangian:
\[
\textstyle
 L = L^{\mathrm{particles}}
     - \frac12\,\partial_\mu A_\nu\,\partial_\mu A_\nu
     - \alpha\,J_\mu\,A_\mu  \ .
\]
Let us suppose that the corresponding equations of motion are
solvable and trajectories of particles and field $A_\mu(x)$ that
make action stationary are found. Let us also suppose that
adiabatic turning on (off) of interaction is made in so far past
(future) that all particles are far from each other and the field
$A_\mu(k)^{\,\underline{\mathrm{in}}}$
($A_\mu(k)^{\,\underline{\mathrm{out}}}$) is far from all
particles and the field
$A_\mu(k)^{\,\underline{\smash{\mathrm{gauge}}}}$ is also far from
all particles.

Let us add to the field $A_\mu(x)$ the field
$A_\mu(x)^{\,\underline{\smash{\mathrm{gauge}}}}$. Or, in other
words, let us make the gauge transformation:
\begin{equation}
 A_\mu(x) \to A_\mu(x) + \partial_\mu\Lambda(x) \ ,
\label{GaugeTrans}
\end{equation}
where $\Lambda(x)$ is a function whose Fourier transform is
connected with $\lambda(k)$ by relation:
\[
 \widetilde\Lambda(k) = 2\pi\,\delta(k^2)\cdot\lambda(k)  \ .
\]

In the region where interaction is turned on and off this addition
can not influence the stationarity of action. Indeed, so far as
the particles are outside of the region where this addition is
present, variations of terms $L^{\mathrm{particles}}$ and
$-\,\alpha\,J_\mu\,A_\mu$ can not change. Furthermore, the term
$-\,\frac12\,\partial_\mu A_\nu\,\partial_\mu A_\nu$ differs from
the gauge-invariant by the full divergence, which can not change
the variation, and by the term $-\,\frac12\,(\partial_\mu
A_\mu)^2$, which has a zero variation, because $\partial_\mu
A_\mu=0$.

Consider now the region, where interaction is fully turned on. The
transformation~(\ref{GaugeTrans}) can not change the variations of
the terms $L^{\mathrm{particles}}$ and
 $-\,\frac12\,\partial_\mu A_\nu\,\partial_\mu A_\nu$ for the same
reasons that in the case of the regions where the interaction is
turned on and off. There remains the term $-\,J_\mu\,A_\mu$. Under
the gauge transformation~(\ref{GaugeTrans}) it becomes
 $-\,J_\mu\,A_\mu -\,J_\mu\,\partial_\mu\Lambda(x)$. Therefore, in
the variation of the action we get the additional term $-\,\delta
J_\mu\,\partial_\mu\Lambda(x)$. Integrating it by parts and using
conservation of the current we get that it does not add anything
to the variation.

So, under the transformation~(\ref{GaugeTrans}) motions that
satisfy the condition of stationarity of action are transformed to
the motions that also satisfy this condition. It is possible to
say that the space~$Z^\parallel$ does not participate in
scattering at all.

It is natural to suppose that the scattered states of
electromagnetic field are observed only by their influence on
charged particles. Then any two vectors from $Z^\perp$ that differ
by vector from $Z^\parallel$ are physically indistinguishable.
Therefore, instead of using the space $Z^\perp$, we can use the
factor-space $Z^{\perp/\parallel}=Z^\perp/Z^\parallel$.

Notice now that in the Fourier representation the
space~$Z^{\perp/\parallel}$ can be considered as the set of
functions $a_\mu(k)$ on the light cone $k^2=0$ satisfying the
Lorentz condition $-\,i\,k_\mu\,a_\mu(k)=0$ and considered
accurate up to the gauge transformation:
\[
 a_\mu(k) \ \sim\ a_\mu(k) - i\,k_\mu\,\lambda(k)  \ .
\]
Therefore, we can establish the natural one-to-one correspondence
between elements of the spaces $Z^{\perp/\parallel}$ and
$Z^\mathrm{phys}$.

And what is more, the space~$Z^{\perp/\parallel}$ naturally
inherits from the full space~$Z^4$ the symplectic
structure~(\ref{symmom2}), and this structure obviously is the
same as the symplectic structure of the space~$Z^\mathrm{phys}$
given by the formula~(\ref{semmom}). So, $Z^{\perp/\parallel}$ and
$Z^\mathrm{phys}$ are also naturally identified as {\it
symplectic} spaces.


\myparagraph{About the Poisson bracket on $Z^{\perp/\parallel}$.}
\label{Poisspp}
 Consider now on the space~$Z^4$ a linear function
$\widetilde F_{\mu\rho}(k)^{\,\underline\cdot}$ (or
$F_{\mu\rho}(x)^{\,\underline\cdot}$). And let us suppose $k$ (or,
correspondingly, $x$) to be fixed. As we have said in the
section~\ref{InvHamForm}, this function is a generator of a vector
field on~$Z^4$.

Notice, first, that so far as the function under consideration is
linear on~$Z^4$, its differential does not depend on the point of
the space~$Z^4$ (and is identical with the function). Therefore
the obtained vector field is constant on the whole space~$Z^4$.

Second, differential of the function under consideration, like the
function, is zero on~$Z^\parallel$. Therefore, as it is seen from
the formula for symplectic structure~(\ref{symmom2}), the vector
of the obtained constant vector field belongs to the
subspace~$Z^\perp$.

From these two statements we get that the obtained field can be
naturally restricted to the factor-space~$Z^{\perp/\parallel}$.

On the other hand, the space~$Z^{\perp/\parallel}$ has its own
symplectic structure. If we restrict the function
 $\widetilde F_{\mu\rho}(k)^{\,\underline\cdot}$ (or
 $F_{\mu\rho}(x)^{\,\underline\cdot}$) to this factor-space,
the obtained function will be a generator of some vector field
in~$Z^{\perp/\parallel}$.

But, so far as the symplectic structure in~$Z^{\perp/\parallel}$
is inherited from~$Z^4$ we can see that both vector fields
in~$Z^{\perp/\parallel}$ are identical.

From here we also get that the formulas~(\ref{PoissFmom})
and~(\ref{PoissFcoord}) obtained in the section~\ref{nonphys} for
the ``non-physical'' field work for the ``physical'' field also.




\newcounter{parnum}
\newlength{\picwidth}
\newcommand*{\myparagraph}[1]{\refstepcounter{parnum}%
\paragraph{\arabic{parnum}. #1}}


\title{О квантовании электромагнитного поля.\\
       I. Классическая электродинамика.}
\author{Д.~А.~Арбатский\footnote{http:\dslash d-a-arbatsky.narod.ru/}}
\date{\today}
\maketitle

\begin{abstract}
Развивается методика исследования классических полей на основе
инвариантного гамильтонова формализма. Электромагнитное поле,
наряду со скалярным, выступает как частный пример применения
общего метода. Вычисляются скобки Пуассона для этих полей.
Разъясняется необходимость введения \rql нефизических\rqr{}
степеней свободы для электромагнитного поля.
\end{abstract}


\myparagraph{Вводные замечания.}
 Данная статья является первой в серии из шести статей,
объединённых общим заголовком. Ссылки на эти статьи будут здесь
даваться римскими цифрами: [I],~\dots,~[VI].

Одним из поводов для настоящего исследования явилась попытка
ответить на вопрос о том, как устроено пространство состояний
квантованного электромагнитного поля с точки зрения
функционального анализа.

\begin{enumerate}

\item
При изложении схемы квантования Гупты-Блейлера современные
учебники молчаливо предполагают, что вопрос о топологии квантового
пространства состояний у электромагнитного поля решается в полной
аналогии с полем скалярным. При этом электромагнитное поле
квантуется либо в обычном гильбертовом пространстве (такая
конструкция не обладает даже неявной релятивистской
инвариантностью [VI]), либо в \rql гильбертовом пространстве с
индефинитной метрикой\rqr{} (при таком подходе вопрос о топологии
до сих пор вообще не был исследован на должном уровне строгости;
при этом фактически оказывается, что пространство состояний вообще
не определялось конструктивно).

В любом случае, основываясь на различных аналогиях, обычно
считают, что пространство состояний, с {\it топологической} точки
зрения, {\it должно} быть гильбертовым. Из нашего исследования
станет ясно, что такая точка зрения ошибочна. Причём этот вывод
является достаточно общим: он заставляет по-новому взглянуть и на
квантование полей, с квантованием которых до сих пор всё как-будто
было в порядке (например, скалярного).

\item
Корень проблемы, как будет видно, лежит не в функциональном
анализе, а в алгебре. Процесс квантования в пространстве Фока до
сих пор не был описан удовлетворительно с {\it алгебраической}
точки зрения. Дело в том, что, следуя за Фоком~\cite{Fock1932},
для построения квантованного поля в качестве отправной
алгебраической структуры берут одночастичное квантовое
пространство. Но такая алгебраическая структура слишком бедна: не
существует алгебраического процесса, позволяющего сконструировать
на её основе поле квантовое (именно, опираясь только на структуру
одночастичного подпространства, невозможно ввести локальные
операторы поля (см., например, \cite{Weinberg1996}, гл.~5)).

Как будет видно из настоящего исследования, подходящей
алгебраической структурой для построения квантованного поля
является классическое поле, описанное на языке инвариантного
гамильтонова формализма.

\item
После решения алгебраических вопросов (т.~е. построения
удовлетворительной схемы квантования) решение вопроса о топологии
оказывается в значительной степени чисто технической проблемой.
При этом важным оказывается осознание того факта, что при
построении пространства состояний электромагнитного поля в любом
случае приходится от гильбертова пространства отказаться. В статье
[VI] для решения этой проблемы вводятся новые классы
функциональных пространств.

\item
Вопрос о релятивистской инвариантности предложенной здесь
конструкции квантованного поля не встаёт. Дело в том, что эта
конструкция не просто релятивистски-инвариантна: требование
релятивистской инвариантности составляет основу самой конструкции.

Если использовать приведённую здесь конструкцию для квантования
гармонического осциллятора, то требование инвариантности
квантования  по отношению к некоторой группе в известном смысле
заменяет требование о том, чтобы квантовое пространство состояний
было гильбертовым. При этом квантований оказывается два: одно в
гильбертовом пространстве и одно~--- в пространстве с индефинитным
скалярным произведением.

Таким образом, мы даже не делаем попытки как-либо обобщать теорему
Стоуна-фон~Неймана на случай релятивистских полей: вместо этого
предложена принципиально иная концепция квантования, которая даже
в случае квантования гармонического осциллятора работает {\it
иначе}.

\item
Поскольку наш подход к квантованию последовательно ориентирован на
явное отслеживание симметрии (в случае релятивистских полей~--- по
отношению к группе Пуанкаре), удаётся ясно проследить, каким
образом действие группы симметрии переносится с классического
пространства состояний на квантовое. Не будет преувеличением
сказать, что именно в настоящих статьях впервые (на строгом
уровне) раскрывается, каким образом у квантовой системы
оказываются интегралы движения того же типа, что и у
соответствующей классической системы (и связанные с наличием
группы симметрии).

\item
Ещё со времён классической работы Бора и Розенфельда
\cite{BohrRos1933} известно, что при рассмотрении полевых
операторов в некоторой фиксированной точке пространства-времени
возникают проблемы. В современных книгах, претендующих на
математическую строгость, принято говорить, что мы имеем в этом
случае дело с \rql операторным распределением\rqr. Однако,
насколько мне известно, никакой содержательной математической
теории таких распределений (подобной теории обобщённых функций)
дано не было. Поэтому термин \rql операторное распределение\rqr{}
до сих пор всего лишь отражал тот факт, что символам вида $\hat
A_\mu(x)$ вообще {\it никакого} смысла придать нельзя.

При использовании инвариантного гамильтонова формализма в качестве
основы для построения квантовых полей можно видеть, что проблема
эта может быть исследована уже в классической механике, причём с
гораздо большей ясностью.

С точки зрения формальной, можно было бы полностью разобрать этот
вопрос уже в данной статье. Тем более что строгий подход к
вычислению скобок Пуассона всё равно требует обсуждения топологии
инвариантного фазового пространства. Тем не менее этот вопрос я
предпочёл оставить до последней статьи. Дело в том, что выбор
топологии для инвариантного фазового пространства может быть ясно
и просто {\it мотивирован} только после рассмотрения
релятивистских полей с точки зрения теории групп [IV] и обсуждения
квантования [VI].

\item
Как известно, к лагранжиану поля можно добавлять некоторую
дивергенцию, не меняя при этом уравнений движения. В статье [II]
показано, что при этом не меняется и инвариантное гамильтоново
описание поля. Поскольку предлагаемая схема квантования целиком
базируется на структуре инвариантного гамильтонова формализма,
получаемое в результате квантование также к замене лагранжиана
нечувствительно.

\item
Поскольку основой для построения квантованных полей у нас
выступает классическое поле, а не одночастичное квантовое
пространство, теория Вигнера-Макки (об унитарных представлениях
группы Пуанкаре) у нас не играет фундаментальной роли. Аналогом
этой теории является теория симплектических представлений группы
Пуанкаре (в статье [IV] я попытался описать некоторые основы).

Теория симплектических представлений группы Пуанкаре до известной
степени похожа на теорию унитарных. Однако, пример
электромагнитного поля как раз показывает, что полной аналогии
здесь нет.

\item
В своём изложении мы будем ориентироваться преимущественно на
линейные вещественные бозе-поля. Требование линейности здесь
принципиально: нелинейные поля мы квантовать не умеем. Что
касается требований вещественности и того, чтобы поле было
бозевским, то эти ограничения приняты здесь только для того, чтобы
не усложнять обозначения и формулировки. \rql Комплексность\rqr{}
поля означает просто наличие дополнительной комплексной структуры.
Что касается фермиевских полей, то для их описания требуется
использование грассмановых переменных (см.,
например,~\cite{OgiMez1975}). Используются ли обычные числа или
грассмановы, для нашего рассмотрения не принципиально.

\end{enumerate}

В данной статье даётся описание классического электромагнитного
поля. Основная цель~--- изложить общеизвестные факты на новом
языке. Впрочем, данный подход позволяет взглянуть на некоторые
вопросы совершенно иначе: например, выясняется, что вопрос об
энергии электростатического поля до сих пор трактовался
неправильно.


\myparagraph{Обозначения.}
 В тензорных обозначениях мы будем записывать векторы
пространства Минковского как $a_\mu$, $b_\nu$ и~т.~п. При этом
будут использоваться только {\it контравариантные\/} компоненты
тензоров; индексы же, нумерующие их, будут писаться всегда
 {\it внизу}. Система координат всегда будет предполагаться
ортонормированной относительно скалярного произведения
$g(\,\cdot\,,\,\cdot\,)$, т.~е. такой, что метрический тензор
имеет стандартный вид:
 $g_{\mu\nu} = \mathrm{diag}\,(+1,-1,-1,-1)_{\mu\nu}$.
По повторяющимся тензорным индексам суммирование всегда проводится
с учётом знаков. Например, скалярное произведение будет
записываться как
\[
 a\,b = a_\mu\, b_\mu = g_{\mu\nu}\, a_\mu\, b_\nu =
 a_0\, b_0 - a_1\, b_1 - a_2\, b_2 - a_3\, b_3 \ .
\]

Производные по пространственно-временным координатам обозначаются
символом $\partial_\mu$:
\[
 \partial_\mu =
   \left(\begin{array}{c|ccc}
          \frac\partial{\partial x_0} &
         -\frac\partial{\partial x_1} &
         -\frac\partial{\partial x_2} &
         -\frac\partial{\partial x_3}
   \end{array}\right)_\mu                   \ ,
\]
т.~е. индекс у производной, как и все индексы, является
контравариантным.

Для сокращения обозначений введён символ антисимметризации вида
${}^{[\mu\nu]}$. Он означает, что по ближайшим следующим за ним
индексам~$\mu$ и~$\nu$ проводится антисимметризация. Например,
тензор электромагнитного поля записывается как
 $F_{\mu\nu}=\partial_\mu A_\nu - \partial_\nu A_\mu =
 {}^{[\mu\nu]}\,2\,\partial_\mu A_\nu$.

Оператор Даламбера задаётся формулой:
$\partial^2=\partial_\mu\partial_\mu$.

Преобразование Фурье имеет всюду вид:
\begin{equation}
 \widetilde\varphi(k) = \int d^4\!x\,e^{+ikx}\,\varphi(x)  \ ,
 \qquad
 \varphi(x) = \int \frac{d^4\!k}{(2\pi)^4}\,e^{-ikx}\,
              \widetilde\varphi(k) \ .
\label{Fourier}
\end{equation}

Система единиц всюду такова, что $\hbar=c=1$.


\myparagraph{Инвариантный гамильтонов формализм.}
\label{InvHamForm}
 В работах~\cite{Witten1986,Crnkovic1987,Zucker1987} было
показано, что базовым понятиям гамильтонова формализма (а именно,
фазовому пространству и симплектической структуре на нём) можно
придать релятивистски инвариантный смысл. Там же были получены
формулы для симплектической структуры наиболее важных систем (в
координатном представлении). Тем самым была заложена основа схемы,
которую можно назвать инвариантным гамильтоновым формализмом.
Подробное изложение имеется в оригинальных
обзорах~\cite{CrWitt1987,Crnkovic1988}. Вопрос об эквивалентности
этого подхода и обычной гамильтоновой механики в случае систем со
связями обсуждался также в~\cite{BHS1991}.

Здесь я напомню, однако, некоторые основные понятия. Это нужно для
того, чтобы ввести наши собственные обозначения.

Пусть поле $\varphi_i(x)$ описывается лагранжианом
$L(\varphi_i,\partial_\mu\varphi_i)$. Лагранжиан мы здесь будем
считать зависящим только от самих полевых переменных $\varphi_i$ и
их производных\footnote{Нам здесь достаточно будет ограничиться
лагранжианами, зависящими только от производных первого порядка. В
общем случае можно рассматривать и лагранжианы с высшими
производными~\cite{BHS1991}.}, и не зависящим от точки~$x$. Кроме
того, мы будем иметь дело только с линейными полями, и лагранжиан
будет считаться квадратичной функцией.

Принцип наименьшего действия приводит к уравнениям
Эйлера-Лагранжа:
\begin{equation}
 \frac{\partial L}{\partial\,\varphi_i} -
 \partial_\mu
    \frac{\partial L}{\partial\,(\partial_\mu\varphi_i)} = 0 \ .
\label{EuLag}
\end{equation}

Инвариантное фазовое пространство~$Z$ определяется как множество
всех функций\footnote{Точнее, будем пока считать, что речь идёт о
гладких функциях. Кроме того, будем считать, что эти функции
отличны от нуля только на множестве, пересечение которого с любой
пространственно-подобной плоскостью ограничено. В статье~[VI] мы
обсудим вопрос о естественной топологии пространства~$Z$.}
$\varphi_i$, которые удовлетворяют уравнениям~(\ref{EuLag}).
Поскольку мы предполагаем уравнение~(\ref{EuLag}) линейным,
пространство~$Z$ обладает естественной линейной структурой.

Элементы пространства~$Z$ будем обозначать подчёркнутыми наборами
символов, например, $\underline c\in Z$. Функции на этом
пространстве будем записывать как $f^{\,\underline c}$. Например,
значение функции $\varphi$, отвечающей элементу $\underline c$, в
точке $x$ записывается как $\varphi(x)^{\,\underline c}$.

Мы будем, как правило, отождествлять элементы касательного
расслоения~$TZ$ с элементами самого~$Z$. Если, однако, необходимо
указать, что касательный вектор $\underline c$ приложен в точке
$\underline b\in Z$, то можно его записать
как~$\underline{\smash{c\,[b]}}$. Элементы кокасательного
расслоения~$T^*\!Z$ будут, аналогично, отождествляться с
элементами сопряжённого пространства~$Z^*$. Дифференциалы линейных
функций при этом отождествляются с самими функциями, т.~е. вместо
$df$ пишем просто: $f$.

На инвариантном фазовом пространстве~$Z$, так же, как и на
обычном, имеется симплектическая структура~$\omega$:
\begin{equation}
 \omega = \int\limits_\Sigma d\sigma_\mu\,\delta j_\mu(x) \ ,
 \qquad \mbox{где} \ \
 \delta j_\mu(x) =
       \frac{\partial L}{\partial\,(\partial_\mu\varphi_i)}
            \wedge \varphi_i  \ .
\label{symplstr}
\end{equation}
Интегрирование ведётся по любой пространственно-подобной
гиперповерхности~$\Sigma$, не слишком плохо ведущей себя на
бесконечности. 2-форма~$\delta j_\mu$ называется симплектическим
током. Поскольку симплектический ток сохраняется, результат от
выбора поверхности~$\Sigma$ не зависит.

Произвол в выборе лагранжиана и возникающая отсюда неоднозначность
симплектической структуры будут рассмотрены в статье~[II]. В
разумном смысле можно сказать, что неоднозначность в выборе
лагранжиана не влияет на симплектическую структуру.

Симплектическая структура~$\omega$, как обычно~\cite{Arnold1989},
задаёт изоморфизм $I:T^*\!Z\to TZ$ кокасательного $T^*\!Z$ и
касательного $TZ$ расслоений\footnote{Строго говоря, о таком
изоморфизме можно говорить только после обсуждения топологии
инвариантного фазового пространства. Без такого обсуждения даже
понятие кокасательного расслоения не имеет ясного смысла. По
причинам, объяснённым в \rql Вводных замечаниях\rqr{} обсуждение
этого вопроса для случая релятивистских полей мы оставляем до
статьи [VI]. \par Также следует иметь в виду, что употребляемые
здесь величины типа $\varphi(x)$ при введении надлежащей топологии
оказываются не принадлежащими кокасательному расслоению фазового
пространства. Этот недостаток легко устраним, но до обсуждения
топологии мы предпочитаем просто не обращать на него внимания.}.
Он сопоставляет произвольной 1-форме $l^{\,\underline\cdot}\in
T^*\!Z$ касательный вектор $\underline{Il}\in TZ$ такой, что для
любого касательного вектора $\underline c\in TZ$ выполняется
равенство:
\begin{equation}
 \omega^{\,\underline c\,;\,\underline{Il}} = l^{\,\underline c} \ .
\label{cotanisom}
\end{equation}

Если форма $l^{\,\underline\cdot}$ является дифференциалом
некоторой функции $g^{\,\underline\cdot}$, т.~е.
$l^{\,\underline\cdot}=dg^{\,\underline\cdot}$, то мы говорим, что
$g^{\,\underline\cdot}$ является генератором векторного поля
$\underline{Il}=\underline{\smash{I\,dg}}$.

Скобка Пуассона двух функций $f^{\,\underline\cdot}$ и
$g^{\,\underline\cdot}$ определяется равенством:
\begin{equation}
 \{\,f\,,\,g\,\} = df^{\,\underline{\smash{I\,dg}}} \ .
\label{Poisdef}
\end{equation}
Если же функции $f^{\,\underline\cdot}$ и $g^{\,\underline\cdot}$
линейны, то это определение можно записать просто как:
\begin{equation}
 \{\,f\,,\,g\,\} = f^{\,\underline{\smash{Ig}}} \ .
\label{Poislin}
\end{equation}

Здесь нам достаточно будет ограничиться вещественными полями,
поэтому пространство~$Z$ будет вещественным линейным
пространством. Однако, оказывается полезным рассматривать
комплексные функции на фазовом пространстве. Их дифференциалы
принадлежат комплексифицированному сопряжённому пространству. По
этой причине вместо пространства~$Z^*$, мы будем всегда
использовать его комплексификацию~$Z^*_{\mathbb C}$. Кроме того,
при определении скобки Пуассона комплексных функций по
формуле~(\ref{Poisdef}) приходится рассматривать комплексные
векторы $\underline{\smash{I\,dg}}$. Этого можно было бы избежать,
определив скобки Пуассона несколько более
абстрактно\footnote{Комплексификация~$Z$ представляется
нежелательной, т.~к. это пространство можно рассматривать как
множество {\it физических\/} состояний, существующих в природе.
Пространство~$Z^*_{\mathbb C}$ при этом можно рассматривать как
множество {\it математических\/} величин, используемых для
описания физических состояний.}.


\myparagraph{Скалярное поле.} \label{ScalF}
 Рассмотрим скалярное поле. Его лагранжиан таков:
\[
\textstyle
 L = \frac12 \, \partial_\nu \varphi \, \partial_\nu \varphi -
     \frac{m^2}2 \, \varphi^2  \ .
\]

Уравнение движения~(\ref{EuLag}) принимает вид (уравнение
Клейна-Фока-Гор\-до\-на):
\begin{equation}
 \left(\partial^2\! + m^2\right)\varphi = 0        \ .
\label{KlFock}
\end{equation}

Симплектическая структура~(\ref{symplstr}) на инвариантном фазовом
пространстве~$Z^\mathrm{scal}$ в координатном представлении имеет
вид (указанный ещё в~\cite{Witten1986}):
\begin{equation}
 \omega =
 \int\limits_\Sigma d\sigma_\nu \,\partial_\nu\varphi\wedge\varphi \ .
\label{symplKF}
\end{equation}

Совершим теперь преобразование Фурье функции $\varphi(x)$ по
формулам~(\ref{Fourier}). Поскольку функция $\varphi(x)$
удовлетворяет уравнению~(\ref{KlFock}), её фурье-образ
$\widetilde\varphi(k)$ можно представить в виде:
\begin{equation}
 \widetilde\varphi(k) = 2\pi\,\delta(k^2-m^2) \cdot a(k)  \ ,
\label{defa}
\end{equation}
где $a(k)$~--- обычная функция, заданная на массовой поверхности
$k^2=m^2$.

Преобразуя теперь симплектическую структуру~(\ref{symplKF}) в
фурье-пред\-ста\-вле\-ние, имеем:
\begin{equation}
 \omega^{\,\underline\cdot\,;\,\underline\cdot} =
 \int d\mu_m \cdot i\,\varepsilon(k) \cdot
            a(-k)^{\,\underline\cdot} \, a(k)^{\,\underline\cdot} \ ,
 \qquad \mbox{где} \ \
 d\mu_m = \frac{d^4\!k}{(2\pi)^4} \cdot 2\pi\,\delta(k^2-m^2) \ .
\label{sKFmom}
\end{equation}
Под знаком интеграла стоит обычное произведение\footnote{Если
ввести обозначение $d\mu_m^+=d\mu_m \cdot \theta(k)$, то
формулу~(\ref{sKFmom}) можно переписать и в терминах внешнего
произведения форм:
\[
 \omega = \int d\mu_m^+ \cdot i\, a^*(k) \wedge a(k) \ .
\]}.
Подразумевается, что в правой части формулы аргументы
подставляются в том же порядке, в котором они следуют слева.

Чтобы теперь вычислить скобку Пуассона двух полевых величин
$\{\,\widetilde\varphi(k)\,,\,\widetilde\varphi(k')\,\}$, согласно
формуле~(\ref{Poislin}), рассмотрим вектор
$\underline{\smash{I\widetilde\varphi(k')}}$. Из
формул~(\ref{cotanisom}) и~(\ref{sKFmom}) имеем, что для любого
вектора~$\underline c$:
\[
 \int d\mu_m \cdot i\,\varepsilon(k)\cdot a(-k)^{\,\underline c\,}
      a(k)^{\,\underline{\smash{I\widetilde\varphi(k')}}} =
 \widetilde\varphi(k')^{\,\underline c}   \ .
\]
Используя линейную независимость формы~$a(k)^{\,\underline\cdot}$
при различных~$k$, $k^2=m^2$, а также определение этой
формы~(\ref{defa}), отсюда получаем, что
вектор~$\underline{\smash{I\widetilde\varphi(k')}}$
характеризуется соотношением:
\[
 a(k)^{\,\underline{\smash{I\widetilde\varphi(k')}}} =
 -\,i\,\varepsilon(k) \cdot (2\pi)^4\delta(k+k') \ .
\]

Отсюда, по формуле~(\ref{Poislin}), искомая скобка Пуассона равна:
\begin{equation}
 \{\,\widetilde\varphi(k)\,,\,\widetilde\varphi(k')\,\} =
 -\, \Big[ \,i\,\varepsilon(k) \cdot 2\pi\,\delta(k^2-m^2) \,\Big]
     \cdot (2\pi)^4\delta(k+k')   \ .
\label{PoisKF}
\end{equation}

Иногда бывает полезно иметь аналогичное соотношение в координатном
представлении. Чтобы его получить, нужно в формуле~(\ref{PoisKF})
совершить преобразование Фурье по переменным~$k$ и~$k'$:
\[
 \{\,\varphi(x)\,,\,\varphi(x')\,\} = - D_m(x-x') \ .
\]
Здесь функция~$D_m(y)$ имеет вид:
\begin{equation}
 D_m(y)=\int \frac{d^4\!k}{(2\pi)^4} \, e^{-iky} \cdot
        \Big[\, i\,\varepsilon(k) \cdot
                2\pi\, \delta(k^2-m^2) \,\Big]  \ .
\label{Dfunc}
\end{equation}


\myparagraph{\protect\rql Физическое\rqr{} электромагнитное поле.}
\label{emfield}
 Выберем лагранжиан электромагнитного поля
ка\-ли\-бро\-воч\-но-инвариантным:
\begin{equation}
\textstyle
 L = -\, \frac14\, F_{\mu\nu}\, F_{\mu\nu} =
     -\,{}^{[\mu\nu]}\,\partial_\mu A_\nu \, \partial_\mu A_\nu \ ,
 \qquad \mbox{где} \ \
 F_{\mu\nu}={}^{[\mu\nu]}\,2\,\partial_\mu A_\nu \ .
\label{emlag}
\end{equation}

Уравнение движения~(\ref{EuLag}) принимает вид (второе уравнение
Максвелла):
\begin{equation}
 \partial_\mu F_{\mu\nu}=0 \ .
\label{Maxwell}
\end{equation}

Под инвариантным фазовым пространством~$Z^\mathrm{phys}$ в данном
случае удобно понимать множество функций $F_{\mu\nu}(x)$,
удовлетворяющих уравнению~(\ref{Maxwell}). При этом значение
векторного потенциала~$A_\mu$ в данной точке~$x$ не является
хорошо определённой функцией на~$Z^\mathrm{phys}$: запись
вида~$A_\mu(x)^{\,\underline\cdot}$ не имеет смысла, пока не
указано, какая калибровка имеется в виду. По этой причине никакой
однозначной формулы для скобок Пуассона от векторного потенциала
написать нельзя.

Симплектическая структура~(\ref{symplstr}) в данном случае имеет
вид (указанный ещё в~\cite{Crnkovic1987,CrWitt1987}):
\begin{equation}
 \omega =
 - \int\limits_\Sigma d\sigma_\mu\, F_{\mu\nu} \wedge A_\nu  \ .
\label{symplem}
\end{equation}

Здесь уместно отметить, что хотя форма~$\omega$ выражена через
ка\-ли\-бро\-воч\-но-неинвариантную величину~$A_\nu$, сама она,
разумеется, ка\-ли\-бро\-воч\-но-инвариантна.

Выберем теперь векторный потенциал~$A_\mu$ так, чтобы он
удовлетворял условию Лоренца $\partial_\nu A_\nu=0$. При этом,
согласно уравнению~(\ref{Maxwell}), он будет удовлетворять
уравнению Даламбера: $\partial^2\! A_\mu=0$. Разобьём
выражение~(\ref{symplem}) на два слагаемых:
\[
 \omega =
 - \int\limits_\Sigma d\sigma_\mu\, \partial_\mu A_\nu \wedge A_\nu +
   \int\limits_\Sigma d\sigma_\mu\, \partial_\nu A_\mu \wedge A_\nu   \ .
\]
Второй из этих двух интегралов равен нулю. В этом можно убедиться
следующим образом. Во-первых, из условия Лоренца и антисимметрии
внешнего умножения следует, что подынтегральное выражение является
сохраняющимся током:
 $\partial_\mu(\partial_\nu A_\mu \wedge A_\nu)=0$.
Поэтому во втором интеграле, независимо от первого, можно перейти
от интегрирования по поверхности~$\Sigma$ к интегрированию по
поверхности $t=0$. Получаем интеграл:
\[
 \int d^3\!\vect x\, \partial_\nu A_0 \wedge A_\nu \ .
\]
Вследствие условия Лоренца его можно переписать так:
\[
 \int d^3\!\vect x\, \partial_\nu (A_0 \wedge A_\nu) \ .
\]
Но $A_0\wedge A_0=0$. Поэтому суммирование по индексу~$\nu$ можно
фактически проводить лишь от $1$ до $3$. Получаем:
\[
 \int d^3\!\vect x\, \partial_n (A_0 \wedge A_n) \ .
\]
Это интеграл от дивергенции трёхмерного вектора $A_0 \wedge A_n$.
По теореме Остроградского-Гаусса, он равен потоку этого вектора
через удалённую двумерную замкнутую поверхность. Этот поток равен
нулю, поскольку векторный потенциал предполагается выбранным так,
чтобы он там обращался в нуль.

Таким образом, при выполнении условия Лоренца
 $\partial_\nu A_\nu=0$, симплектическая структура~(\ref{symplem})
может быть записана в виде:
\begin{equation}
 \omega =
 - \int\limits_\Sigma d\sigma_\mu\, \partial_\mu A_\nu \wedge A_\nu  \ .
\label{symplemlor}
\end{equation}

Произведём теперь над функциями $A_\mu(x)$ и $F_{\mu\nu}(x)$
преобразование Фурье~(\ref{Fourier}). Точно так же, как в случае
скалярного поля, представим эти фурье-образы в виде:
\[
 \widetilde A_\mu(k) = 2\pi\,\delta(k^2) \cdot a_\mu(k)  \ , \qquad
 \widetilde F_{\mu\nu}(k) = 2\pi\,\delta(k^2) \cdot f_{\mu\nu}(k)  \ .
\]
Здесь $a_\mu(k)$ и $f_{\mu\nu}(k)$~--- обычные функции, заданные
на световом конусе $k^2=0$.

Симплектическая структура~(\ref{symplemlor}) в фурье-представлении
принимает вид\footnote{В терминах внешнего произведения её можно
записать так:
\[
 \omega = - \int d\mu_m^+ \,i\, a_\nu^*(k) \wedge a_\nu(k) \ .
\]}:
\begin{equation}
 \omega^{\,\underline\cdot\,;\,\underline\cdot} =
 - \int d\mu_m \cdot i\,\varepsilon(k)\cdot
       a_\nu(-k)^{\,\underline\cdot} \, a_\nu(k)^{\,\underline\cdot} \ .
\label{semmom}
\end{equation}

Формулы~(\ref{symplemlor}) и~(\ref{semmom}) внешне очень похожи на
соответствующие формулы~(\ref{symplKF}) и~(\ref{sKFmom}) для
скалярного поля. Точнее, они выглядят так же, как формулы для
симплектической структуры четырёх скалярных полей. Следует,
однако, иметь в виду, что инвариантное фазовое
пространство~$Z^\mathrm{phys}$, введённое в этом пункте, не
является прямой суммой инвариантных фазовых пространств четырёх
скалярных полей. По этой причине, как уже отмечалось, скобку
Пуассона для векторного потенциала вычислить нельзя.

Рассмотрим теперь, как устроено пространство~$Z^\mathrm{phys}$ в
фурье-пред\-ста\-вле\-нии. Это пространство можно рассматривать
как множество функций $a_\mu(k)$ на световом конусе $k^2=0$,
удовлетворяющих условию Лоренца $-\,i\,k_\mu\,a_\mu(k)=0$ и
рассматриваемых с точностью до калибровочного преобразования:
\[
 a_\mu(k) \ \sim\ a_\mu(k) - i\,k_\mu\,\lambda(k)  \ .
\]
Отсюда, из формулы~(\ref{semmom}), очевидна невырожденность
симплектической структуры на пространстве~$Z^\mathrm{phys}$.

Теперь ясно, что хотя и нельзя вычислить скобки Пуассона для
компонент векторного потенциала, можно вычислить скобки
 $\{\, \widetilde F_{\mu\rho}(k) \, , \,
       \widetilde F_{\nu\sigma}(k')\,\}$ и
$\{\,F_{\mu\rho}(x)\,,\,F_{\nu\sigma}(x')\,\}$. Эту задачу мы,
однако, отложим до пункта~\ref{Poisspp}.


\myparagraph{Добавка Гейзенберга-Паули.} \label{HeisPau}
 При обсуждении квантовой теории электромагнитного поля со времён
работы Гейзенберга и Паули~\cite{HeisPau1929}
лагранжиан~(\ref{emlag}) принято видоизменять, добавляя к нему
дополнительный член $\frac\varepsilon2\,(\partial_\mu A_\mu)^2$.
Для такого шага в современных учебниках даётся множество различных
\rql объяснений\rqr. В основном они делятся на два класса.

К первому классу относятся \rql объяснения\rqr, апеллирующие к
формальным трудностям построения гамильтонова формализма и
отслеживания релятивистской инвариантности. Как было показано в
пункте~\ref{emfield}, если тут и имелись какие-то трудности, то с
построением инвариантного гамильтонова формализма они исчезают.

Ко второму классу относятся \rql объяснения\rqr, так или иначе
апеллирующие к квантовой теории. Наша ближайшая задача~---
объяснить, что причины модификации лагранжиана присутствуют уже в
теории классической.

На первый взгляд, добавка
 $\frac\varepsilon2\,(\partial_\mu A_\mu)^2$
приводит только к фиксации калибровки: если условие Лоренца
выполнялось в прошлом, то оно будет выполняться и в будущем. В
действительности же имеется и более глубокое следствие: при
построении теории рассеяния и теории возмущений появляется
возможность избавиться от такой фикции, как разделение поля на
\rql собственное\rqr{} поле частиц и на \rql излучение\rqr{}.

Действительно, на практике оказывается невозможным решить
динамическую задачу точно. Поэтому вначале вычисляется какое-либо
\rql первое приближение\rqr{}, а к нему уже далее вычисляются \rql
радиационные по\-прав\-ки\rqr{}. Чтобы учесть радиационную отдачу,
приходится тем или иным способом делить поле на \rql
собственное\rqr{} и \rql радиационное\rqr{}. Сделать это вполне
корректно, однако, невозможно даже в простейших случаях. Это не
удивительно, поскольку в самом лагранжевом формализме никакого
намёка на такое разделение не содержится.

Можно, однако, подходить к задачам рассеяния иначе. Рассмотрим для
примера рассеяние двух частиц, взаимодействующих друг с другом
посредством электромагнитного поля. Будем исходить из какой-нибудь
наивной механической аналогии: скажем, двух тяжёлых шариков,
скользящих по гладкой упругой мембране. Электромагнитному полю при
этом соответствует деформированная мембрана.

Если не делить поле на собственное и радиационное, приходится
считать, что всё поле частицами \rql со\-зда\-ёт\-ся\rqr. Чтобы
это как-то формализовать, нужно ввести процедуру включения и
выключения взаимодействия.

Процесс рассеяния частиц и электромагнитного поля будет при этом
описываться так:
\begin{enumerate}
\item
В далёком прошлом частицы являются свободными и с полем не
взаимодействуют. Присутствует также свободное поле, находящееся
далеко от частиц.
\item
Затем, пока частицы ещё далеко друг от друга, взаимодействие с
электромагнитным полем адиабатически включается. Частицы при этом
облачаются в собственное поле.
\item
После того, как взаимодействие полностью включено, частицы (и
поле) подлетают друг к другу близко, взаимодействуют и снова
разлетаются.
\item
Когда они достаточно далеко разлетятся, взаимодействие
адиабатически выключается.
\item
После этого остаются голые частицы, в окрестности которых никакого
поля нет, и свободное поле, с частицами не связанное.
\end{enumerate}

У шариков, скользящих по упругой мембране, включение и выключение
взаимодействия осуществляется очевидным образом: нужно включить и
выключить гравитационное поле. В случае же электромагнитного поля
так просто поступить нельзя. Рассмотрим, например, взаимодействие
электромагнитного поля $A_\mu(x)$ с заданным током $J_\mu(x)$.
Лагранжиан в этом случае имеет вид:
\begin{equation}
\textstyle
 L = -\, \frac14\, F_{\mu\nu}\, F_{\mu\nu} - \alpha\,J_\mu\,A_\mu \ .
\label{gaugeinvlag}
\end{equation}
Здесь $\alpha(x)$~--- функция, включающая и выключающая
взаимодействие. В дальнейшем, чтобы формулы записывались короче,
мы будем использовать обозначение:
\[
 J^\alpha_\mu(x) = \alpha(x)\,J_\mu(x)    \ .
\]

Из лагранжиана~(\ref{gaugeinvlag}) вытекает уравнение движения:
\begin{equation}
 \partial_\mu F_{\mu\nu}=J^\alpha_\nu \ .
\label{emeq}
\end{equation}
Применяя теперь к обеим частям операцию~$\partial_\nu$,
убеждаемся, что при $\partial_\nu J^\alpha_\nu\neq0$
уравнение~(\ref{emeq}) решений не имеет. Таким образом,
взаимодействие включить и выключить невозможно.

Если же к лагранжиану добавить член
$\frac\varepsilon2\,(\partial_\mu A_\mu)^2$, то есть перейти к
лагранжиану
\[
\textstyle
 L = -\, \frac14\, F_{\mu\nu}\, F_{\mu\nu} +
     \frac\varepsilon2\,(\partial_\mu A_\mu)^2 -
     J^\alpha_\mu\,A_\mu \ ,
\]
то уравнение движения принимает вид:
\begin{equation}
 \partial_\mu F_{\mu\nu} -
 \varepsilon\,\partial_\nu\partial_\mu A_\mu =
 J^\alpha_\nu \ ,
\label{modeq}
\end{equation}
и решение~$A_\mu(x)$ легко отыскивается для любой
функции~$\alpha(x)$.

Далее мы будем всюду полагать $\varepsilon=-1$. При этом
уравнение~(\ref{modeq}) принимает особенно простой вид:
\begin{equation}
 \partial^2\! A_\nu = J^\alpha_\nu \ .
\label{DAlabA}
\end{equation}
Его решение~--- это потенциал Лиенара-Вихерта. По этой причине
указанную \rql калибровку\rqr{} следовало бы назвать калибровкой
Лиенара-Вихерта. В квантовой электродинамике она, однако,
называется калибровкой Фейнмана.

Если от обеих частей уравнения (\ref{DAlabA}) взять четырёхмерную
дивергенцию, получим уравнение:
\[
 \partial^2(\partial_\nu A_\nu) = \partial_\nu J^\alpha_\nu    \ .
\]
Отсюда нетрудно показать, что если взаимодействие включается
адиабатически, и условие Лоренца $\partial_\nu A_\nu=0$
выполнялось в далёком прошлом, то оно останется ненарушенным
всюду.

Отсюда также следует, что дополнительный член
$\frac12\,(\partial_\mu A_\mu)^2$ в лагранжиане будет вносить
нулевой вклад в вариацию действия. Поэтому в той области, где
взаимодействие полностью включено, система из частиц и поля будет
удовлетворять тем же уравнениям, что и система без добавки в
лагранжиане.

Таким образом, введение в лагранжиан дополнительного члена не
меняет в существенном картину рассеяния. Но при этом отпадает
необходимость выделять каким-либо способом \rql собственное\rqr{}
поле частиц, и появляется возможность для построения локальной
теории возмущений.

Сейчас представляется затруднительным сказать, насколько далеко
можно продвинуться в построении теории возмущений для классической
электродинамики. Ясно только, что без введения дополнительного
члена в лагранжиан ситуация была бы совершенно безнадёжной.


\myparagraph{\protect\rql Нефизическое\rqr{} электромагнитное
поле.} \label{nonphys}
 Рассмотрим теперь снова свободное электромагнитное поле, но в
его лагранжиан уже сразу включим добавку
$-\,\frac12\,(\partial_\mu A_\mu)^2$. Лагранжиан тогда можно
переписать в следующем виде:
\[
\textstyle
 L = -\, \frac14\, F_{\mu\nu}\, F_{\mu\nu} -
     \frac12\,(\partial_\mu A_\mu)^2 =
     -\, \frac12\,\partial_\mu A_\nu\,\partial_\mu A_\nu +
     \frac12\,\partial_\mu(A_\nu\,\partial_\nu A_\mu -
                           A_\mu\,\partial_\nu A_\nu)          \ .
\]

Как будет показано в статье~[II], член, являющийся полной
дивергенцией, можно отбросить. После этого лагранжиан принимает
замечательно простой вид:
\[
\textstyle
 L = -\, \frac12\,\partial_\mu A_\nu\,\partial_\mu A_\nu   \ ,
\]
т.~е. он по сути является лагранжианом четырёх скалярных
полей\footnote{Член, отвечающий полю с индексом $\nu=0$, входит с
отрицательным знаком.} с $m=0$, рассмотренных в
пункте~\ref{ScalF}. Уравнение движения имеет вид:
\[
 \partial^2\! A_\mu = 0 \ .
\]

Инвариантное фазовое пространство теперь является прямой суммой
четырёх фазовых пространств скалярных полей. Обозначим его
как~$Z^4$. Симплектическая структура в координатном и
фурье-представлениях даётся формулами:
\begin{equation}
\omega =
 - \int\limits_\Sigma d\sigma_\mu\, \partial_\mu A_\nu \wedge
                                                 A_\nu         \ .
\label{symcoor2}
\end{equation}
\begin{equation}
 \omega^{\,\underline\cdot\,;\,\underline\cdot} =
 - \int d\mu_m \cdot i\,\varepsilon(k)\cdot
          a_\nu(-k)^{\,\underline\cdot} \,
          a_\nu(k)^{\,\underline\cdot} \ .
\label{symmom2}
\end{equation}

Формулы~(\ref{symcoor2}) и~(\ref{symmom2}) выглядят точно так же,
как соответствующие формулы из пункта~\ref{emfield}. Однако,
разница в содержании этих формул существенна: теперь мы можем
вычислить скобку Пуассона для векторного потенциала. В
фурье-представлении она имеет вид:
\begin{equation}
 \{\,\widetilde A_\mu(k)\,,\,\widetilde A_\nu(k')\,\} =
 g_{\mu\nu}
 \cdot \Big[\, i\,\varepsilon(k)
               \cdot 2\pi\,\delta(k^2) \,\Big]
 \cdot (2\pi)^4\delta(k+k')       \ .
\label{PoissAmom}
\end{equation}
В координатном представлении:
\begin{equation}
 \{\,A_\mu(x)\,,\,A_\nu(x')\,\} = g_{\mu\nu}\,D_0(x-x')       \ .
\label{PoissAcoord}
\end{equation}
Здесь $D_0(y)$~--- функция~(\ref{Dfunc}) при $m=0$. В этом случае
её можно записать просто как:
\[
 D_0(y)=\frac1{2\pi}\,\varepsilon(y)\,\delta(y^2)  \ .
\]

Подействуем теперь на обе части формулы~(\ref{PoissAmom}) символом
 ${}^{[\rho\mu]}\,2\,(-\,i\,k_\rho) \,
  {}^{[\sigma\nu]}\,2\,(-\,i\,k'_\sigma)$.
Получаем формулу для скобки Пуассона тензора электромагнитного
поля в фурье-представлении:
\begin{equation}
 \{\,\widetilde F_{\mu\rho}(k)\,,\,\widetilde F_{\nu\sigma}(k')\,\} =
 {}^{[\mu\rho]}\, {}^{[\nu\sigma]}\,
     4\, g_{\mu\nu}\, k_\rho\, k_\sigma \cdot
 \Big[\, i\,\varepsilon(k) \cdot 2\pi\,\delta(k^2) \,\Big]
 \cdot (2\pi)^4\delta(k+k')       \ .
\label{PoissFmom}
\end{equation}

В координатном представлении это соотношение переписывается так:
\begin{equation}
 \{\,F_{\mu\rho}(x)\,,\,F_{\nu\sigma}(x')\,\} =
 - \,{}^{[\mu\rho]}\, {}^{[\nu\sigma]}\, 4\, g_{\mu\nu}\,
     \partial_\rho\, \partial_\sigma D_0(x-x')     \ .
\label{PoissFcoord}
\end{equation}
Здесь обе производные берутся по переменной без штриха.

Отметим сразу, что формулы~(\ref{PoissFmom}) и (\ref{PoissFcoord})
совпадают с формулами для скобок Пуассона \rql физического\rqr{}
электромагнитного поля, описанного в пункте~\ref{emfield}. Это
будет доказано в пункте~\ref{Poisspp}.


\myparagraph{Энергия электростатического поля.}
 В статье~[III] получена формула, с помощью которой
можно в рамках инвариантного гамильтонова формализма получать
генераторы группы Пуанкаре. С её помощью легко получаем формулу
для импульса \rql нефизического\rqr{} электромагнитного поля:
\begin{equation}
 P_\nu = - \int d\mu^+_m \cdot k_\nu\, a^*_\rho(k)\,a_\rho(k) \ .
\label{emmom}
\end{equation}

Из этой формулы видно, что энергия \rql нефизического\rqr{} поля
не является положительно-определённой. Более того, из этой формулы
следует, что энергия, запасённая в электростатическом поле,
созданном системой неподвижных зарядов, всегда {\it отрицательна}.

Здесь мы обсудим возникший парадокс чуть подробнее. Это нужно
сделать потому, что он обычно неправильно трактовался в рамках
{\it квантовой\/} теории (см., например,
\cite{Gupta1950,Gupta1957,Nagy1969,Haag1961}).

Проблема отрицательности энергии \rql временн\'ых\rqr{} фотонов в
квантовой электродинамике была замечена давно. Считалось, что
квантование в пространстве с индефинитным скалярным произведением
позволяет переменить знак этой энергии. И это рассматривалось как
один из аргументов для введения индефинитного скалярного
произведения. Так казалось потому, что при индефинитном скалярном
произведении в квантовом случае не вполне ясно, что следует
называть энергией состояния: среднее значение гамильтониана или
его собственное число. Эта неясность, конечно, происходила от
того, что, во-первых, отсутствовала ясная концепция квантования, а
во-вторых, классическая теория не была сформулирована должным
образом.

Из проведённого рассмотрения, однако, видно, что обсуждаемая
энергия {\it должна\/} быть отрицательной: только такая величина
представляется естественной с точки зрения релятивистской теории.
А индефинитность метрики в квантовом случае здесь вообще ни при
чём.

Стоит также сразу сказать, что понятие вакуума поля мы вовсе не
связываем с минимальностью энергии. У классического поля
вакуум~--- это просто нулевой вектор пространства~$Z$. Понятие же
вакуума поля квантового мы введём при описании операции
квантования в статье~[VI].

Как же быть с положительностью электростатической энергии, хорошо
известной в обычной электростатике? Дело в том, что в это понятие
традиционно вкладывают иной смысл: под ним подразумевают работу,
которую совершит система, когда заряды разобьют на бесконечно
малые кусочки и медленно разнесут эти кусочки далеко друг от
друга. Формула же~(\ref{emmom}) даёт энергию, которая останется у
поля, если мгновенно выключить его взаимодействие с зарядами.

Более подробно этот вопрос мы здесь обсуждать не будем. Заметим
лишь вот что. В электростатике имеется следующая формула для
энергии:
\[
 E^\mathrm{\,full}={\textstyle\frac12}\int\varphi\,dQ   \ .
\]
Множитель $\frac12$ при этом обычно объясняют тем, что если
использовать формулу для энергии во внешнем поле, то взаимная
энергия любых двух элементов заряда учтётся дважды.

В рамках же той теории, которую мы здесь рассматриваем, поле
всегда рассматривается как внешнее. Поэтому можно сказать, что
энергия зарядов в электростатическом поле всегда даётся формулой:
\[
 E^\mathrm{\,charges} = \int\varphi\,dQ \ .
\]
К ней нужно ещё добавить энергию поля:
\[
 E^\mathrm{\,field} = \,- \,{\textstyle\frac12}\int\varphi\,dQ  \ .
\]
В сумме $E^\mathrm{\,charges}$ и $E^\mathrm{\,field}$ как раз дают
$E^\mathrm{\,full}$.


\myparagraph{Состояния рассеяния \protect\rql нефизического\rqr{}
поля.} \label{ScatStat}
 Рассмотрим теперь рассеяние \rql нефизического\rqr{} поля на
заданном токе. Как объяснено в пункте~\ref{HeisPau}, будем
считать, что взаимодействие с электромагнитным полем включается и
выключается адиабатически, а частицы, составляющие ток, ускоряются
только в той области пространства-времени, где взаимодействие
полностью включено, т.~е. $\alpha(y)=1$.

В произвольной точке пространства-времени поле будет состоять из
двух компонент: из поля, которое существовало в удалённом прошлом
и из поля, которое создали частицы. Это последнее добавочное поле
можно вычислить по формуле Лиенара-Вихерта:
\begin{equation}
 A_\mu^\mathrm{ret}(x) = \int d^4\!x'\,
      D_0^\mathrm{c}(x-x')\,J^\alpha_\mu(x')  \ ,
 \qquad \mbox{где} \ \
 D_0^\mathrm{c}(y)=\theta(y)\,D_0(y)  \ .
\label{LienWiech}
\end{equation}

В инвариантном гамильтоновом формализме под состоянием поля
подразумевается решение соответствующего однородного уравнения. В
данном случае это уравнение $\partial^2\!A_\mu=0$. Поэтому, когда
мы говорим о состоянии поля в удалённом прошлом или в удалённом
будущем, естественно при этом функцию $A_\mu(x)$ продолжить с
помощью однородного уравнения на всё пространство. Таким образом
мы приходим к определению in- и out- полей:
 $A_\mu(x)^{\,\underline{\mathrm{in}}}$ и
 $A_\mu(x)^{\,\underline{\mathrm{out}}}$. Здесь уже
 $\underline{\mathrm{in}}$ и $\underline{\mathrm{out}}$~--- векторы
инвариантного фазового пространства~$Z^4$. Эти векторы естественно
назвать in- и out- состояниями. Разность этих двух векторов
отвечает излучённому полю:
\[
 \underline{\mathrm{out}} - \underline{\mathrm{in}} =
 \underline{\mathrm{rad}} \ .
\]

Чтобы найти поле $A_\mu(x)^{\,\underline{\mathrm{rad}}}$, нужно
формулу Лиенара-Вихерта~(\ref{LienWiech}) слегка подправить:
\[
 A_\mu(x)^{\,\underline{\mathrm{rad}}} =
 \int d^4\!x'\, D_0(x-x')\,J^\alpha_\mu(x')  \ .
\]
В фурье-представлении эта формула записывается особенно просто:
\[
 \widetilde A_\mu(k)^{\,\underline{\mathrm{rad}}} =
 i\,\varepsilon(k)\cdot 2\pi\,\delta(k^2)\cdot
    \widetilde J^\alpha_\mu(k) \ ,
\]
или, с использованием несингулярной функции $a_\mu(k)$ на световом
конусе, ещё короче:
\begin{equation}
 a_\mu(k)^{\,\underline{\mathrm{rad}}} =
 i\,\varepsilon(k)\,\widetilde J^\alpha_\mu(k) \ .
\label{myLienWiech}
\end{equation}

Видно, что в приведённых формулах можно перейти к адиабатическому
пределу. При этом  множество всех возможных векторов
$\underline{\mathrm{rad}}$ образует линейное подпространство
инвариантного фазового пространства~$Z^4$. Обозначим его
как~$Z^\perp$. Как было разъяснено в пункте~\ref{HeisPau}, при
адиабатическом включении и выключении взаимодействия условие
Лоренца не нарушается. Поэтому пространство~$Z^\perp$ не совпадает
со всем~$Z^4$. С другой стороны, из формулы~(\ref{myLienWiech})
следует, что кроме условия Лоренца никаким другим условием
пространство~$Z^\perp$ не ограничено. Таким образом, $Z^\perp$~---
это подпространство пространства~$Z^4$, выделяемое условием
Лоренца:
\begin{equation}
 \underline{\mathrm{rad}} \in Z^\perp
 \ \ \Longleftrightarrow \ \
   -\,i\,k_\mu\, a_\mu(k)^{\,\underline{\mathrm{rad}}} = 0 \ .
\label{Lorcond0}
\end{equation}
Этим объясняется обозначение~$Z^\perp$.

Естественно считать, что все возможные in-состояния когда-то были
созданы из вакуума при взаимодействии с токами. Тогда векторы in-
и out- полей также принадлежат подпространству~$Z^\perp$.
Поскольку все возможные поля излучения заполняют всё~$Z^\perp$, то
все возможные in- и out- поля его тоже заполняют.

Подставляя условие~(\ref{Lorcond0}) в формулу~(\ref{emmom}) для
вектора энергии-импульса, замечаем, что состояния из
пространства~$Z^\perp$ обладают неотрицательной энергией.


\myparagraph{Остатки калибровочной инвариантности.}
 Рассмотрим теперь такой вектор
 $\underline{\smash{\mathrm{gauge}}}\in Z$, что величину
 $a_\mu(k)^{\,\underline{\smash{\mathrm{gauge}}}}$
можно записать в виде:
\[
 a_\mu(k)^{\,\underline{\smash{\mathrm{gauge}}}} =
 -\,i\,k_\mu\,\lambda(k) \ ,
\]
где $\lambda(k)$~--- некоторая функция на световом конусе.
Множество векторов такого вида образует линейное подпространство
в~$Z^4$. Обозначим это подпространство~$Z^\parallel$. Очевидно,
$Z^\parallel$ содержится в~$Z^\perp$.

Рассмотрим опять рассеяние частиц и поля, когда движение частиц
заранее не задано, и присутствует, вообще говоря, ненулевое
in-поле. Это рассеяние описывается лагранжианом:
\[
\textstyle
 L = L^{\mathrm{particles}}
     - \frac12\,\partial_\mu A_\nu\,\partial_\mu A_\nu
     - \alpha\,J_\mu\,A_\mu  \ .
\]
Будем считать, что соответствующие уравнения движения разрешимы, и
траектории частиц и поле $A_\mu(x)$, для которых действие
стационарно, найдены. И будем также считать, что адиабатическое
включение (выключение) взаимодействия происходит в столь удалённом
прошлом (будущем), что все частицы находятся далеко друг от друга,
поле $A_\mu(k)^{\,\underline{\mathrm{in}}}$
($A_\mu(k)^{\,\underline{\mathrm{out}}}$) находится в основном
далеко от всех частиц и поле
 $A_\mu(k)^{\,\underline{\smash{\mathrm{gauge}}}}$ также находится
далеко от всех частиц.

Добавим теперь к полю $A_\mu(x)$ поле
$A_\mu(x)^{\,\underline{\smash{\mathrm{gauge}}}}$. Иными словами,
совершим калибровочное преобразование:
\begin{equation}
 A_\mu(x) \to A_\mu(x) + \partial_\mu\Lambda(x) \ ,
\label{GaugeTrans}
\end{equation}
где $\Lambda(x)$~--- функция, фурье-образ которой связан с
$\lambda(k)$ соотношением:
\[
 \widetilde\Lambda(k) = 2\pi\,\delta(k^2)\cdot\lambda(k)  \ .
\]

В области, где взаимодействие включается или выключается, эта
добавка на стационарность действия повлиять не может. В самом
деле, поскольку частицы находятся вне области, где эта добавка
присутствует, вариации членов $L^{\mathrm{particles}}$ и
$-\,\alpha\,J_\mu\,A_\mu$ измениться не могут. Член же
$-\,\frac12\,\partial_\mu A_\nu\,\partial_\mu A_\nu$ отличается от
калибровочно-инвариантного на полную дивергенцию, которая изменить
вариацию не может, и на член $-\,\frac12\,(\partial_\mu A_\mu)^2$,
вариация которого равна нулю, т.~к. $\partial_\mu A_\mu=0$.

Рассмотрим теперь область, где взаимодействие полностью включено.
При преобразовании~(\ref{GaugeTrans}) вариации членов
$L^{\mathrm{particles}}$ и
 $-\,\frac12\,\partial_\mu A_\nu\,\partial_\mu A_\nu$ измениться
не могут по тем же самым причинам, что и в случае областей, где
взаимодействие включается или выключается. Остаётся член
$-\,J_\mu\,A_\mu$. При калибровочном
преобразовании~(\ref{GaugeTrans}) он переходит в
 $-\,J_\mu\,A_\mu -\,J_\mu\,\partial_\mu\Lambda(x)$. Следовательно,
в вариации действия появляется дополнительный член
 $-\,\delta J_\mu\,\partial_\mu\Lambda(x)$. Интегрируя его по
частям и учитывая сохранение тока, получаем, что он
дополнительного вклада в вариацию не даёт.

Таким образом, при преобразовании~(\ref{GaugeTrans}) движения,
удовлетворяющие условию стационарности действия, переходят в
движения также удовлетворяющие этому условию. Можно сказать, что
подпространство~$Z^\parallel$ в процессе рассеяния фактически
никак не участвует.

Естественно предположить, что состояния рассеяния
электромагнитного поля наблюдаются только по их влиянию на
заряженные частицы. Тогда любые два вектора из $Z^\perp$,
отличающиеся на вектор из $Z^\parallel$, физически неразличимы.
Следовательно, вместо пространства $Z^\perp$ можно рассматривать
факторпространство $Z^{\perp/\parallel}=Z^\perp/Z^\parallel$.

Заметим теперь, что в фурье-представлении
пространство~$Z^{\perp/\parallel}$ можно рассматривать как
множество функций $a_\mu(k)$ на световом конусе $k^2=0$,
удовлетворяющих условию Лоренца $-\,i\,k_\mu\,a_\mu(k)=0$ и
рассматриваемых с точностью до калибровочного преобразования:
\[
 a_\mu(k) \ \sim\ a_\mu(k) - i\,k_\mu\,\lambda(k)  \ .
\]
Следовательно, можно установить естественное взаимно-однозначное
соответствие между элементами пространств~$Z^{\perp/\parallel}$ и
$Z^\mathrm{phys}$.

Более того, пространство~$Z^{\perp/\parallel}$ естественным
образом наследует из полного пространства~$Z^4$ симплектическую
структуру~(\ref{symmom2}), и эта структура, очевидно, совпадает с
симплектической структурой пространства~$Z^\mathrm{phys}$,
даваемой формулой~(\ref{semmom}). Таким образом,
$Z^{\perp/\parallel}$ и $Z^\mathrm{phys}$ естественно
отождествляются и как {\it симплектические} пространства.


\myparagraph{О скобке Пуассона на $Z^{\perp/\parallel}$.}
\label{Poisspp}
 Рассмотрим теперь на пространстве~$Z^4$ линейную функцию
$\widetilde F_{\mu\rho}(k)^{\,\underline\cdot}$ (или
$F_{\mu\rho}(x)^{\,\underline\cdot}$). При этом $k$ (или,
соответственно, $x$) считается фиксированным. Как было сказано в
пункте~\ref{InvHamForm}, эта функция является генератором
векторного поля на~$Z^4$.

Заметим, во-первых, что поскольку рассматриваемая функция линейна
на~$Z^4$, её дифференциал не зависит от точки пространства~$Z^4$
(и совпадает с самой этой функцией). Следовательно, полученное
векторное поле постоянно на всём пространстве~$Z^4$.

Во-вторых, дифференциал рассматриваемой функции, как и сама
функция, обращается в нуль на~$Z^\parallel$. Поэтому, как видно из
формулы для симплектической структуры~(\ref{symmom2}), вектор
полученного постоянного поля лежит в подпространстве~$Z^\perp$.

Из этих двух утверждений следует, что полученное поле естественным
образом сужается на факторпространство~$Z^{\perp/\parallel}$.

С другой стороны, пространство~$Z^{\perp/\parallel}$ само наделено
симплектической структурой. Если сузить функцию
 $\widetilde F_{\mu\rho}(k)^{\,\underline\cdot}$ (или
 $F_{\mu\rho}(x)^{\,\underline\cdot}$) на это факторпространство,
то полученная функция сама будет генератором некоторого векторного
поля в~$Z^{\perp/\parallel}$.

Однако, поскольку симплектическая структура
в~$Z^{\perp/\parallel}$ наследуется из~$Z^4$, легко видеть, что
оба полученных векторных поля в~$Z^{\perp/\parallel}$ совпадают.

Отсюда также следует, что формулы~(\ref{PoissFmom})
и~(\ref{PoissFcoord}), полученные в пункте~\ref{nonphys} для \rql
нефизического\rqr{} поля, годятся и для \rql физического\rqr.




\newcounter{parnum}
\newlength{\picwidth}
\newcommand*{\myparagraph}[1]{\refstepcounter{parnum}%
\paragraph{\arabic{parnum}. #1}}


\title{On quantization of electromagnetic field.\\
       II. Arbitrariness in choice of Lagrangian.}
\author{D.~A.~Arbatsky\footnote{http:\dslash d-a-arbatsky.narod.ru/}}
\date{\today}
\maketitle

\begin{abstract}
Here we show that addition to Lagrangian a divergence of a
function does not change the symplectic structure on invariant
phase space.
\end{abstract}


 It is well-known that the same dynamics can be obtained by
different choices of Lagrangian. The simplest replacement of this
type is a linear substitution:
\[
 L\,\longrightarrow\,L'=\alpha L   \ .
\]
So far as in these papers we study the questions of quantization
and normalize action by Planck's constant, $\hbar=1$, this
arbitrariness is absent in our research.

But there is another essential arbitrariness which is the addition
to the Lagrangian a full divergence\footnote{In principle, there
are other, more complicated substitutions which usually do not
play a practical role because of their complexity.}:
\begin{equation}
 L\,\longrightarrow\,L'=L+\partial_\mu F_\mu   \ ,
\label{LPlusDiv}
\end{equation}
where $F_\mu$ is some vector function of the coordinate $x$ and
field values $\varphi_i(x)$.

In general case the Lagrangian may contain derivatives of the
field with respect to coordinates of the order higher than first.
Here we will take into account such a possibility and we will
accept that the function $F_\mu$ may also depend on the
derivatives of the field with respect to coordinates.

So far as the substitution (\ref{LPlusDiv}) does not change
dynamics, it leaves the invariant phase space $Z$ unchanged. Now
we will show that the symplectic structure on $Z$ also remains
unchanged. In fact, this means that the whole invariant
Hamiltonian formalism does not feel such an addition. And so far
as quantization is based on invariant Hamiltonian formalism, we
get that the construction of the quantized field does not change
under the substitution~(\ref{LPlusDiv}).

It seems to be difficult to prove the invariability of the
symplectic structure using the formula for the symplectic current
as a starting point. In this connection we will remind here how
symplectic structure is defined directly in the variational terms
\cite{Witten1986,Crnkovic1987,Zucker1987,CrWitt1987,%
Crnkovic1988,BHS1991}.

Consider the action in an arbitrary region $\Omega$:
\begin{equation}
 S = \int\limits_\Omega d^4\!x\,
     L(x,\varphi_i,\partial_\mu\varphi_i,\dots,
       \partial_{\mu_1}\dots\partial_{\mu_k}\varphi_i) \ .
\label{SInOmega}
\end{equation}
Its variation, if the region is unchanged, is:
\begin{equation}
 \delta S = \int\limits_\Omega d^4\!x\,
 \left(\frac{\delta L}{\delta\varphi_i}\,\delta\varphi_i +
       \partial_\mu j_\mu\right)   \ ,
\label{VarSInOmega}
\end{equation}
Here
\[
 j_\mu =
 \frac{\delta L}{\delta(\partial_\mu\varphi_i)}\,\delta\varphi_i +
 \dots +
 \frac{\delta L}
      {\delta(\partial_\mu\partial_{\mu_2}\dots
              \partial_{\mu_k}\varphi_i)}\,
 \partial_{\mu_2}\dots\partial_{\mu_k}\delta\varphi_i   \ .
\]
Variational derivative here is understood as
\[
 \frac{\delta}{\delta a} =
 \frac{\partial}{\partial a} -
 \partial_\mu\frac{\partial}{\partial(\partial_\mu a)} +
 \partial_\mu\partial_\nu
    \frac{\partial}{\partial(\partial_\mu\partial_\nu a)} -
 \dots    \ .
\]
The action (\ref{SInOmega}) can be considered as a functional on
the set of functions $\varphi_i$. Using the identity $\delta^2=0$
we get:
\[
\int\limits_\Omega d^4\!x\,
 \left(\delta\left(\frac{\delta L}{\delta\varphi_i}\,
                   \delta\varphi_i\right) +
       \partial_\mu\delta j_\mu\right) = 0  \ .
\]
If we consider here variations only for functions $\varphi_i$ for
which action is stationary, then the first term in the last
equation becomes zero. From here we get that the symplectic
structure defined on $Z$ as
\[
 \omega = \int\limits_\Sigma d\sigma_\mu\,\delta j_\mu(x) \
\]
does not depend on the choice of the space-like surface $\Sigma$
(it is implied that $\Sigma$ behaves well enough at the infinity).

Let us consider now how the above equations are changed under the
substitution (\ref{LPlusDiv}).

Variation of the new action $S'$ can be written like in the
formula (\ref{VarSInOmega}):
\[
 \delta S' = \int\limits_\Omega d^4\!x\,
 \left(\frac{\delta L'}{\delta\varphi_i}\,\delta\varphi_i +
       \partial_\mu j'_\mu\right) =
 \int\limits_\Omega d^4\!x\,
 \left(\frac{\delta L}{\delta\varphi_i}\,\delta\varphi_i +
       \frac{\delta(\partial_\mu F_\mu)}{\delta\varphi_i}\,
              \delta\varphi_i +
       \partial_\mu j'_\mu\right)      \ .
\]
But $\delta(\partial_\mu F_\mu)/\delta\varphi_i = 0$. Therefore,
\begin{equation}
 \delta S' = \int\limits_\Omega d^4\!x\,
 \left(\frac{\delta L}{\delta\varphi_i}\,\delta\varphi_i +
       \partial_\mu j'_\mu\right)      \ .
\label{VarSDash1}
\end{equation}
On the other hand, this variation can be written in the terms of
the old action:
\begin{equation}
  \delta S' = \delta S +
     \int\limits_\Omega d^4\!x\,\partial_\mu\delta F_\mu    \ .
\label{VarSDash2}
\end{equation}
Combining (\ref{VarSDash1}) and (\ref{VarSDash2}) with
(\ref{VarSInOmega}) we get:
\[
 \int\limits_\Omega d^4\!x\,
  (\partial_\mu j'_\mu -
    \partial_\mu j_\mu - \partial_\mu\delta F_\mu) = 0 \ .
\]
So far as in this equation all three functions $j'_\mu$, $j_\mu$,
and $\delta F_\mu$ are local (i.~e. they become zero in regions
where $\delta\varphi_i=0$) we have:
\[
 \int\limits_\Sigma d\sigma_\mu\,
 (j'_\mu - j_\mu - \delta F_\mu) = 0 \ .
\]
The value $F_\mu$ can be considered as a functional on the set of
functions $\varphi_i$. Using the identity $\delta^2=0$ we get:
\[
 \int\limits_\Sigma d\sigma_\mu\,\delta j'_\mu(x) =
 \int\limits_\Sigma d\sigma_\mu\,\delta j_\mu(x)   \ .
\]
I.~e. the substitution (\ref{LPlusDiv}) leaves the symplectic
structure on $Z$ unchanged:
\[
 \omega' = \omega  \ .
\]




\newcounter{parnum}
\newlength{\picwidth}
\newcommand*{\myparagraph}[1]{\refstepcounter{parnum}%
\paragraph{\arabic{parnum}. #1}}


\title{О квантовании электромагнитного поля.\\
       II. Неоднозначность выбора лагранжиана.}
\author{Д.~А.~Арбатский\footnote{http:\dslash d-a-arbatsky.narod.ru/}}
\date{\today}
\maketitle

\begin{abstract}
Показано, что добавление к лагранжиану дивергенции некоторой
функции не меняет симплектической структуры на инвариантном
фазовом пространстве.
\end{abstract}


 Как известно, к одной и той же динамике могут приводить
различные выборы лагранжиана. Простейшей подстановкой такого рода
является линейная замена:
\[
 L\,\longrightarrow\,L'=\alpha L   \ .
\]
Поскольку в данном цикле статей мы изучаем вопросы квантования и
нормируем действие на постоянную Планка, $\hbar=1$, то указанный
произвол в нашем рассмотрении отсутствует.

Есть, однако, другой существенный произвол, заключающийся в
добавлении к лагранжиану полной дивергенции\footnote{В принципе,
существуют и другие, более сложные подстановки, которые ввиду
своей сложности практической роли обычно не играют.}:
\begin{equation}
 L\,\longrightarrow\,L'=L+\partial_\mu F_\mu   \ ,
\label{LPlusDiv}
\end{equation}
где $F_\mu$~--- некоторая векторная функция от координаты $x$ и
полевых переменных $\varphi_i(x)$.

В общем случае лагранжиан может содержать производные от поля по
координатам выше первого порядка. Мы будем здесь иметь в виду
такую возможность и будем считать, что функция $F_\mu$ также может
зависеть от производных поля по координатам.

Поскольку подстановка (\ref{LPlusDiv}) не меняет динамики, она
оставляет инвариантное фазовое пространство $Z$ неизменным. Сейчас
мы покажем, что при этом также остаётся неизменной и
симплектическая структура на $Z$. Это фактически приводит к тому,
что весь инвариантный гамильтонов формализм к такой добавке
нечувствителен. А поскольку квантование осуществляется на основе
инвариантного гамильтонова формализма, то отсюда последует, что и
конструкция квантованного поля при подстановке (\ref{LPlusDiv}) не
меняется.

Доказать неизменность симплектической структуры, исходя из формулы
для симплектического тока, представляется затруднительным. В связи
с этим мы напомним здесь, как определяется симплектическая
структура непосредственно в вариационных терминах
\cite{Witten1986,Crnkovic1987,Zucker1987,CrWitt1987,%
Crnkovic1988,BHS1991}.

Рассмотрим действие в произвольной области $\Omega$:
\begin{equation}
 S = \int\limits_\Omega d^4\!x\,
     L(x,\varphi_i,\partial_\mu\varphi_i,\dots,
       \partial_{\mu_1}\dots\partial_{\mu_k}\varphi_i) \ .
\label{SInOmega}
\end{equation}
Его вариация, если область не меняется, равна:
\begin{equation}
 \delta S = \int\limits_\Omega d^4\!x\,
 \left(\frac{\delta L}{\delta\varphi_i}\,\delta\varphi_i +
       \partial_\mu j_\mu\right)   \ ,
\label{VarSInOmega}
\end{equation}
Здесь
\[
 j_\mu =
 \frac{\delta L}{\delta(\partial_\mu\varphi_i)}\,\delta\varphi_i +
 \dots +
 \frac{\delta L}
      {\delta(\partial_\mu\partial_{\mu_2}\dots
              \partial_{\mu_k}\varphi_i)}\,
 \partial_{\mu_2}\dots\partial_{\mu_k}\delta\varphi_i   \ .
\]
Вариационная производная при этом понимается как
\[
 \frac{\delta}{\delta a} =
 \frac{\partial}{\partial a} -
 \partial_\mu\frac{\partial}{\partial(\partial_\mu a)} +
 \partial_\mu\partial_\nu
    \frac{\partial}{\partial(\partial_\mu\partial_\nu a)} -
 \dots    \ .
\]
Действие (\ref{SInOmega}) может рассматриваться как функционал на
множестве функций $\varphi_i$. Используя тождество $\delta^2=0$,
получаем:
\[
\int\limits_\Omega d^4\!x\,
 \left(\delta\left(\frac{\delta L}{\delta\varphi_i}\,
                   \delta\varphi_i\right) +
       \partial_\mu\delta j_\mu\right) = 0  \ .
\]
Если здесь рассматривать варьирование только функций $\varphi_i$,
для которых действие стационарно, то первый член  в последнем
уравнении зануляется. Отсюда получаем, что симплектическая
структура, задаваемая на $Z$ как
\[
 \omega = \int\limits_\Sigma d\sigma_\mu\,\delta j_\mu(x) \ ,
\]
не зависит от выбора пространственно-подобной поверхности $\Sigma$
(предполагается, что $\Sigma$ ведёт себя достаточно хорошо на
бесконечности).

Посмотрим теперь, как изменятся приведённые уравнения при
подстановке (\ref{LPlusDiv}).

Вариация нового действия $S'$ может быть записана аналогично
формуле (\ref{VarSInOmega}):
\[
 \delta S' = \int\limits_\Omega d^4\!x\,
 \left(\frac{\delta L'}{\delta\varphi_i}\,\delta\varphi_i +
       \partial_\mu j'_\mu\right) =
 \int\limits_\Omega d^4\!x\,
 \left(\frac{\delta L}{\delta\varphi_i}\,\delta\varphi_i +
       \frac{\delta(\partial_\mu F_\mu)}{\delta\varphi_i}\,
              \delta\varphi_i +
       \partial_\mu j'_\mu\right)      \ .
\]
Но $\delta(\partial_\mu F_\mu)/\delta\varphi_i = 0$.
Следовательно,
\begin{equation}
 \delta S' = \int\limits_\Omega d^4\!x\,
 \left(\frac{\delta L}{\delta\varphi_i}\,\delta\varphi_i +
       \partial_\mu j'_\mu\right)      \ .
\label{VarSDash1}
\end{equation}
С другой стороны, эта же вариация может быть записана в терминах
старого действия:
\begin{equation}
  \delta S' = \delta S +
     \int\limits_\Omega d^4\!x\,\partial_\mu\delta F_\mu    \ .
\label{VarSDash2}
\end{equation}
Комбинируя (\ref{VarSDash1}) и (\ref{VarSDash2}) с
(\ref{VarSInOmega}), получаем:
\[
 \int\limits_\Omega d^4\!x\,
  (\partial_\mu j'_\mu -
    \partial_\mu j_\mu - \partial_\mu\delta F_\mu) = 0 \ .
\]
Поскольку в этом уравнении все три функции $j'_\mu$, $j_\mu$ и
$\delta F_\mu$ локальные (т.~е. они обращаются в ноль в тех
областях, где $\delta\varphi_i=0$), отсюда следует:
\[
 \int\limits_\Sigma d\sigma_\mu\,
 (j'_\mu - j_\mu - \delta F_\mu) = 0 \ .
\]
Величину $F_\mu$ можно рассматривать как функционал на множестве
функций $\varphi_i$. Используя тождество $\delta^2=0$, получаем:
\[
 \int\limits_\Sigma d\sigma_\mu\,\delta j'_\mu(x) =
 \int\limits_\Sigma d\sigma_\mu\,\delta j_\mu(x)   \ .
\]
Т.~е. преобразование (\ref{LPlusDiv}) оставляет симплектическую
структуру на $Z$ неизменной:
\[
 \omega' = \omega  \ .
\]




\newcounter{parnum}
\newlength{\picwidth}
\newcommand*{\myparagraph}[1]{\refstepcounter{parnum}%
\paragraph{\arabic{parnum}. #1}}


\title{On quantization of electromagnetic field.\\
       III. Formula for generators of infinitesimal\\
       linear canonical transformations.}
\author{D.~A.~Arbatsky\footnote{http:\dslash d-a-arbatsky.narod.ru/}}
\date{\today}
\maketitle

\begin{abstract}
Here we suggest a formula for generators of infinitesimal linear
symplectic transformations of invariant phase space. We discuss
applications of this formula to classical and quantum field
theory. We show the existence of generators of the symmetry group
for quantum case.
\end{abstract}


\myparagraph{Formula.}
 In the paper [I] we used a formula that gives generators of
infinitesimal linear transformations for invariant Hamiltonian
formalism. In view of importance and generality of this formula we
will give here its derivation and discussion.

So, consider some linear field which is described by the invariant
Hamiltonian formalism. Let us for definiteness look for the
generators of the Poincare group.

So far as the symplectic structure~$\omega$ of the invariant phase
space is supposed to be Poincare-invariant, the Poincare group
acts on~$Z$ as continuous linear symplectic group. For any
infinitesimal transformation from this group\footnote{I.~e. an
element of its Lie algebra.} we have a Hamiltonian flow on~$Z$.
The Hamiltonian of this flow is a function on~$Z$. Let us denote
it $G^{\,\underline .}$. Now we will find the explicit formula for
this function.

First, it is immediately seen that this function is defined up to
addition of a constant. We will fix this constant by the condition
$G^{\,\underline 0}=0$, i.~e. the function is equal to zero for
the zero vector from space~$Z$. Speaking in a more physical way,
we will suppose that the vacuum of a classical field has zero
energy, momentum etc.

Let us fix now in the space~$Z$ a point $\underline c$. The vector
of velocity of the flow in this point we denote
as~$\underline{\delta c}$. Let us draw the straight line segment
from the point~$\underline0$ to the point $\underline c$, i.~e.
let us consider the set of points like $\alpha\,\underline c$,
$\alpha\in[0;1]$. So far as transformations of the group under
consideration are linear, in the point $\alpha\,\underline c$ the
velocity of flow is equal to $\alpha\,\underline{\delta c}$. On
the other hand, the vector of the velocity of flow is connected
with the Hamiltonian by the relation:
\begin{equation}
 dG\big|_{\alpha\underline c}{}^{\underline b} =
 \omega^{\,\underline b\,;\,\alpha\underline{\delta c}} \ ,
\label{defg}
\end{equation}
which is true for any $\underline b\in Z$. Here the left side
denotes the differential of the function $G^{\,\underline .}$
calculated in the point ${\alpha\,\underline c}$ and taken on the
vector $\underline b$.

As a vector~$\underline b$ in the equation~(\ref{defg}) we can
take the vector $d\alpha\,\underline c$:
\[
 dG\big|_{\alpha\underline c}{}^{d\alpha\underline c} =
 \omega^{\,d\alpha\underline c\,;\,\alpha\underline{\delta c}} \ .
\]
Using bilinearity of the form $\omega$ we can write it as:
\[
 dG\big|_{\alpha\underline c}{}^{d\alpha\underline c} =
 \omega^{\,\underline c\,;\,\underline{\delta c}}\ \alpha\,d\alpha \ .
\]
Integrating this relation with respect to $\alpha$ in the limits
from $0$ to $1$ we get the formula for the generator:
\begin{equation}
 \begin{array}{|ccc|}
 \hline
 \phantom{1} & & \phantom{1}\\
  & G^{\,\underline c} =
              \frac12\,
              \omega^{\,\underline c\,;\,\underline{\delta c}} & \\
 \phantom{1} & & \phantom{1}\\
 \hline
 \end{array}
\label{basf}
\end{equation}


\myparagraph{Application to classical field theory.}
 It should not be thought that the formula~(\ref{basf}) is just
other form of writing the Noether formula. First, it works only in
the linear case. Second, it is not supposed to be used for looking
for {\it integrals of motion}, because they are supposed to be
known for linear fields (the equations of the motion are solved).
Nevertheless, the formula~(\ref{basf}) turns out to be very
convenient in practical application for many reasons.

First, it is applied much easier. Second, in contrast to the
Noether formula, it is not connected with coordinate
representation. If we know a formula for the symplectic
structure~$\omega$, for example, in the Fourier-representation and
we know how in this representations the group under consideration
acts (for example, Poincare group) then we get the generators as
easily as in the coordinate representation.

Furthermore, as it was shown in the paper~[II], the structure of
the invariant Hamiltonian formalism does not change, if we add to
the Lagrangian a full divergence. So far as the
formula~(\ref{basf}) is written in the terms of this formalism
only, the independence of generators from this substitution
becomes apparent.

As a simplest example, consider the scalar field. In the paper~[I]
we have shown that in the Fourier representation the symplectic
structure of this field is given by the formula:
\begin{equation}
 \omega^{\,\underline\cdot\,;\,\underline\cdot} =
 \int d\mu_m \cdot i\,\varepsilon(k) \cdot
        a(-k)^{\,\underline\cdot} \, a(k)^{\,\underline\cdot} \ .
\label{sympl}
\end{equation}

Now consider a space-time shift of the field by infinitesimal
vector $\delta\varepsilon_\nu$. A state $\underline c$ changes to
 $\underline c + \underline{\delta c}$.
And:
\[
 a(k)^{\,\underline{\delta c}} =
     i\,\delta\varepsilon_\nu\,k_\nu\,a(k)^{\,\underline c}\ .
\]
Substituting this expression to the formula~(\ref{basf}), and
taking into account the formula~(\ref{sympl}), we get:
\[
 G^{\,\underline c} =
   -\,{\textstyle\frac12}\,\delta\varepsilon_\nu
   \int d\mu_m \cdot \varepsilon(k) \cdot
   k_\nu\,a(-k)^{\,\underline c}\,a(k)^{\,\underline c} =
  {}- \delta\varepsilon_\nu \displaystyle\int d\mu^+_m \cdot
   k_\nu\, a^*(k)^{\,\underline c}\,a(k)^{\,\underline c} \ .
\]

Let us denote the coefficient of $\delta\varepsilon_\nu$ as
$-P_\nu$. Omitting the argument $\underline c$ we get the formula
for the vector of energy and momentum:
\begin{equation}
 P_\nu =
   {\textstyle\frac12}
   \int d\mu_m \cdot \varepsilon(k) \cdot k_\nu\,a(-k)\,a(k) =
   \int d\mu^+_m \cdot k_\nu\, a^*(k)\,a(k) \ .
\label{MomScalF}
\end{equation}


\myparagraph{About generators in quantum field theory.}
 In the paper~[VI] we will describe how construction of quantum
fields is performed. Let us discuss here how generators are found
in the quantum case.

As far as I know, there was not suggested any analog of Noether
theorem for the quantum field theory\footnote{Some authors allege
opposite. I allow myself just to ignore their opinion on this
question.}.

Note now, that in the case of invariant quantization of fields, in
accordance with the scheme given in the paper~[VI], the action of
group of invariance is naturally transferred to the quantum space
of states $\mathcal H$. And this action is unitary (in the case of
indefinite scalar product~--- pseudounitary).

If the scalar product in $\mathcal H$ is positive-definite, then
in accordance with the Stone theorem, we have the self-adjoint
quantum generator. If the scalar product in $\mathcal H$ is
indefinite, then everything is similar, but we need to introduce
some special mathematical terminology.

So, generators of the group of invariance can be rigorously
introduced in the quantum case also.

It seems to be of interest to get some useful formulas for quantum
generators also.

So far as the vacuum $|\,0\,\rangle$ of the quantum field remains
invariant under the action of the symmetry group, then, speaking
formally, we just need to write expressions like (\ref{MomScalF})
making normal ordering of operators in the integrand. In the case
of Poincare-invariant quantization of the scalar field that leads
to the positive-definite scalar product we get:
\begin{equation}
 \hat P_\nu = \int d\mu^+_m \cdot k_\nu\, \hat a^*(k)\,\hat a(k) \ .
\label{QuantMomScalF}
\end{equation}

But, nevertheless, the strict mathematical meaning of the obtained
expression is not fully clear\footnote{It will be seen from
discussion of quantization and topological questions in the paper
[VI] that symbols like $\hat a^*(k)$ have sense only after a
proper averaging. But in the given formula we need to integrate
{\it the product} of such symbols.}. It seems to be topical to
develop a theory of integrals like (\ref{QuantMomScalF}) and
prove that if the integrands are normally ordered, we get exactly
the generators introduced here.






\newcounter{parnum}
\newlength{\picwidth}
\newcommand*{\myparagraph}[1]{\refstepcounter{parnum}%
\paragraph{\arabic{parnum}. #1}}


\title{О квантовании электромагнитного поля.\\
       III. Формула для генераторов инфинитезимальных\\
       линейных канонических преобразований.}
\author{Д.~А.~Арбатский\footnote{http:\dslash d-a-arbatsky.narod.ru/}}
\date{\today}
\maketitle

\begin{abstract}
Предложена формула для генераторов бесконечно малых линейных
симплектических преобразований инвариантного фазового
пространства. Обсуждаются применения этой формулы к классической и
к квантовой теории поля. Показано существование генераторов группы
симметрии для квантового случая.
\end{abstract}


\myparagraph{Формула.}
 В статье [I] использовалась формула, дающая генераторы бесконечно
малых линейных преобразований для инвариантного гамильтонова
формализма. Ввиду важности и общности этой формулы, здесь будут
даны её вывод и обсуждение.

Итак, рассмотрим какое-нибудь линейное поле, описываемое
инвариантным гамильтоновым формализмом. Будем, для определённости,
искать для него генераторы группы Пуанкаре.

Поскольку симплектическая структура~$\omega$ инвариантного
фазового пространства~$Z$ предполагается
пу\-ан\-ка\-ре-инвариантной, группа Пуанкаре действует на~$Z$ как
непрерывная линейная симплектическая группа. Каждому бесконечно
малому преобразованию этой группы\footnote{То есть элементу её
алгебры Ли.} соответствует гамильтонов поток на~$Z$. Гамильтониан
этого потока является функцией на~$Z$. Обозначим её
$G^{\,\underline .}$. Найдём явный вид этой функции.

Во-первых, сразу ясно, что эта функция определена с точностью до
аддитивной константы. Эту константу мы зафиксируем условием
$G^{\,\underline 0}=0$, т.~е. функция равна нулю на нулевом
векторе пространства~$Z$. Выражаясь более физическим языком, мы
будем полагать, что вакуум классического поля обладает нулевыми
энергией, импульсом и~т.~п.

Зафиксируем теперь в пространстве~$Z$ точку $\underline c$. Вектор
скорости потока в этой точке обозначим~$\underline{\delta c}$.
Проведём из точки~$\underline0$ в точку $\underline c$ отрезок,
т.~е. рассмотрим множество точек вида $\alpha\,\underline c$,
$\alpha\in[0;1]$. Поскольку преобразования рассматриваемой группы
линейны, в точке $\alpha\,\underline c$ скорость потока равна
$\alpha\,\underline{\delta c}$. С~другой стороны, вектор скорости
потока связан с гамильтонианом соотношением:
\begin{equation}
 dG\big|_{\alpha\underline c}{}^{\underline b} =
 \omega^{\,\underline b\,;\,\alpha\underline{\delta c}} \ ,
\label{defg}
\end{equation}
выполняющимся для любого $\underline b\in Z$. Здесь левая часть
обозначает дифференциал функции $G^{\,\underline .}$, вычисленный
в точке ${\alpha\,\underline c}$ и взятый на векторе
 $\underline b$.

В качестве вектора~$\underline b$ в уравнении~(\ref{defg}) можно
взять вектор $d\alpha\,\underline c$:
\[
 dG\big|_{\alpha\underline c}{}^{d\alpha\underline c} =
 \omega^{\,d\alpha\underline c\,;\,\alpha\underline{\delta c}} \ .
\]
Используя билинейность формы $\omega$, перепишем это так:
\[
 dG\big|_{\alpha\underline c}{}^{d\alpha\underline c} =
 \omega^{\,\underline c\,;\,\underline{\delta c}}\ \alpha\,d\alpha \ .
\]

Интегрируя это соотношение по $\alpha$ в пределах от $0$ до $1$,
получаем формулу для генератора:
\begin{equation}
 \begin{array}{|ccc|}
 \hline
 \phantom{1} & & \phantom{1}\\
  & G^{\,\underline c} =
              \frac12\,
              \omega^{\,\underline c\,;\,\underline{\delta c}} & \\
 \phantom{1} & & \phantom{1}\\
 \hline
 \end{array}
\label{basf}
\end{equation}


\myparagraph{Применение к классической теории поля.}
 Не следует думать, что формула~(\ref{basf}) является просто
другой записью формулы Нётер. Во-первых, она работает только в
линейном случае. Во-вторых, она вовсе не предназначается для
поиска {\it интегралов движения}, ибо у линейных полей они и так
известны (уравнения движения решены). Несмотря на это,
формула~(\ref{basf}) оказывается очень удобной при практическом
применении по многим причинам.

Во-первых, её применять значительно проще. Во-вторых, в отличие от
формулы Нётер, она вовсе не привязана к координатному
представлению. Если вид симплектической структуры~$\omega$
известен, например, в фурье-представлении, и известно, как в этом
представлении действует рассматриваемая группа (например, группа
Пуанкаре), то генераторы получаются столь же просто, как и в
координатном представлении.

Кроме того, в статье~[II] мы видели, что структура инвариантного
гамильтонова формализма не меняется, если к лагранжиану добавить
полную дивергенцию. Поскольку формула~(\ref{basf}) записана
исключительно в терминах этого формализма, независимость
генераторов от указанной подстановки оказывается очевидной.

В качестве простейшего примера рассмотрим скалярное поле. В
статье~[I] мы видели, что в Фу\-рье-пред\-ста\-вле\-нии
симплектическая структура этого поля даётся формулой:
\begin{equation}
 \omega^{\,\underline\cdot\,;\,\underline\cdot} =
 \int d\mu_m \cdot i\,\varepsilon(k) \cdot
        a(-k)^{\,\underline\cdot} \, a(k)^{\,\underline\cdot} \ .
\label{sympl}
\end{equation}

Рассмотрим далее пространственно-временной сдвиг поля на
бесконечно малый вектор $\delta\varepsilon_\nu$. Состояние
$\underline c$ заменяется при этом на
 $\underline c + \underline{\delta c}$. При этом:
\[
 a(k)^{\,\underline{\delta c}} =
     i\,\delta\varepsilon_\nu\,k_\nu\,a(k)^{\,\underline c}\ .
\]
Подставляя это выражение в формулу~(\ref{basf}), с учётом
формулы~(\ref{sympl}), получаем:
\[
 G^{\,\underline c} =
   -\,{\textstyle\frac12}\,\delta\varepsilon_\nu
   \int d\mu_m \cdot \varepsilon(k) \cdot
   k_\nu\,a(-k)^{\,\underline c}\,a(k)^{\,\underline c} =
  {}- \delta\varepsilon_\nu \displaystyle\int d\mu^+_m \cdot
   k_\nu\, a^*(k)^{\,\underline c}\,a(k)^{\,\underline c} \ .
\]

Коэффициент при $\delta\varepsilon_\nu$ обозначим теперь как
$-P_\nu$. Опуская аргумент $\underline c$, имеем формулу для
вектора энергии-импульса:
\begin{equation}
 P_\nu =
   {\textstyle\frac12}
   \int d\mu_m \cdot \varepsilon(k) \cdot k_\nu\,a(-k)\,a(k) =
   \int d\mu^+_m \cdot k_\nu\, a^*(k)\,a(k) \ .
\label{MomScalF}
\end{equation}


\myparagraph{О генераторах в квантовой теории поля.}
 В статье~[VI] мы опишем, каким образом осуществляется построение
квантовых полей. Обсудим здесь, каким образом обстоят дела с
поиском генераторов в квантовом случае.

Аналога теоремы Нётер в квантовой теории поля, насколько мне
известно, до сих пор не предложено\footnote{Многие очень
авторитетные авторы утверждают обратное. Я позволю себе их мнение
по данному вопросу проигнорировать.}.

Заметим, однако, что в случае инвариантного квантования полей в
соответствии со схемой, изложенной в статье~[VI], действие группы
инвариантности естественно переносится на квантовое пространство
состояний $\mathcal H$. Причём это действие унитарно (в случае
индефинитного скалярного произведения~--- псевдоунитарно).

Если скалярное произведение в $\mathcal H$ является
положительно-определённым, то в соответствии с теоремой Стоуна
имеем самосопряжённый квантовый генератор. Если же скалярное
произведение в $\mathcal H$ индефинитно, то дело обстоит столь же
просто, но нужно вводить специальную математическую терминологию.

Таким образом, генераторы группы инвариантности могут быть строго
введены и в квантовом случае.

Представляет интерес получить для квантовых генераторов также
какие-нибудь удобные формулы.

Поскольку вакуум $|\,0\,\rangle$ квантованного поля остаётся при
вышеуказанном действии группы симметрии инвариантным, то чисто
формально, классические выражения типа (\ref{MomScalF}) нужно
просто записать, произведя нормальное упорядочение соответствующих
операторов под знаком интеграла. В случае, если речь идёт о
пуанкаре-инвариантном квантовании скалярного поля, приводящем к
положительно-определённому скалярному произведению, получаем:
\begin{equation}
 \hat P_\nu = \int d\mu^+_m \cdot k_\nu\, \hat a^*(k)\,\hat a(k) \ .
\label{QuantMomScalF}
\end{equation}

При этом, однако, точный математический смысл приведённого
выражения не вполне ясен\footnote{Как будет ясно из обсуждения
квантования и топологических вопросов в статье [VI] символы вида
$\hat a^*(k)$ имеют ясный смысл после подходящего усреднения. В
приведённой же формуле производится интегрирование {\it
произведения} таких символов.}. Представляется актуальным
построить теорию интегралов вида (\ref{QuantMomScalF}) и показать,
что если подынтегральные выражения нормально упорядочены, то
получаются в точности введённые нами генераторы.






\newcounter{parnum}
\newlength{\picwidth}
\newcommand*{\myparagraph}[1]{\refstepcounter{parnum}%
\paragraph{\arabic{parnum}. #1}}


\title{On quantization of electromagnetic field.\\
       IV. Theory of field representations.}
\author{D.~A.~Arbatsky\footnote{http:\dslash d-a-arbatsky.narod.ru/}}
\date{\today}
\maketitle

\begin{abstract}
We introduce a notion  of induced symplectic representation of the
Poincare group. Classical relativistic fields are considered as
such representations. We describe the method of investigation of
these fields in the sense of their reducibility. We introduce the
notion  of the field oscillator as an inducing Hamiltonian system.
\end{abstract}


\myparagraph{Symplectic representations.}
 By Poincare group we will call a connected continuous
ten-parameter group of transformations of space-time including
both homogeneous transformations (the Lorentz group) and
space-time shifts. We will denote the Poincare group by the
symbol~$\mathcal P$.

So far as in these papers we study fields that are relativistic
invariant, every solution of equations of the motion is
transformed to a solution by a transformation from the Poincare
group. So, this defines the action of the Poincare group on the
invariant phase space~$Z$.

In a more concrete way this action can be described in the
following way. Let we have a transformation $g\in\mathcal P$. It
acts on points of space-time in the following way:
\[
 x \ \to\ x'=gx \ .
\]
In the space, where the values of the field function~$\varphi_i$
lie, the transformation~$g$ acts as:
\[
 \varphi_i \ \to \ \varphi'_i= \Lambda_{ij}\,\varphi_j \ .
\]
The action of $g$ in the space~$Z$ can be described by the
formulas:
\begin{equation}
 \underline c \,\longrightarrow\,\underline{\smash{gc}} \ ,
 \qquad
 \varphi_i(x)^{\,\underline{\smash{gc}}} =
          \Lambda_{ij}\,\varphi_j(g^{-1}x)^{\,\underline c}    \ .
\label{Poin}
\end{equation}

As we have said in the paper~[I], we will consider here only
linear fields. The action of the Poincare group~$\mathcal P$
obviously preserve the linear structure in the space~$Z$. So, in
the space~$Z$ we have a linear representation of the Poincare
group.

Furthermore, the Poincare group preserves in the space~$Z$ the
symplectic structure. So, we come to linear {\it symplectic}
representations of the Poincare group $\mathcal P$. The connection
of symplectic representations with unitary will become clear after
we define the operation of quantization in the paper~[VI]. It will
be shown there that in the field theory symplectic representations
apparently play a role that is not less important than that of
unitary. Anyway, they are connected with the construction of
quantized field more closely.


\myparagraph{``Reduction'' of field representation.}
\label{reduction}
 Consider now the question of reducibility of field
representations of the Poincare group~$\mathcal P$. We will always
imply here reducibility in the complex sense. In order to not
complexify the space~$Z$ let us consider the conjugate
space~$Z^*_{\mathbb C}$. In~$Z^*_{\mathbb C}$, obviously, acts the
conjugate representation of the group~$\mathcal P$. In order to
define this action we just need to read the formula~(\ref{Poin}) a
little differently:
\begin{equation}
 \varphi_i(x)^{\,\underline \cdot} \,\longrightarrow\,
     \big(g\,\varphi_i(x)\big)^{\,\underline \cdot} \ ,
 \qquad
 \big(g\,\varphi_i(x)\big)^{\,\underline c} =
          \Lambda_{ij}\,\varphi_j(g^{-1}x)^{\,\underline c}    \ ,
\label{Poinadj}
\end{equation}
i.~e. we connect the symbol~$g$ not with an element of~$Z$, but
with an element of~$Z^*_{\mathbb C}$. Below by a field
representation we will always imply the action of the Poincare
group by formula~(\ref{Poinadj}) in the complexified conjugate
space~$Z^*_{\mathbb C}$.

For infinite-dimensional representations the notion of reduction
requires specification. Now we will explain the structure of field
representations under consideration and simultaneously specify in
what sense we will talk about their reducibility.

In the case of unitary representations the main tool of
investigation is Mackey theorem (see, for example,
\cite{Mackey1968}). This theorem brings investigation of a
representation of the Poincare group~$\mathcal P$ to investigation
of the corresponding little Lorentz group~$\mathcal L_k$. The
construction of Mackey induction can be generalized to any linear
representations. But a general linear representation (including
symplectic) of the group~$\mathcal P$ can be not induced from the
little group. Nevertheless, representations induced in Mackey
sense constitute a very wide class, and in fact, we will restrict
ourself with them.

Let us make a Fourier transformation of the field and write the
result in the form:
\begin{equation}
 \widetilde\varphi_i(k) = 2\pi\,\delta(k^2-m^2)\cdot a_i(k)  \ .
\label{fourdec}
\end{equation}
It is natural to call the scalar~$m$ a mass of the field, like in
quantum theory.

On the grounds of the formula~(\ref{fourdec}) it is natural to say
that the field representation decomposes into the direct sum of
positive- and negative-frequency subrepresentations:
\[
 Z^*_{\mathbb C} = Z^{*\,(+)}_{\mathbb C} \oplus
                   Z^{*\,(-)}_{\mathbb C} \ .
\]
These two representations we will also call field representations.
The projectors to the positive- and negative-frequency subspaces
in the Fourier representation are just operators of multiplication
by the functions $\theta(+k)$ and $\theta(-k)$, correspondingly.

So far as we consider here real fields, $a_i(-k)=a^*_i(k)$. So, it
is natural to suppose that both subspaces $Z^{*\,(+)}_{\mathbb C}$
and $Z^{*\,(-)}_{\mathbb C}$ are reducible in the same degree. In
this connection we can restrict ourself with consideration of the
positive-frequency subspace.

Consider now a fixed vector~$k^{(0)}$ on the mass surface:
$(k^{(0)})^2=m^2$. And consider the subgroup of the Lorentz group
that leaves this vector unchanged, i.~e. the so called little
group of this vector. Let us denote this
 group~$\mathcal L_{k^{(0)}}$. Obviously, the set of values
$a_i(k^{(0)})$ corresponding to different values of the index~$i$
is transformed linearly under action of transformations
from~$\mathcal L_{k^{(0)}}$. This complex representation of the
little group we will suppose to be finite-dimensional. Like in the
unitary case, we will say that the field representation under
consideration is induced by this representation of the little
group.

Let us, in analogy with the unitary case, suppose that the further
reducibility of the field representation (i.~e. reduction of its
positive-frequency part) is fully defined by reducibility of the
inducing representation.

Let us note here also that in the case of massive fields, i.~e.
when $m>0$, the little group~$\mathcal L_{k^{(0)}}$ is the group
of three-dimensional rotations $SO(3)$. This group is compact, and
therefore, its representation can be made unitary by introducing a
proper scalar product. Therefore, the inducing representation
turns out to be fully reducible, and the equivalence classes of
irreducible components, like in the unitary case, are defined by a
one integer or half-integer number. It is natural to call this
number, like in the quantum case, a {\it spin} of the irreducible
component under consideration.

In the case of zero-mass field the little group, as we know, is
the group of motions of a Euclidian plane~$E(2)$. This group is
not compact, and situation becomes much more complicated than in
the unitary case. We will see it soon with example of
electromagnetic field.


\myparagraph{Field oscillator as inducing Hamiltonian system.}
 It is easy to notice that there is a great similarity in
description of the harmonic oscillator and the scalar field. Now
we will research this phenomenon in detail.

Consider any real field that is written in the Fourier
representation as~(\ref{fourdec}). Let us suppose that the Poisson
brackets of the field values in the Fourier representation have
the form:
\begin{equation}
 \{\,\widetilde\varphi_i(k)\,,\,\widetilde\varphi_j(k')\,\} =
 B_{ij}(k)\cdot
 \Big[\,i\,\varepsilon(k)\cdot 2\pi\,\delta(k^2-m^2)\,\Big]\cdot
 (2\pi)^4\delta(k+k') \ .
\label{generalPoiss}
\end{equation}
Here $B_{ij}(k)$ is some tensor function. In the cases of the
scalar field and the non-physical electromagnetic field, in
accordance with the formulas of the paper [I], this function is
just a constant. But in general case it is not so. So, let us
consider its properties with more details.

First, so far as $B_{ij}(k)$ is multiplied by
$2\pi\,\delta(k^2-m^2)$, we can admit that it is defined only on
the mass surface $k^2=m^2$. Second, using antisymmetry of the
Poisson bracket, we get:
\begin{equation}
 B_{ij}(k)=B_{ji}(-k) \ .
\label{asymmpoiss}
\end{equation}

Furthermore, requirements of relativistic invariance make very
hard restrictions for $B_{ij}(k)$. Let us fix some point~$k^{(0)}$
on the mass surface. If we know the value of the function
$B_{ij}(k)$ in this point, then, using relativistic invariance, we
can define it in the other points. The value $B_{ij}(k^{(0)})$ is
not arbitrary also: it must be invariant under action of the
little group $\mathcal L_{k^{(0)}}$.

The fact that the Poisson bracket is a complexified real bracket
makes another restriction for $B_{ij}(k)$:
$B^*_{ij}(k)=B_{ij}(-k)$. Taking into account~(\ref{asymmpoiss})
we can write it also as $B^*_{ij}(k)=B_{ji}(k)$.

Consider now the values $a_i(+k^{(0)})$ and $a_i(-k^{(0)})$. As it
was shown in the section~\ref{reduction}, they form a
representation of the little group $\mathcal L_{k^{(0)}}$. It is
obvious also that these values form a representation of the group
of shifts\footnote{Speaking more generally, these values form a
representation of the seven-parameter subgroup of the Poincare
group that includes the little group $\mathcal L_{k^{(0)}}$ and
shifts.}. A shift by 4-vector $l$ can be described by the
formulas:
\[
 a_i(k^{(0)})  \,\to\, a_i(k^{(0)}) \cdot e^{+ik^{(0)}l} \ ,
 \qquad
 a_i(-k^{(0)}) \,\to\, a_i(-k^{(0)})\cdot e^{-ik^{(0)}l} \ .
\]
Let us introduce a notation $\tau=k^{(0)}l$. Then we can think
that the values $a_i(+k^{(0)})$ and $a_i(-k^{(0)})$ form a
representation of the group $\mathbb R\times\mathcal L_{k^{(0)}}$,
where $\mathbb R$ is the additive group of real numbers,
parameterized by the number~$\tau$.

Let us consider now the values $a_i(+k^{(0)})$ and $a_i(-k^{(0)})$
as dynamical variables of some new system. We will call this
system a {\it field oscillator}. In order to economize notations
we will write variables of the field oscillator with the same
symbols, but we will write as an argument not the vector~$k^{(0)}$
but a real number~$\omega$; and $\omega=+1$ corresponds to
$k=+k^{(0)}$, and $\omega=-1$ corresponds to $k=-k^{(0)}$. The
phase space of the field oscillator can be constructed by
realification of the space of inducing representation. The Poisson
bracket can be defined by the formulas:
\[
 \{\,a_i(+1)\,,\,a_j(+1)\,\} = \{\,a_i(-1)\,,\,a_j(-1)\,\} = 0 \ ,
\]
\[
 \{\,a_i(+1)\,,\,a_j(-1)\,\} =\,iB_{ij}(+1)   \ .
\]
Assuming that the Poisson bracket is not degenerate we can, as
usually, calculate the symplectic structure.

The action of the group $\mathbb R\times\mathcal L_{k^{(0)}}$ on
the phase space of the field oscillator leaves the Poisson
brackets invariant. Therefore it is symplectic.

Now we can interpret the action of a transformation from $\mathbb
R$, characterized by the parameter~$\tau$, as a time shift by the
time period~$\tau$ (the ``time'' of the field oscillator is a
non-dimensional parameter).

We can even define ``coordinates'' of the field oscillator at
``time'' $t$:
\[
 \textstyle
 \varphi_i(t)=
    \frac1{\sqrt2}\,
    \big(\,  a_i(+1)\cdot e^{-it} +
             a_i(-1)\cdot e^{+it}    \,\big) \ .
\]
We see that the oscillator of a real field has real coordinates.

If we consider the little group $\mathcal L_{k^{(0)}}$ as an
abstract group, then it does not depend on the choice of the
vector $k^{(0)}$. If we describe the oscillator by internal
notions of the Hamiltonian formalism (phase space, symplectic
structure, symplectic action of the group $\mathbb R\times\mathcal
L_{k^{(0)}}$), then its construction does not depend on the choice
of the vector $k^{(0)}$ also. So, for any $\mathcal P$-invariant
field we can uniquely juxtapose
 $\mathbb R\times\mathcal L_{k^{(0)}}$- invariant field
oscillator.


\myparagraph{Scalar field representation.} \label{sfrepr}
 Let us apply now the described scheme to the scalar field. The
inducing representation in this case is the one-dimensional
identity representation. The field oscillator is just the usual
real oscillator with one degree of freedom.

So, with respect to the action of the Poincare group, the
space~$Z^*_{\mathbb C}$ decomposes into the direct sum of the
positive- and the negative-frequency reducing subspaces, and each
of these subspaces is irreducible. These two subspaces we will
denote $Z^{*\,(+)}_{\mathbb C}$ and $Z^{*\,(-)}_{\mathbb C}$. The
projections of elements of the space~$Z^*_{\mathbb C}$ to these
two spaces we will provide with labels~${}^{(+)}$ or ${}^{(-)}$,
correspondingly:
\[
 \widetilde\varphi^{\,(\pm)}(k) =
       \theta(\pm k)\,\widetilde\varphi(k) \ ,
 \qquad
 \varphi^{\,(\pm)}(x)=
       \int \frac{d^4\!k}{(2\pi)^4}\,e^{-ikx}\,
       \widetilde\varphi^{\,(\pm)}(k)   \ .
\]

Using formulas from the paper [I], we can calculate the Poisson
brackets of these projections:
\[
 \{\,\widetilde\varphi^{\,(\pm)}(k) \,,\,
     \widetilde\varphi^{\,(\pm)}(k')\,\} = 0 \ ,
\]
\[
 \{\,\widetilde\varphi^{\,(\pm)}(k) \,,\,
     \widetilde\varphi^{\,(\mp)}(k')\,\} =
 -\, \Big[\, i\,\theta(\pm k)\,\varepsilon(k)
             \cdot 2\pi\,\delta(k^2-m^2) \,\Big]
     \cdot (2\pi)^4\delta(k+k')       \ .
\]

In the coordinate representation we have, correspondingly:
\[
 \{\,\varphi^{\,(\pm)}(x)\,,\,\varphi^{\,(\pm)}(x')\,\} = 0 \ ,
\]
\[
 \{\,\varphi^{\,(\pm)}(x)\,,\,\varphi^{\,(\mp)}(x')\,\} =
   - D^{\,(\pm)}_m(x-x') \ .
\]
Here symbols $D^{\,(+)}_m(y)$ and $D^{\,(-)}_m(y)$ denote the
positive and the negative-frequency parts of the function
$D_m(y)$:
\begin{equation}
 D^{\,(\pm)}_m(y)=\int \frac{d^4\!k}{(2\pi)^4} \, e^{-iky} \cdot
 \Big[\, i\,\theta(\pm k)\,\varepsilon(k)
         \cdot 2\pi\,\delta(k^2-m^2) \,\Big]     \ .
\label{Dpmfunc}
\end{equation}

The given formulas turn out to be very useful for calculation of
propagators of the quantized field. But they naturally follow from
the classical theory.


\myparagraph{Electromagnetic field representation.}
\label{emrepres}
 According to the section~\ref{reduction}, the representation of
the non-physical electromagnetic field decomposes also into the
direct sum of positive- and negative-frequency representations.
The formulas of the section~\ref{sfrepr} can be rewritten in the
following way:
\[
 \widetilde A_\mu^{\,(\pm)}(k) = \theta(\pm k)\,\widetilde A_\mu(k) \ ,
 \qquad
 A_\mu^{\,(\pm)}(x)= \int \frac{d^4\!k}{(2\pi)^4}\,e^{-ikx}\,
              \widetilde A_\mu^{\,(\pm)}(k)   \ ,
\]
\[
 \{\,\widetilde A_\mu^{\,(\pm)}(k) \,,\,
     \widetilde A_\nu^{\,(\pm)}(k')\,\} = 0 \ ,
\]
\[
 \{\,\widetilde A_\mu^{\,(\pm)}(k) \,,\,
     \widetilde A_\nu^{\,(\mp)}(k')\,\} =
 g_{\mu\nu}\cdot
 \Big[\, i\,\theta(\pm k)\,\varepsilon(k)
         \cdot 2\pi\,\delta(k^2) \,\Big]
 \cdot (2\pi)^4\delta(k+k')       \ ,
\]
\[
 \{\,A_\mu^{\,(\pm)}(x)\,,\,A_\nu^{\,(\pm)}(x')\,\} = 0 \ ,
\]
\[
 \{\,A_\mu^{\,(\pm)}(x)\,,\,A_\nu^{\,(\mp)}(x')\,\} =
   g_{\mu\nu}\, D^{\,(\pm)}_0(x-x') \ .
\]

Here $D^{\,(\pm)}_0(y)$ are functions~(\ref{Dpmfunc}) with $m=0$.
In this case they can be written also as:
\[
 D^{\,(\pm)}_0(y) =
    \frac1{i\,(2\pi)^2} \cdot
    \frac1{\pm\, y^2-i\,0\cdot\varepsilon(y)} =
    \frac1{4\pi}\,\varepsilon(y)\,\delta(y^2) \pm
       \frac1{i\,(2\pi)^2}\, \mathcal P \,\frac1{y^2}       \ .
\]

The inducing representation in this case is the complex vector
representation of the little group~$\mathcal L_{k^{(0)}}$ with
$(k^{(0)})^2=0$. A more detailed description of this
representation is given in the paper~[V]. It has two composition
series:
\[
 \{0\}= M^0_{\mathbb C} \subset M^\parallel_{\mathbb C} \subset
 M^{(+1)}_{\mathbb C} \subset M^\perp_{\mathbb C} \subset
 M^4_{\mathbb C} = M_{\mathbb C} \ ,
\]
\[
 \{0\}= M^0_{\mathbb C} \subset
 M^\parallel_{\mathbb C} \subset
 M^{(-1)}_{\mathbb C} \subset
 M^\perp_{\mathbb C} \subset
 M^4_{\mathbb C} = M_{\mathbb C} \ .
\]
Under induction these two composition series turn into composition
series of the field representation.
\[
 \{0\}= Z^{*\,(+)\,0}_{\mathbb C}         \subset
        Z^{*\,(+)\,\parallel}_{\mathbb C} \subset
        Z^{*\,(+)\,(+1)}_{\mathbb C}      \subset
        Z^{*\,(+)\,\perp}_{\mathbb C}     \subset
        Z^{*\,(+)\,4}_{\mathbb C} =
               Z^{*\,(+)}_{\mathbb C} \ ,
\]
\[
 \{0\}= Z^{*\,(+)\,0}_{\mathbb C}         \subset
        Z^{*\,(+)\,\parallel}_{\mathbb C} \subset
        Z^{*\,(+)\,(-1)}_{\mathbb C}      \subset
        Z^{*\,(+)\,\perp}_{\mathbb C}     \subset
        Z^{*\,(+)\,4}_{\mathbb C} =
               Z^{*\,(+)}_{\mathbb C} \ .
\]
Here the notations for the subspaces are in natural concordance
with the notations of the paper [I].

The field oscillator in this case turns out to be a system with
eight-dimensional phase space.

In connection with the division of the field into the positive-
and the negative-frequency parts, let us notice also that so far
as electromagnetic field is real, the Lorentz condition for
scattered states can be written in three equivalent forms:
\[
 -\,i\,k_\mu\, a_\mu(k)^{\,\underline{\mathrm{rad}}} = 0 \ ,
\]
\[
 -\,i\,k_\mu\, a_\mu^{(+)}(k)^{\,\underline{\mathrm{rad}}} = 0 \ ,
\]
\[
 -\,i\,k_\mu\, a_\mu^{(-)}(k)^{\,\underline{\mathrm{rad}}} = 0 \ .
\]

So far as positive- and negative-frequency parts of the field play
different roles in quantization, in the paper~[VI] we will see
that in quantum theory there is an analog of only the second form
of this condition.




\newcounter{parnum}
\newlength{\picwidth}
\newcommand*{\myparagraph}[1]{\refstepcounter{parnum}%
\paragraph{\arabic{parnum}. #1}}


\title{О квантовании электромагнитного поля.\\
       IV. Теория полевых представлений.}
\author{Д.~А.~Арбатский\footnote{http:\dslash d-a-arbatsky.narod.ru/}}
\date{\today}
\maketitle

\begin{abstract}
Вводится понятие об индуцированном симплектическом представлении
группы Пуанкаре. Классические релятивистские поля рассматриваются
как такие представления. Описывается методика исследования этих
представлений в смысле их приводимости. Вводится понятие полевого
осциллятора как индуцирующей гамильтоновой системы.
\end{abstract}


\myparagraph{Симплектические представления.}
 Группой Пуанкаре мы будем называть связную непрерывную
десятипараметрическую группу преобразований пространства-времени,
включающую в себя как однородные преобразования (группу Лоренца),
так и пространственно-временные сдвиги. Мы будем обозначать группу
Пуанкаре символом~$\mathcal P$.

Поскольку в данном цикле статей мы изучаем поля, которые
релятивистски инвариантны, всякое решение уравнений движения под
действием преобразования из группы Пуанкаре переходит в решение.
Таким образом, определяется действие группы Пуанкаре на
инвариантном фазовом пространстве~$Z$.

Более конкретно это действие можно описать так. Пусть задано
преобразование $g\in\mathcal P$. На точки про\-стран\-ства-времени
оно действует следующим образом:
\[
 x \ \to\ x'=gx \ .
\]
В пространстве, где лежат значения полевой функции~$\varphi_i$,
преобразование~$g$ действует так:
\[
 \varphi_i \ \to \ \varphi'_i= \Lambda_{ij}\,\varphi_j \ .
\]
Действие~$g$ в пространстве~$Z$ характеризуется формулами:
\begin{equation}
 \underline c \,\longrightarrow\,\underline{\smash{gc}} \ ,
 \qquad
 \varphi_i(x)^{\,\underline{\smash{gc}}} =
          \Lambda_{ij}\,\varphi_j(g^{-1}x)^{\,\underline c}    \ .
\label{Poin}
\end{equation}

Как уже было сказано в статье~[I], мы здесь будем иметь дело с
линейными полями. Действие группы Пуанкаре~$\mathcal P$, очевидно,
сохраняет линейную структуру в пространстве~$Z$. Таким образом, в
пространстве~$Z$ реализовано линейное представление группы
Пуанкаре.

Кроме того, группа Пуанкаре сохраняет в пространстве~$Z$
симплектическую структуру. Таким образом, мы приходим к линейным
{\it симплектическим\/} представлениям группы Пуанкаре
 $\mathcal P$. Связь симплектических представлений с унитарными
станет ясна, когда мы определим операцию квантования поля в
статье~[VI]. Из проведённого там рассмотрения станет ясно, что
симплектические представления, по-видимому, играют для теории поля
роль не менее важную, чем унитарные. Во-всяком случае, с
конструкцией квантованного поля они связаны более непосредственно.


\myparagraph{\protect\rql Приведение\rqr{} полевых представлений.}
\label{reduction}
 Рассмотрим теперь вопрос о приводимости полевых представлений
группы Пуанкаре~$\mathcal P$. Приводимость будет здесь пониматься
в комплексном смысле. Чтобы не комплексифицировать
пространство~$Z$, будем рассматривать комплексифицированное
сопряжённое пространство~$Z^*_{\mathbb C}$. В~$Z^*_{\mathbb C}$,
очевидно, действует сопряжённое представление группы~$\mathcal P$.
Чтобы определить это действие, нужно формулу~(\ref{Poin})
прочитать немного иначе:
\begin{equation}
 \varphi_i(x)^{\,\underline \cdot} \,\longrightarrow\,
     \big(g\,\varphi_i(x)\big)^{\,\underline \cdot} \ ,
 \qquad
 \big(g\,\varphi_i(x)\big)^{\,\underline c} =
          \Lambda_{ij}\,\varphi_j(g^{-1}x)^{\,\underline c}    \ ,
\label{Poinadj}
\end{equation}
т.~е. отнести символ~$g$ не к элементу~$Z$, а к
элементу~$Z^*_{\mathbb C}$. В дальнейшем под полевым
представлением мы будем понимать действие группы Пуанкаре по
формуле~(\ref{Poinadj}) в комплексифицированном сопряжённом
пространстве~$Z^*_{\mathbb C}$.

Для бесконечномерных представлений понятие приводимости требует
уточнения. Сейчас мы разъясним, как устроены полевые
представления, которые мы здесь будем рассматривать, и
одновременно уточним, в каком смысле мы будем понимать
приводимость.

В случае унитарных представлений основным инструментом
исследования является теорема Макки (см., например,
\cite{Mackey1968}). Эта теорема сводит исследование представления
группы Пуанкаре~$\mathcal P$ к исследованию соответствующего
представления малой группы Лоренца~$\mathcal L_k$. Конструкцию
индуцирования по Макки можно распространить и на произвольные
линейные представления. Однако, общее линейное представление (в
том числе симплектическое) группы~$\mathcal P$ может и не быть
индуцированным с малой группы. Тем не менее, представления,
индуцированные по Макки, составляют весьма широкий класс и мы
здесь фактически ими ограничимся.

Произведём преобразование Фурье поля и представим фурье-образ в
виде:
\begin{equation}
 \widetilde\varphi_i(k) = 2\pi\,\delta(k^2-m^2)\cdot a_i(k)  \ .
\label{fourdec}
\end{equation}
Величину~$m$ естественно назвать, как и в квантовой теории, массой
поля.

На основании формулы~(\ref{fourdec}) естественно считать, что
полевое представление распадается в прямую сумму положительно- и
отрицательно-частотных подпредставлений:
\[
 Z^*_{\mathbb C} = Z^{*\,(+)}_{\mathbb C} \oplus
                   Z^{*\,(-)}_{\mathbb C} \ .
\]
Эти два представления мы также будем называть полевыми. Проекторы
на положительно- и отрицательно-частотные подпространства в
фурье-пред\-ста\-вле\-нии устроены как операторы умножения на
функции $\theta(+k)$ и $\theta(-k)$, соответственно.

Далее, поскольку мы имеем здесь дело с вещественными полями,
$a_i(-k)=a^*_i(k)$. Поэтому естественно считать, что каждое из
подпространств $Z^{*\,(+)}_{\mathbb C}$ и $Z^{*\,(-)}_{\mathbb C}$
приводимо ровно в той же мере, что и другое. В связи с этим можно
ограничиться положительно-частотным подпространством.

Зафиксируем теперь вектор~$k^{(0)}$ на массовой поверхности:
$(k^{(0)})^2=m^2$. Рассмотрим подгруппу группы Лоренца,
оставляющую этот вектор неизменным, т.~е. так называемую малую
группу данного вектора. Обозначим эту
 группу~$\mathcal L_{k^{(0)}}$. Очевидно, набор величин
$a_i(k^{(0)})$, соответствующий различным значениям индекса~$i$,
преобразуется линейно под действием преобразований
 из~$\mathcal L_{k^{(0)}}$. Это комплексное представление
малой группы мы будем считать конечномерным. Аналогично унитарному
случаю, будем говорить, что рассматриваемое полевое представление
индуцировано данным представлением малой группы.

Будем далее по аналогии с унитарным случаем считать, что
дальнейшая приводимость полевого представления (т.~е. приведение
его положительно-частотной части) целиком определяется
приводимостью индуцирующего представления.

Заметим здесь ещё, что в случае массивных полей, т.~е. когда
$m>0$, малая группа~$\mathcal L_{k^{(0)}}$ является группой
трёхмерных вращений $SO(3)$. Эта группа компактна, и,
следовательно, её представление может быть сделано унитарным путём
введения надлежащего скалярного произведения. Поэтому индуцирующее
представление оказывается вполне приводимым, и классы
эквивалентности неприводимых компонент, так же как и в унитарном
случае, определяются единственным целым или полуцелым числом. Это
число, имея в виду связь с квантовым случаем, естественно назвать
{\it спином} рассматриваемой неприводимой компоненты.

В случае же безмассового поля малая группа, как известно, является
группой движений евклидовой плоскости~$E(2)$. Эта группа
некомпактна, и ситуация существенно усложняется по сравнению с
унитарным случаем. Скоро мы это увидим на примере
электромагнитного поля.


\myparagraph{Полевой осциллятор как индуцирующая гамильтонова
система.}
 Нетрудно заметить, что между описаниями гармонического
осциллятора и скалярного поля существует большое сходство. Сейчас
мы изучим это явление более детально.

Рассмотрим произвольное вещественное поле, которое в
фурье-пред\-ста\-вле\-нии записывается в виде~(\ref{fourdec}).
Пусть скобки Пуассона полевых величин, взятых в
фурье-представлении, имеют вид:
\begin{equation}
 \{\,\widetilde\varphi_i(k)\,,\,\widetilde\varphi_j(k')\,\} =
 B_{ij}(k)\cdot
 \Big[\,i\,\varepsilon(k)\cdot 2\pi\,\delta(k^2-m^2)\,\Big]\cdot
 (2\pi)^4\delta(k+k') \ .
\label{generalPoiss}
\end{equation}
Здесь $B_{ij}(k)$~--- некоторая тензорная функция. В случае
скалярного поля и нефизического электромагнитного поля, как
следует из формул статьи [I], эта функция просто является
константой. В общем случае это, однако, не так. Рассмотрим поэтому
её свойства более подробно.

Во-первых, поскольку $B_{ij}(k)$ домножается на
$2\pi\,\delta(k^2-m^2)$, можно считать, что она задана только на
массовой поверхности $k^2=m^2$. Во-вторых, используя антисимметрию
скобки Пуассона, получаем:
\begin{equation}
 B_{ij}(k)=B_{ji}(-k) \ .
\label{asymmpoiss}
\end{equation}

Далее, требования релятивистской инвариантности накладывают на
функцию $B_{ij}(k)$ очень жёсткие ограничения. Зафиксируем на
массовой поверхности какую-нибудь точку~$k^{(0)}$. Если известно
значение функции $B_{ij}(k)$ в этой точке, то, используя
лоренц-инвариантность, можно определить её значения во всех
остальных точках. Значение же $B_{ij}(k^{(0)})$ также не является
произвольным: оно должно быть инвариантным по отношению к действию
малой группы $\mathcal L_{k^{(0)}}$.

То, что скобка Пуассона является комплексификацией вещественной,
накладывает на $B_{ij}(k)$ ещё одно условие:
$B^*_{ij}(k)=B_{ij}(-k)$. С учётом~(\ref{asymmpoiss}) его также
можно записать в виде $B^*_{ij}(k)=B_{ji}(k)$.

Рассмотрим теперь величины $a_i(+k^{(0)})$ и $a_i(-k^{(0)})$. Как
было указано в пункте~\ref{reduction}, они образуют представление
малой группы $\mathcal L_{k^{(0)}}$. Очевидно также, что указанные
величины образуют представление группы сдвигов\footnote{Более общо
можно сказать, что указанные величины образуют представление
семипараметрической подгруппы группы Пуанкаре, включающей малую
группу $\mathcal L_{k^{(0)}}$ и сдвиги.}. Сдвиг на 4-вектор $l$
описывается формулами:
\[
 a_i(k^{(0)})  \,\to\, a_i(k^{(0)}) \cdot e^{+ik^{(0)}l} \ ,
 \qquad
 a_i(-k^{(0)}) \,\to\, a_i(-k^{(0)})\cdot e^{-ik^{(0)}l} \ .
\]
Введём обозначение $\tau=k^{(0)}l$. Можно тогда считать, что
величины $a_i(+k^{(0)})$ и $a_i(-k^{(0)})$ образуют представление
группы $\mathbb R\times\mathcal L_{k^{(0)}}$, где $\mathbb R$~---
аддитивная группа вещественных чисел, параметризованная
числом~$\tau$.

Будем теперь рассматривать величины $a_i(+k^{(0)})$ и
$a_i(-k^{(0)})$ как динамические переменные некоторой новой
системы. Эту систему мы будем называть {\it полевым осциллятором}.
С целью экономии обозначений будем записывать переменные полевого
осциллятора теми же символами, но в качестве аргумента будем
писать не вектор~$k^{(0)}$, а вещественное число~$\omega$; причём
$\omega=+1$ соответствует $k=+k^{(0)}$, и $\omega=-1$
соответствует $k=-k^{(0)}$. Фазовое пространство полевого
осциллятора получается овеществлением пространства индуцирующего
представления. Скобку Пуассона зададим формулами:
\[
 \{\,a_i(+1)\,,\,a_j(+1)\,\} = \{\,a_i(-1)\,,\,a_j(-1)\,\} = 0 \ ,
\]
\[
 \{\,a_i(+1)\,,\,a_j(-1)\,\} =\,iB_{ij}(+1)   \ .
\]
Предполагая невырожденность скобки Пуассона, можно, как обычно,
вычислить и симплектическую структуру.

Действие группы $\mathbb R\times\mathcal L_{k^{(0)}}$ на фазовом
пространстве полевого осциллятора оставляет неизменной скобку
Пуассона. Следовательно, оно является симплектическим.

При этом действие преобразования из $\mathbb R$, характеризуемого
параметром~$\tau$, мы можем интерпретировать как временной сдвиг
на промежуток времени~$\tau$ (\rql время\rqr{} полевого
осциллятора~--- параметр безразмерный).

Можно даже определить \rql координаты\rqr{} полевого осциллятора в
\rql момент времени\rqr{} $t$:
\[
 \textstyle
 \varphi_i(t)=
    \frac1{\sqrt2}\,
    \big(\,  a_i(+1)\cdot e^{-it} +
             a_i(-1)\cdot e^{+it}    \,\big) \ .
\]
Видно, что у осциллятора вещественного поля координаты
вещественны.

Далее, малая группа $\mathcal L_{k^{(0)}}$, как абстрактная
группа, от выбора вектора $k^{(0)}$ не зависит. Если осциллятор
описывать во внутренних терминах гамильтонова формализма (фазовое
пространство, симплектическая структура, симплектическое действие
группы $\mathbb R\times\mathcal L_{k^{(0)}}$), то его конструкция
также от выбора $k^{(0)}$ не зависит. Таким образом, каждому
$\mathcal P$-инвариантному полю однозначно сопоставляется
 $\mathbb R\times\mathcal L_{k^{(0)}}$-инвариантный полевой
осциллятор.


\myparagraph{Представление скалярного поля.} \label{sfrepr}
 Применим теперь изложенную схему к скалярному полю. Индуцирующее
представление в этом случае является одномерным тождественным
представлением. Полевой осциллятор~--- обычный вещественный
осциллятор с одной степенью свободы.

Таким образом, по отношению к действию группы Пуанкаре
пространство~$Z^*_{\mathbb C}$ распадается в прямую сумму
положительно- и отрицательно-час\-тот\-ных приводящих
подпространств, каждое из которых неприводимо. Эти два
подпространства мы обозначим $Z^{*\,(+)}_{\mathbb C}$ и
$Z^{*\,(-)}_{\mathbb C}$. Проекции элементов
пространства~$Z^*_{\mathbb C}$ на эти два подпространства будем
снабжать значками~${}^{(+)}$ или ${}^{(-)}$, соответственно:
\[
 \widetilde\varphi^{\,(\pm)}(k) =
       \theta(\pm k)\,\widetilde\varphi(k) \ ,
 \qquad
 \varphi^{\,(\pm)}(x)=
       \int \frac{d^4\!k}{(2\pi)^4}\,e^{-ikx}\,
       \widetilde\varphi^{\,(\pm)}(k)   \ .
\]

Используя формулы из статьи [I], можно вычислить скобки Пуассона
этих проекций:
\[
 \{\,\widetilde\varphi^{\,(\pm)}(k) \,,\,
     \widetilde\varphi^{\,(\pm)}(k')\,\} = 0 \ ,
\]
\[
 \{\,\widetilde\varphi^{\,(\pm)}(k) \,,\,
     \widetilde\varphi^{\,(\mp)}(k')\,\} =
 -\, \Big[\, i\,\theta(\pm k)\,\varepsilon(k)
             \cdot 2\pi\,\delta(k^2-m^2) \,\Big]
     \cdot (2\pi)^4\delta(k+k')       \ .
\]

В координатном представлении имеем, соответственно:
\[
 \{\,\varphi^{\,(\pm)}(x)\,,\,\varphi^{\,(\pm)}(x')\,\} = 0 \ ,
\]
\[
 \{\,\varphi^{\,(\pm)}(x)\,,\,\varphi^{\,(\mp)}(x')\,\} =
   - D^{\,(\pm)}_m(x-x') \ .
\]
Здесь символами $D^{\,(+)}_m(y)$ и $D^{\,(-)}_m(y)$ обозначены
положительно- и отри\-ца\-тель\-но-частотная части функции
$D_m(y)$:
\begin{equation}
 D^{\,(\pm)}_m(y)=\int \frac{d^4\!k}{(2\pi)^4} \, e^{-iky} \cdot
 \Big[\, i\,\theta(\pm k)\,\varepsilon(k)
         \cdot 2\pi\,\delta(k^2-m^2) \,\Big]     \ .
\label{Dpmfunc}
\end{equation}

Приведённые формулы оказываются весьма полезными при вычислении
пропагаторов квантованного поля. При этом сами по себе они
естественно вытекают из классической теории.


\myparagraph{Представление электромагнитного поля.}
\label{emrepres}
 Представление нефизического электромагнитного поля, согласно
пункту~\ref{reduction}, также распадается в прямую сумму
положительно- и отрицательно-частотных подпредставлений. Формулы
пункта~\ref{sfrepr} переписываются следующим образом:
\[
 \widetilde A_\mu^{\,(\pm)}(k) = \theta(\pm k)\,\widetilde A_\mu(k) \ ,
 \qquad
 A_\mu^{\,(\pm)}(x)= \int \frac{d^4\!k}{(2\pi)^4}\,e^{-ikx}\,
              \widetilde A_\mu^{\,(\pm)}(k)   \ ,
\]
\[
 \{\,\widetilde A_\mu^{\,(\pm)}(k) \,,\,
     \widetilde A_\nu^{\,(\pm)}(k')\,\} = 0 \ ,
\]
\[
 \{\,\widetilde A_\mu^{\,(\pm)}(k) \,,\,
     \widetilde A_\nu^{\,(\mp)}(k')\,\} =
 g_{\mu\nu}\cdot
 \Big[\, i\,\theta(\pm k)\,\varepsilon(k)
         \cdot 2\pi\,\delta(k^2) \,\Big]
 \cdot (2\pi)^4\delta(k+k')       \ ,
\]
\[
 \{\,A_\mu^{\,(\pm)}(x)\,,\,A_\nu^{\,(\pm)}(x')\,\} = 0 \ ,
\]
\[
 \{\,A_\mu^{\,(\pm)}(x)\,,\,A_\nu^{\,(\mp)}(x')\,\} =
   g_{\mu\nu}\, D^{\,(\pm)}_0(x-x') \ .
\]

Здесь $D^{\,(\pm)}_0(y)$~--- функции~(\ref{Dpmfunc}) при $m=0$. В
этом случае эти функции также можно записать в виде:
\[
 D^{\,(\pm)}_0(y) =
    \frac1{i\,(2\pi)^2} \cdot
    \frac1{\pm\, y^2-i\,0\cdot\varepsilon(y)} =
    \frac1{4\pi}\,\varepsilon(y)\,\delta(y^2) \pm
       \frac1{i\,(2\pi)^2}\, \mathcal P \,\frac1{y^2}       \ .
\]

Индуцирующее представление в этом случае~--- это комплексное
векторное представление малой группы~$\mathcal L_{k^{(0)}}$ при
$(k^{(0)})^2=0$. Подробное описание этого представления имеется в
статье~[V]. Оно имеет два композиционных ряда:
\[
 \{0\}= M^0_{\mathbb C} \subset M^\parallel_{\mathbb C} \subset
 M^{(+1)}_{\mathbb C} \subset M^\perp_{\mathbb C} \subset
 M^4_{\mathbb C} = M_{\mathbb C} \ ,
\]
\[
 \{0\}= M^0_{\mathbb C} \subset
 M^\parallel_{\mathbb C} \subset
 M^{(-1)}_{\mathbb C} \subset
 M^\perp_{\mathbb C} \subset
 M^4_{\mathbb C} = M_{\mathbb C} \ .
\]
При индуцировании эти композиционные ряды переходят в
композиционные ряды полевого представления:
\[
 \{0\}= Z^{*\,(+)\,0}_{\mathbb C}         \subset
        Z^{*\,(+)\,\parallel}_{\mathbb C} \subset
        Z^{*\,(+)\,(+1)}_{\mathbb C}      \subset
        Z^{*\,(+)\,\perp}_{\mathbb C}     \subset
        Z^{*\,(+)\,4}_{\mathbb C} =
               Z^{*\,(+)}_{\mathbb C} \ ,
\]
\[
 \{0\}= Z^{*\,(+)\,0}_{\mathbb C}         \subset
        Z^{*\,(+)\,\parallel}_{\mathbb C} \subset
        Z^{*\,(+)\,(-1)}_{\mathbb C}      \subset
        Z^{*\,(+)\,\perp}_{\mathbb C}     \subset
        Z^{*\,(+)\,4}_{\mathbb C} =
               Z^{*\,(+)}_{\mathbb C} \ .
\]
Здесь обозначения для подпространств естественно согласованы с
обозначениями статьи [I].

Полевой осциллятор в данном случае оказывается системой с
восьмимерным фазовым пространством.

В связи с разделением поля на положительно- и
отрицательно-частотные части, отметим также, что поскольку
электромагнитное поле вещественно, условие Лоренца на состояния
рассеяния можно записать тремя эквивалентными способами:
\[
 -\,i\,k_\mu\, a_\mu(k)^{\,\underline{\mathrm{rad}}} = 0 \ ,
\]
\[
 -\,i\,k_\mu\, a_\mu^{(+)}(k)^{\,\underline{\mathrm{rad}}} = 0 \ ,
\]
\[
 -\,i\,k_\mu\, a_\mu^{(-)}(k)^{\,\underline{\mathrm{rad}}} = 0 \ .
\]

Поскольку при квантовании положительно- и отрицательно-частотные
части поля играют различные роли, то, как станет ясно из
статьи~[VI], в квантовой теории имеется аналог лишь второго
варианта этого условия.




\newcounter{parnum}
\newlength{\picwidth}
\newcommand*{\myparagraph}[1]{\refstepcounter{parnum}%
\paragraph{\arabic{parnum}. #1}}


\title{On quantization of electromagnetic field.\\
       V. Vector representation of little Lorentz group\\
       for light-like momentum.}
\author{D.~A.~Arbatsky\footnote{http:\dslash d-a-arbatsky.narod.ru/}}
\date{\today}
\maketitle

\begin{abstract}
Using elementary geometric methods we prove the isomorphism of
the little Lorentz group for light-like momentum and the group of
motions of a Euclidian plane. In accordance with
Jordan-H\"older-Noether theorem we perform ``reduction'' or the
real and complex vector representations of this group. We also
prove indecomposability of these representations.
\end{abstract}


\myparagraph{Preliminary remarks.}
 There are quite many books devoted to the theory of the Lorentz
group~$\EuScript L$ and the theory of representations of this
group. Among them there are monographs specially devoted  to this
topic; for
example,~\cite{Naymark1958,GelMinShap1958,Fyodorov1979}. It is
mentioned in all them that the little Lorentz group~$\EuScript
L_k$ for a light-like momentum\footnote{The word ``momentum'' is
used because we keep in mind applications to the theory of
zero-mass fields; in particular electromagnetic. While we perform
purely mathematical investigation, we can suppose that we talk
about any isotropic vector~$k$: $k^2=0$.} $k$ is isomorphic to the
group of motions of a Euclidian plane~$E(2)$. The proof that is
usually provided is usually based on an analytic investigation of
the group $SL(2,\mathbb C)$ which is the universal covering group
of the Lorentz group.

Such an approach seems to be optimal if we want to study then
arbitrary (including two-valued) representations of the
group~$\EuScript L_k$. At the same time it turns out to be very
difficult to realize the geometric sense of the isomorphism of
$\EuScript L_k$ and $E(2)$\footnote{In particular,
Ryder~\cite{Ryder1987} noted about it that the physical meaning of
this result is unclear.}.

Let us specify here what is usually implied when we talk about
``geometric sense''. From the point of view of physical
applications the Lorentz group is most naturally defined by its
vector representation\footnote{Such a representation is sometimes
called ``fundamental''.}. And the little group $\EuScript L_k$ is
defined just as the stationary subgroup of the vector~$k$. The
group $E(2)$ in its turn is defined as the group of motions of a
Euclidean plane. Taking into account these definitions of the
groups $\EuScript L_k$ and $E(2)$ and using purely geometric
constructions we will find some two-dimensional plane, possessing
natural Euclidean structure, where the group $\EuScript L_k$ acts
as the group of motions.

Another question that we will study here (and which is the main
motivation for publishing of this paper) is the ``reduction'' of
vector representation of the group~$\EuScript L_k$, i.~e. we will
expose what irreducible representations it consists of. From the
point of view of the general representation theory, this case is
very particular and does not contain any specific difficulties.
Nevertheless, taking into account its great practical importance,
it seems to be useful to perform its investigation with using
constructions that are closest to our geometric intuition.

\section*{Real vector representation}

\myparagraph{Notations and terminology.}
 {\it The Minkowski space\/} $M$ is a four-dimensional real linear
space~$\mathbb R^4$, where we have real symmetrical bilinear form
$g(\,\cdot\,,\,\cdot\,)$ with the signature $(+,-,-,-)$. This form
is called {\it scalar product}. For short, instead of writing
$g(a,b)$, we will write just~$a b$ or~$a\cdot b$.

Consider now the group of all linear transformations of the
Minkowski space~$M$ preserving the scalar product. The connected
component of this group containing neutral element is a group
also. This latter group we will call {\it the Lorentz group\/} and
we will denote it~$\EuScript L$.

In this paper we will consider only the vector representation of
the Lorentz group, i.~e. the representation in the Minkowski
space. This representation will be denoted as~$T$.

{\it The little group of a vector\/}~$k$ is the stationary
subgroup of the vector~$k$, i.~e. the subgroup of the Lorentz
group leaving the vector~$k$ unchanged. It will be denoted
as~$\EuScript L_k$.

A vector~$k$ is called {\it time-like}, if $k^2>0$; {\it
light-like\/} or {\it isotropic}, if $k^2=0$; and {\it
space-like}, if $k^2<0$.

Consider also the group of all transformations of a Euclidean
plane preserving distances. The connected component of this group
containing neutral element is a group also. We will call it {\it
the group of motions of a Euclidean plane\/} and denote it~$E(2)$.


\myparagraph{Reducing subspaces.}
 Following the Jordan-H\"older-Noether theorem let us immediately
guess a composition series of the representation~$T$, i.~e. such a
monotonic set of reducing subspaces of the representation~$T$ that
corresponding successive factor-representations are irreducible:
\begin{equation}
 \{0\}= M^0 \subset M^\parallel \subset M^\perp \subset M^4=M \ .
\label{admset}
\end{equation}
Here
 $M^0$ is the zero subspace in~$M$;
 $M^\parallel$ is the subspace containing all vectors parallel
     to~$k$;
 $M^\perp$ is the subspace containing all vectors orthogonal
     to~$k$;
 $M^4$ is another notation for the Minkowski space.

Each of these subspaces is obviously invariant with respect to the
action of the group~$\EuScript L_k$. The subrepresentations in the
subspaces $M^\parallel$ and $M^\perp$ we denote $T^\parallel$ and
$T^\perp$, correspondingly.

For the set~(\ref{admset}) we have the corresponding sequence of
the three successive factor-representations in the factor-spaces
 $M^\parallel/M^0=M^\parallel$,
 $M^\perp/M^\parallel=M^{\perp/\parallel}$ and
 $M^4/M^\perp=M^{4/\perp}$.
The first of these factor-representations is a subrepresentation.
We have already denoted it as $T^\parallel$. The second and the
third factor-representations we denote $T^{\perp/\parallel}$ and
$T^{4/\perp}$, correspondingly.

The irreducibility of the factor-representations $T^\parallel$ and
$T^{4/\perp}$ is undoubted because they are one-dimensional. As
regards the factor-representation~$T^{\perp/\parallel}$, its
irreducibility (real) will be proved in the
section~\ref{tperppar}.

In the sections~\ref{unt} and~\ref{tperp} we will prove two
lemmas that have as a consequence that besides the
subspaces~(\ref{admset}) the representation $T$ does not have any
other reducing subspaces. So, the composition
series~(\ref{admset}) is the only one. Therefore in this case the
statement of the Jordan-H\"older-Noether theorem that all
composition series are of the same length and corresponding
factor-representations of these series are equivalent turns out to
be trivial.


\myparagraph{Homomorphism $\EuScript L_k$ into $E(2)$.}
 \label{homom}
 Consider now in the Minkowski space a three-dimensional
hyperplane~$N$ defined by the equation:
\begin{equation}
 k\cdot x = const \neq 0  \ .
\label{plane}
\end{equation}
So far as the group~$\EuScript L_k$ leaves the vector~$k$
unchanged and preserves the scalar product, the hyperplane~$N$ is
invariant with respect to the action of the group~$\EuScript L_k$.

Let us introduce now an equivalence relation~$\sim$ for vectors
from~$N$:
\begin{equation}
 a\sim b \ \ \Longleftrightarrow \ \  a-b \,\in\, M^\parallel \ ,
\label{equiv}
\end{equation}
In other words, two vectors are equivalent if their ends belong to
a line parallel to~$k$. Because of~(\ref{admset}) and
(\ref{plane}) every such a line completely lies in the
hyperplane~$N$. Making factorization of~$N$ with respect to the
equivalence relation~(\ref{equiv}) we get some two-dimensional
plane\footnote{The plane $\widetilde N$ is not {\it imbedded\/}
into the Minkowski space.}~$\widetilde N$. We will denote points
from~$\widetilde N$ and lines corresponding to them on~$N$ by the
same symbols: $\widetilde a$, $\widetilde c$ etc.

The plane~$\widetilde N$ has natural Euclidean structure. Really,
let us take two arbitrary points $\widetilde a$ and $\widetilde c$
on~$\widetilde N$. Both of them are equivalence classes of points
on~$N$. Let us take a one arbitrary representative for each of the
two classes: $a$ and $c$, correspondingly. Let us calculate the
value
\begin{equation}
 \rho\,(a,c)=\sqrt{(a-c)^2} \ .
\label{distN}
\end{equation}
This value does not depend on the choice of the representatives
$a$ and $c$. Really, let us substitute, for example, $a$ with
$a+\alpha k$. Taking into account~(\ref{plane}), we have:
\[
 \rho\,(a+\alpha k,c) = \sqrt{(a-c+\alpha k)^2} =
 \sqrt{(a-c)^2 + 2\alpha(k a-k c) + \alpha^2k^2} =
 \sqrt{(a-c)^2} = \rho\,(a,c) \ .
\]

What is more, $(a-c)^2$ is always greater than zero and the
function $\rho\,(\,\cdot\,,\,\cdot\,)$ have all properties of
Euclidean metric, because the plane~$\widetilde N$ can be
identified with any of space-like two-dimensional planes lying
in~$N$.

Note now that the equivalence relation~(\ref{equiv}) and metric
$\rho\,(\,\cdot\,,\,\cdot\,)$ are invariant with respect to the
action of the little group~$\EuScript L_k$. So, we have
constructed a homomorphous mapping of the group~$\EuScript L_k$
into the group of motions~$E(2)$ of the Euclidean
plane~$\widetilde N$.


\myparagraph{Inverse mapping.} \label{inverse}
 Later on we will repeatedly use the following two statements that
are true in both the real and the complex cases.

\mystat{}
 {\it If a linear transformation changes some basis
$\{n^i\}_{i=1..4}$ so that scalar products of all elements of the
basis remain unchanged (including squares of the elements of the
basis), then this transformation preserves the scalar product.}

\mystat{}
 {\it If a linear transformation changes some basis
$\{n^i\}_{i=1..4}$ so that all values $(n^i-n^j)^2$ and $(n^i)^2$
remain unchanged, then this transformation preserves the scalar
product.}

The first statement follows from the fact that every vector can be
decomposed into a linear combination of the elements of basis. The
second statement follows from the first, if we take into account
the formula:
\[
 (n^i-n^j)^2=(n^i)^2+(n^j)^2-2n^in^j   \ .
\]

Let us fix now an arbitrary motion of the plane~$\widetilde N$. A
transformation from the little group~$\EuScript L_k$, that
transforms to this motion on~$\widetilde N$ under the homomorphism
constructed in the section~\ref{homom}, we will call {\it
desired}. Let us show that the desired transformation
from~$\EuScript L_k$ exists and it is unique. So, we will prove
that the homomorphism under consideration is really an
isomorphism.

For that consider a family of hypersurfaces defined by the
equations like
\begin{equation}
 x^2=const \ .
\label{hypsurf}
\end{equation}
The intersection of any of them with the hyperplane~$N$ is a
paraboloid. Every such a paraboloid is invariant with respect to
the desired transformation. And every line representing a point
from~$\widetilde N$ intersects such a paraboloid exactly in one
point, and for every point on the paraboloid there is exactly one
such a line that contains this point. Therefore we have defined
the action of the desired transformation on every paraboloid.

Let us take now on one of these paraboloids four points that do
not lie in any two-dimensional plane. The four vectors
$\{n^i\}_{i=1..4}$ that have their ends at these four points form
a basis in~$M$. The transformation on the paraboloid that we have
just defined preserves values $(n^i-n^j)^2$, because each of these
values is a square of the corresponding distance in~$\widetilde
N$. The values~$(n^i)^2$ are also preserved because they are
constant on the paraboloid. Let us linearly continue the
constructed transformation of the vectors from the basis to the
whole space~$M$. It follows from the second statement formulated
in the beginning of this section that we will get a transformation
preserving the scalar product in~$M$. And according to our
construction, this transformation preserves the
plane~(\ref{plane}), and therefore it leaves the vector~$k$
unchanged. So far as the initial motion of the plane~$\widetilde
N$ was supposed to belong to the connected group $E(2)$, the
constructed transformation in~$M$ also belongs to the connected
group~$\EuScript L_k$.

The constructed transformation is the only one that can pretend to
be the desired. On the other hand, the constructed transformation
from~$\EuScript L_k$ under the homomorphism from the
section~\ref{homom} transforms into some motion on~$\widetilde N$.
This motion, in accordance with the construction, acts on the four
points\footnote{The lines containing the ends of the vectors of
basis $\{n^i\}_{i=1..4}$}~$\widetilde n^i\in \widetilde N$ in
exactly the same way as the initial motion in~$\widetilde N$. So
far as a motion in a Euclidean plane is uniquely defined even by
its action on two not-coinciding points, the constructed
transformation from~$\EuScript L_k$ is the desired. So, we have
proved the

\mytheorem{} {\it The groups $\EuScript L_k$ and $E(2)$ are
isomorphic.}


\myparagraph{Factor-representation $T^{\perp/\parallel}$.}
\label{tperppar}
 Similarly to the constructed in the section~\ref{homom}
factor-plane, the factor-space $M^{\perp/\parallel}$ has natural
Euclidean structure\footnote{And what is more,
$M^{\perp/\parallel}$ is linear, and there is the corresponding
scalar product in it:
 $\widetilde a\cdot \widetilde c = a c$.}, this structure is
defined by the same formula~(\ref{distN}). So, the little
group~$\EuScript L_k$ acts in $M^{\perp/\parallel}$ as the group
of rotations of a Euclidean plane over a fixed point. Such a group
is denoted as $SO(2)$.

Obviously, in a Euclidean plane there is no such a direction that
would be invariant under rotations. From this we get the

\mytheorem{}
 {\it The factor-representation $T^{\perp/\parallel}$ is irreducible
(in the real sense).}


\myparagraph{Indecomposability of representation $T$.} \label{unt}
 Let us suppose that the representation~$T$ is decomposable, i.~e.
the space~$M$ can be represented as a direct sum $M=M'\oplus M''$
of two non-trivial reducing spaces~$M'$ and $M''$. One of these
spaces, for example~$M'$, necessarily contains some vector that
does not belong to $M^\perp$. If we properly choose the constant
in the equation~(\ref{plane}), the end of this vector will lie in
the plane~$N$. It was shown in the section~\ref{inverse} that by
transformations from $\EuScript L_k$ this vector can be
transformed into any other vector, if the end of the latter lies
on the same paraboloid. In particular, it can be transformed into
four vectors $\{n^i\}_{i=1..4}$ forming a basis in~$M$. Therefore,
$M'$ must coincide with the whole~$M$. So, we have proved the

\mylemma{}
 {\it Every reducing subspace of the representation~$T$ either
coincides with the whole $M$ or lies in $M^\perp$.}

From this lemma we get the

\mytheorem{} {\it The representation~$T$ is indecomposable.}


\myparagraph{Indecomposability of representation $T^\perp$.}
\label{tperp}
 Consider a cylinder defined by equations:
\begin{equation}
 k\cdot x=0 \ , \qquad x^2 = const \neq 0 \ .
\label{cylinder}
\end{equation}
Let us prove that $\EuScript L_k$ acts transitively on it, i.~e.
every point on the cylinder can be transformed to any other. Let
us denote the vector that have the end in the initial point as
$e^1(0)$, and the vector that have the end in the final point as
$e^1(1)$. So far as the cylinder is connected, these two points
can be connected by a continuous curve $e^1(t)$, $t\in[0;1]$.

Let us introduce now a vector $e^2(t)$ which continuously depends
on~$t$. Let us require that the end of the vector $e^2(t)$ always
lies on the same cylinder of the type~(\ref{cylinder}).
Furthermore, we require that the scalar product $e^1(t)\cdot
e^2(t)=const$ does not depend on $t$. And what is more, we require
that the vectors $k$, $e^1(t)$ and $e^2(t)$ are linearly
independent when $t=0$ and, therefore, they are linearly
independent with any~$t$.

The vector $e^2(t)$ is not defined uniquely, of course. It always
can be substituted by~$e^2(t)+\alpha(t)\,k$, where~$\alpha(t)$ is
an arbitrary continuous function. We will not eliminate this
non-uniqueness in any way.

Let us use the three vectors $k$, $e^1(t)$ and $e^2(t)$ as a basis
in~$M^\perp$. Using the three-dimensional version of the first
statement from the section~\ref{inverse} we get a continuous
family of linear transformations in~$M^\perp$ which preserve the
scalar product in~$M^\perp$.

Let us prove now that these transformations can be completed as
transformations from~$\EuScript L_k$. For this let us take an
arbitrary vector~$n(0)$ that does not belong to the
subspace~$M^\perp$. The vectors $k$, $e^1(t)$, $e^2(t)$ and $n(0)$
form a basis in the Minkowski space~$M$. Let us show now that for
any~$t$ vector $n(t)$ can be chosen so that the following four
values do not depend on~$t$:
\begin{equation}
 k\cdot n(t) =const \ , \qquad e^1(t)\cdot n(t) =const \ ,
 \qquad
 e^2(t)\cdot n(t) =const \ , \qquad (n(t))^2 =const \ .
\label{fourth}
\end{equation}

The first three conditions with fixed~$t$ define in the Minkowski
space a line~$\widetilde n$, which is a point of the factor-plane
$\widetilde N$ constructed in the section~\ref{homom}. As we have
pointed out in the section~\ref{inverse}, on this line there is
exactly one point satisfying the fourth of the
equations~(\ref{fourth}). This is the point where the desired
vector~$n(t)$ has its end.

Using now the first statement from the section~\ref{inverse}, we
see that we have constructed a continuous family of
transformations from $\EuScript L_k$ transforming a given point of
the cylinder~(\ref{cylinder}) to any other given point. So, we
have proved the

\mylemma{}
 {\it Transformations of the little group $\EuScript L_k$ act on
the cylinder}~(\ref{cylinder}) {\it transitively.}

By the way, we have proved the following

\mystat{}
 {\it A linear transformation of the space $M^\perp$, preserving
the scalar product there, can be uniquely completed as a linear
transformation of~$M$, preserving the scalar product in~$M$.}

Let us suppose now that the representation~$T^\perp$ is
decomposable, i.~e. the space~$M^\perp$ decomposes into a direct
sum $M^\perp=M'\oplus M''$ of two reducing subspaces $M'$ and
$M''$. One of these two subspaces, for example~$M'$, necessarily
contain some vector that does not belong to~$M^\parallel$. The end
of this vector lies on the corresponding
cylinder~(\ref{cylinder}). But then, as we have just shown, the
whole cylinder must belong to $M'$. But the linear shell of the
cylinder~(\ref{cylinder}) coincides with~$M^\perp$ and, therefore,
$M'=M^\perp$. So, we have proved a

\mylemma{}
 {\it Every reducing subspace of the representation $T^\perp$
either coincides with~$M^\perp$ or belongs to $M^\parallel$.}

From this we have the

\mytheorem{}
 {\it The representation $T^\perp$ is indecomposable.}

\section*{Complex vector representation}

\myparagraph{Complexification.} \label{compl}
 As we have shown in the paper~[IV], for investigation of
electromagnetic field it is useful to study also the complex
vector representation of the little group $\EuScript L_k$. In
order to get this representation we just should admit that
components of vectors under consideration can be complex numbers
and suppose that representation of the group $\EuScript L_k$ acts
there by the same formulas as in the real case. And the group
$\EuScript L_k$ does not change in any way.

The complexified Minkowski space will be denoted as~$M_{\mathbb
C}$. Generally, later on all spaces and representations will be
labeled by symbol of the corresponding field of scalars. If two
spaces are denoted the same but have different scalar label, then
we imply, that the complex space is the complexification of the
real. For example, the real Minkowski space will be denoted
as~$M_{\mathbb R}$. We will also imply that real spaces are
subsets of corresponding complex spaces. The representation of the
group $\EuScript L_k$ that acts in~$M_{\mathbb C}$ will be denoted
as $T_{\mathbb C}$.

An arbitrary vector $n\in M_{\mathbb C}$ can be represented as a
sum of its real and imaginary parts:
$n=n^\mathrm{Re}+in^\mathrm{Im}$. Here
$n^\mathrm{Re},n^\mathrm{Im}\in M_{\mathbb R}$. So far as
transformations of the group $\EuScript L_k$ are transformations
with real coefficients, we may suppose that a transformation of a
complex vector is a simultaneous transformations of its real and
imaginary parts.

All reducing subspaces and factor-spaces found in the real case
are obviously complexified. But the analogy between the
representations $T_{\mathbb R}$ and $T_{\mathbb C}$ is not full.
Reducibility and decomposability of complexified representations
must be investigated in addition. For proving of
indecomposability of complex representations the following lemma
will play an important role

\mylemma{}
 {\it Let us suppose that we have defined:
a vector~$n\in M_{\mathbb C}$; some two-dimensional
subspace~$P^2_{\mathbb R}$ in~$M_{\mathbb R}$, such that
 $n\in P^2_{\mathbb C}$; and some one-dimensional
subspace~$P^1_{\mathbb R}$ in~$P^2_{\mathbb R}$, such that
 $n\notin P^1_{\mathbb C}$. Then multiplying~$n$ by an appropriate
factor~$\theta\in\mathbb C$ we can make that
 $(\theta n)^\mathrm{Im}\in P^1_{\mathbb R}$ and
 $(\theta n)^\mathrm{Re}\notin P^1_{\mathbb R}$.}

In order to prove this lemma, let us multiply the initial
vector~$n$ by a number like $e^{i\varphi}$, where the
argument~$\varphi$ runs through the set of real numbers. When the
argument~$\varphi$ changes, the vectors $n^\mathrm{Re}(t)$ and
$n^\mathrm{Im}(t)$ will move on some ellipsis in the
plane~$P^2_{\mathbb R}$ (If $n^\mathrm{Re}$ and $n^\mathrm{Im}$
are parallel, the ellipsis degenerates into the segment). This
ellipsis intersects with any one-dimensional
subspace~$P^1_{\mathbb R}$ belonging to~$P^2_{\mathbb R}$. Now it
is obvious that with appropriate~$\varphi$ the imaginary part of
the vector will be in $P^1_{\mathbb R}$ and the real part will not
be there.


\myparagraph{Decomposability of
 $T^{\perp/\parallel}_{\mathbb C}$.} \label{tperpparcompl}
 As we have shown in the section~\ref{tperppar}, in the space
$M^{\perp/\parallel}_{\mathbb R}$ the little group $\EuScript L_k$
acts as the group $SO(2)$. So far as the group $SO(2)$ is Abelian,
as it follows from the Schur lemma, the complex representation
$T^{\perp/\parallel}_{\mathbb C}$ must be reducible. On the other
hand, so far as the group $SO(2)$ is compact, the representation
$T^{\perp/\parallel}_{\mathbb C}$ is equivalent to some unitary
representation and therefore it is fully reducible. The
corresponding subrepresentations we will denote as
$T^{(+1)/\parallel}_{\mathbb C}$ and
 $T^{(-1)/\parallel}_{\mathbb C}$.

In order to make this result more concrete, let us introduce in
the real space $M^{\perp/\parallel}_{\mathbb R}$ an orthonormal
basis~$\{\tilde e^i\}_{i=1,2}$. Then the matrices of the
representation $T^{\perp/\parallel}_{\mathbb R}$ will take the
form:
\[
 \left(\begin{array}{cc}
          \cos\varphi & -\sin\varphi\\
          \sin\varphi & \cos\varphi
 \end{array}\right)
\]

If now in the complex plane
 $M^{\perp/\parallel}_{\mathbb C}$ we use complex ``spiral''
basis~$\{\tilde e^{(\lambda)}\}_{\lambda=\pm1}$:
\[
 \tilde e^{(\pm1)} =
   \frac{\tilde e^1 \pm i\,\tilde e^2}{\sqrt{\,2\,}} \ ,
\]
the matrices of the representation
 $T^{\perp/\parallel}_{\mathbb C}$ become diagonal:
\[
\left(\begin{array}{cc}
          e^{-i\varphi} & 0             \\
          0             & e^{+i\varphi}
\end{array}\right)
\]
So, we have proved the

\mytheorem{}
 {\it The representation $T^{\perp/\parallel}_{\mathbb C}$ has exactly
two reducing subspaces:
 $M^{(+1)/\parallel}_{\mathbb C}$ and
 $M^{(-1)/\parallel}_{\mathbb C}$.
The sum of the corresponding subrepresentations is the
representation $T^{\perp/\parallel}_{\mathbb C}$:
\[
 T^{\perp/\parallel}_{\mathbb C} =
      T^{(+1)/\parallel}_{\mathbb C} \oplus
      T^{(-1)/\parallel}_{\mathbb C} \ .
\]}

For the electromagnetic field (see~[IV]) this theorem means that
there exist plane waves with helicity $+1$ and~$-1$~--- the
well-known fact.

Let us notice also that:
\begin{itemize}
\item[a.]
So far as in the Minkowski space we have not initially introduced
any orientation, there is no orientation on the plane
$M^{\perp/\parallel}_{\mathbb R}$. Therefore we can not say which
space is denoted $M^{(+1)/\parallel}_{\mathbb C}$, and which
$M^{(-1)/\parallel}_{\mathbb C}$. In fact, by fixation of these
notations we introduce orientation.
\item[b.]
If we extend the Lorentz group to the {\it full\/} Lorentz group
by including there space reflections, then the little group also
expands. The group $SO(2)$ acting in the plane
$M^{\perp/\parallel}_{\mathbb R}$ expands to $O(2)$. Under
reflections the complex spaces
 $M^{(+1)/\parallel}_{\mathbb C}$ and
 $M^{(-1)/\parallel}_{\mathbb C}$
will be transformed to each other and the complex representation
in the space $M^{\perp/\parallel}_{\mathbb C}$ will become
irreducible.
\end{itemize}


\myparagraph{Subrepresentations $T^{(+1)}_{\mathbb C}$ and
$T^{(-1)}_{\mathbb C}$.} \label{tpm}
 In accordance with the decomposability of the
factor-rep\-re\-sen\-ta\-tion $T^{\perp/\parallel}_{\mathbb C}$
shown in the section~\ref{tperpparcompl}, in the complex Minkowski
space $M_{\mathbb C}$ we can find two two-dimensional reducing
subspaces of the representation $T_{\mathbb C}$. These two
subspaces we will denote $M^{(+1)}_{\mathbb C}$ and
$M^{(-1)}_{\mathbb C}$. The corresponding one-dimensional
factor-spaces
 $M^{(+1)}_{\mathbb C}/M^\parallel_{\mathbb C} =
  M^{(+1)/\parallel}_{\mathbb C}$ and
 $M^{(-1)}_{\mathbb C}/M^\parallel_{\mathbb C} =
  M^{(-1)/\parallel}_{\mathbb C}$ were introduced in the
section~\ref{tperpparcompl}.

Now it is clear that in the complex case the series~(\ref{admset})
can be thickened in two ways:
\begin{equation}
 \{0\}= M^0_{\mathbb C} \subset
 M^\parallel_{\mathbb C} \subset
 M^{(+1)}_{\mathbb C} \subset
 M^\perp_{\mathbb C} \subset
 M^4_{\mathbb C} = M_{\mathbb C} \ ,
\label{admsetplus}
\end{equation}
\begin{equation}
 \{0\}= M^0_{\mathbb C} \subset
 M^\parallel_{\mathbb C} \subset
 M^{(-1)}_{\mathbb C} \subset
 M^\perp_{\mathbb C} \subset
 M^4_{\mathbb C} = M_{\mathbb C} \ ,
\label{admsetminus}
\end{equation}

It will be shown in the section~\ref{tperpcompl}, that the complex
vector representation does not have other reducing subspaces. So,
besides~(\ref{admsetplus}) and~(\ref{admsetminus}), the
representation $T_{\mathbb C}$ does not have other composition
series.


\myparagraph{Indecomposability of representation
 $T_{\mathbb C}$.} \label{tcompl}
 Let us suppose that the representation $T_{\mathbb C}$ is
decomposable, i.~e. the space $M_{\mathbb C}$ can be represented
as a sum
\begin{equation}
 M_{\mathbb C}=M'_{\mathbb C}\oplus M''_{\mathbb C}
\label{decompt}
\end{equation}
of two reducing subspaces $M'_{\mathbb C}$ and $M''_{\mathbb C}$.
Then one of these subspaces, for example~$M'_{\mathbb C}$,
necessarily contains some vector~$n$ that does not belong
to~$M^\perp_{\mathbb C}$. According to the lemma from the
section~\ref{compl}, without loss of generality we can assume that
$n^\mathrm{Im}\in M^\perp_{\mathbb R}$ and $n^\mathrm{Re}\notin
M^\perp_{\mathbb R}$.

Let us introduce now some real vector $e^1$. In the case if the
vector $n^\mathrm{Im}$ does not belong to
 $M^\parallel_{\mathbb R}$, $e^1$ is just another notation
for $n^\mathrm{Im}$. But if $n^\mathrm{Im}$ belongs to
$M^\parallel_{\mathbb R}$, vector $e^1$ is chosen as any vector
such that $e^1\in M^\perp_{\mathbb R}$ and
 $e^1\notin M^\parallel_{\mathbb R}$.

Consider now the subgroup of the little group $\EuScript L_k$ that
leaves the vector $e^1$ unchanged. According to the definition of
the vector $e^1$, this subgroup in any case leaves unchanged the
vector $n^\mathrm{Im}$. From the results of the
section~\ref{tperp} we get that the elements of this group can be
defined by some vector $e^2$, which always has its end lying on
some cylinder~(\ref{cylinder}), the condition $e^1\cdot e^2=const$
is satisfied, and the vectors $k$, $e^1$ and $e^2$ are linearly
independent. The end of the vector $e^2$ runs through some line
parallel to~$k$: $e^2(t)=e^2(0)+t k$. The equations~(\ref{fourth})
for the vector $n^\mathrm{Re}(t)$ take the form:
\[
 k\cdot n^\mathrm{Re}(t) =const        \ , \qquad
 e^1\cdot n^\mathrm{Re}(t) =const      \ , \qquad
 e^2(t)\cdot n^\mathrm{Re}(t) =const   \ , \qquad
 (n^\mathrm{Re}(t))^2 =const           \ .
\]
The third equation can be written with more details:
\begin{equation}
 (e^2(0)+t k)\cdot n^\mathrm{Re}(t) =const \ .
\label{thirdeq}
\end{equation}
Obviously, if we suppose that
 $(n^\mathrm{Re}(t)-n^\mathrm{Re}(0))\in M^\parallel$, then
the equation~(\ref{thirdeq}) for $t\neq0$ can not be satisfied.
Therefore, the subspace~$M'_{\mathbb C}$ must contain some real
vector belonging to
 $M^\perp_{\mathbb R}$ and not belonging to
 $M^\parallel_{\mathbb R}$. But then, according to the second
lemma in the section~\ref{tperp}, we
 have~$M^\perp_{\mathbb R}\subset M'_{\mathbb C}$ and, therefore,
$M^\perp_{\mathbb C}\subset M'_{\mathbb C}$.

Let us come back again to the complex vector
 $n\in M'_{\mathbb C}$,
for which
 $n^\mathrm{Re}\notin M^\perp_{\mathbb R}$ and
 $n^\mathrm{Im}\in M^\perp_{\mathbb R}$.
So far as we have already proved that
 $M^\perp_{\mathbb C}\subset M'_{\mathbb C}$,
we get that in $M'_{\mathbb C}$ there is also such a vector which
has zero imaginary part and which has the real part equal to the
real part of vector~$n$. In other words, we have proved that
$M'_{\mathbb C}$ contains a real vector that does not belong to
$M^\perp_{\mathbb R}$. But then, according to the lemma of the
section~\ref{unt}, $M_{\mathbb R}\subset M'_{\mathbb C}$, and,
therefore, spaces $M'_{\mathbb C}$ and $M_{\mathbb C}$ coincide.

So, we have proved the

\mylemma{}
 {\it Every reducing subspace of the representation~$T_{\mathbb C}$
either coincides with the whole space $M_{\mathbb C}$ or belongs
to $M^\perp_{\mathbb C}$.}

From this we get the

\mytheorem{}
 {\it The representation~$T_{\mathbb C}$ is indecomposable.}


\myparagraph{Indecomposability of representations
$T^\perp_{\mathbb C}$, $T^{(+1)}_{\mathbb C}$ and
$T^{(-1)}_{\mathbb C}$.} \label{tperpcompl}
 Let us prove the following lemma:

\mylemma{}
 {\it Every subrepresentation of representation $T^\perp_{\mathbb C}$
contains the subrepresentation $T^\parallel_{\mathbb C}$.}

In order to prove this lemma, let us denote the subrepresentation
under consideration as $T'_{\mathbb C}$. The reducing subspace
where it acts we denote $M'_{\mathbb C}$.

Consider some non-zero vector~$n$ from the subspace
 $M'_{\mathbb C}$. If it lies in~$M^\parallel_{\mathbb C}$,
then the statement of the lemma is satisfied. Suppose now that
this vector does not belong to~$M^\parallel_{\mathbb C}$.

If the vectors $k$, $n^\mathrm{Re}$ and $n^\mathrm{Im}$ are
linearly dependent then, according to the lemma from the
section~\ref{compl}, we can without loss of generality think that
 $n^\mathrm{Im}\in M^\parallel_{\mathbb R}$ and
 $n^\mathrm{Re}\notin M^\parallel_{\mathbb R}$.
According to the first lemma from the section~\ref{tperp}, using
transformations of the little group $\EuScript L_k$ we can change
the vector
 $n^\mathrm{Re}$ so that we add to it an arbitrary vector from
 $M^\parallel_{\mathbb R}$. At the same time the vector
 $n^\mathrm{Im}$ remains unchanged. Therefore
 $M^\parallel_{\mathbb R}\subset M'_{\mathbb C}$ and,
therefore, $M^\parallel_{\mathbb C}\subset M'_{\mathbb C}$.

But if the vectors $k$, $n^\mathrm{Re}$ and $n^\mathrm{Im}$ are
linearly dependent, then, according to the statement of the
section~\ref{tperp}, choosing a proper transformation from the
little group $\EuScript L_k$ we can add to the vector
$n^\mathrm{Re}$ a vector parallel to~$k$ so that the vector
$n^\mathrm{Im}$ remains unchanged. Therefore, again
 $M^\parallel_{\mathbb R}\subset M'_{\mathbb C}$ and
 $M^\parallel_{\mathbb C}\subset M'_{\mathbb C}$.
So, the lemma is proved.

From this lemma we get the two theorems:

\mytheorem{}
 {\it The representation $T^\perp_{\mathbb C}$ is
indecomposable.}

\mytheorem{}
 {\it The representations $T^{(+1)}_{\mathbb C}$ and
 $T^{(-1)}_{\mathbb C}$ are indecomposable.}

Furthermore, from the lemma and the theorem of the
section~\ref{tperppar} we get the

\mytheorem{}
 {\it Besides $M^\parallel_{\mathbb C}$, $M^{(+1)}_{\mathbb C}$,
$M^{(-1)}_{\mathbb C}$ (and $M^\perp_{\mathbb C}$) the
representation $T^\perp_{\mathbb C}$ does not have any other
reducing subspaces.}

So, the representation $T_{\mathbb C}$ does not have any other
composition series, besides~(\ref{admsetplus})
and~(\ref{admsetminus}).


\myparagraph{Matrices of representations $T_{\mathbb R}$ and
$T_{\mathbb C}$.}
 Many of the obtained results can be represented by patterns for
matrices of the representations $T_{\mathbb R}$ and
 $T_{\mathbb C}$. If in the spaces $M_{\mathbb R}$ and
$M_{\mathbb C}$ we choose appropriate bases, then these templates
become block-triangular:
\[
 \left(\begin{array}{cccc}      \cline{1-1}
 \multicolumn{1}{|c|}{1} & \ \cdot & \cdot & \cdot \\ \cline{1-3}
  0  & \multicolumn{2}{|c|}{} & \cdot \\
  0  & \multicolumn{2}{|c|}{\smash{\raisebox{1.4ex}{$SO(2)$}}}
     & \cdot \\ \cline{2-4}
  0  &   \ 0   &   0   & \multicolumn{1}{|c|}{1}   \\ \cline{4-4}
 \end{array}\right)
 \qquad
 \left(\begin{array}{cccc}      \cline{1-1}
 \multicolumn{1}{|c|}{1}  & \cdot & \cdot & \cdot \\ \cline{1-2}
  0  & \multicolumn{1}{|c|}{e^{-i\varphi}} & 0 & \cdot \\ \cline{2-3}
  0  &   0   & \multicolumn{1}{|c|}{e^{+i\varphi}} & \cdot \\ \cline{3-4}
  0  &   0   &   0   & \multicolumn{1}{|c|}{1}   \\ \cline{4-4}
 \end{array}\right)
\]
And it is known that on the place of every point we have non-zero
element, in general case.

Studying in what subspaces and factor-spaces different blocks of
these matrices act I leave for the reader.




\newcounter{parnum}
\newlength{\picwidth}
\newcommand*{\myparagraph}[1]{\refstepcounter{parnum}%
\paragraph{\arabic{parnum}. #1}}


\title{О квантовании электромагнитного поля.\\
       V. Векторное представление малой группы Лоренца\\
       для светоподобного импульса.}
\author{Д.~А.~Арбатский\footnote{http:\dslash d-a-arbatsky.narod.ru/}}
\date{\today}
\maketitle

\begin{abstract}
С использованием элементарных геометрических методов
устанавливается изоморфизм малой группы Лоренца для светоподобного
импульса и группы движений евклидовой плоскости. В соответствии с
теоремой Жордана-Гёльдера-Нётер производится \rql приведение\rqr{}
вещественного и комплексного векторных представлений этой группы.
Доказывается неразложимость этих представлений.
\end{abstract}


\myparagraph{Предварительные замечания.}
 Книг, в которых излагается теория группы Лоренца~$\EuScript L$ и
теория представлений этой группы, имеется довольно много. В том
числе имеются монографии, специально посвящённые этой теме;
например,~\cite{Naymark1958,GelMinShap1958,Fyodorov1979}. Всюду
отмечается, что малая группа Лоренца~$\EuScript L_k$ для
светоподобного импульса\footnote{Слово \rql импульс\rqr{}
употребляется, поскольку имеются в виду приложения к теории
безмассовых полей; в частности, электромагнитного. Пока мы
занимаемся чисто математическим исследованием, можно считать, что
речь идёт о произвольном изотропном векторе~$k$: $k^2=0$.} $k$
изоморфна группе движений евклидовой плоскости~$E(2)$.
Доказательство, которое при этом приводится, обычно основывается
на аналитическом изучении группы $SL(2,\mathbb C)$, которая
является универсальной накрывающей группы Лоренца.

Такой подход, по-видимому, является оптимальным, когда в
дальнейшем ставится задача изучения произвольных (в том числе
двузначных) представлений группы~$\EuScript L_k$. При этом,
однако, понять геометрический смысл изоморфизма $\EuScript L_k$ и
$E(2)$ представляется затруднительным\footnote{В частности,
Райдер~\cite{Ryder1987} по этому поводу заметил, что
 \rql физический смысл этого результата не ясен\rqr.}.

Уточним здесь, что обычно понимается под \rql геометрическим
смыслом\rqr. С точки зрения физических приложений, группа Лоренца
наиболее естественно определяется своим векторным
представлением\footnote{Такое представление иногда называют \rql
фундаментальным\rqr.}. При этом малая группа $\EuScript L_k$
определяется просто как стационарная подгруппа вектора~$k$. Группа
$E(2)$, в свою очередь, определяется как группа движений
евклидовой плоскости. Имея в виду эти определения групп
 $\EuScript L_k$ и $E(2)$, с помощью чисто геометрических
построений мы отыщем некоторую двумерную плоскость, обладающую
естественной евклидовой структурой, на которой группа $\EuScript
L_k$ будет действовать как группа движений.

Другой вопрос, которым мы будем здесь заниматься (и который
является главной мотивацией для опубликования данной статьи)~---
это \rql приведение\rqr{} векторного представления
группы~$\EuScript L_k$, т.~е. мы выясним, из каких неприводимых
представлений оно состоит. С точки зрения общей теории
представлений, этот случай является очень частным и не содержит
каких-либо специфических трудностей. Однако, ввиду его большой
практической важности, представляется полезным провести его
рассмотрение с использованием построений, наиболее отвечающих
нашей геометрической интуиции.

\section*{Вещественное векторное представление}

\myparagraph{Обозначения и терминология.}
 {\it Пространством Минковского\/} $M$ называется четырёхмерное
вещественное линейное пространство~$\mathbb R^4$, в котором задана
вещественная симметричная билинейная форма
$g(\,\cdot\,,\,\cdot\,)$, имеющая сигнатуру $(+,-,-,-)$. Эта форма
называется {\it скалярным произведением}. Для краткости вместо
$g(a,b)$ мы будем писать просто~$a b$ или~$a\cdot b$.

Рассмотрим теперь группу всех линейных преобразований пространства
Минковского~$M$, сохраняющих скалярное произведение. Связная
компонента этой группы, содержащая единицу, сама является группой.
Эту последнюю группу мы будем называть {\it группой Лоренца\/} и
обозначать её~$\EuScript L$.

На протяжении всей статьи мы будем иметь дело с векторным
представлением группы Лоренца, то есть с представлением в самом
пространстве Минковского. Это представление будет обозначаться
как~$T$.

{\it Малой группой вектора\/}~$k$ называется стационарная
подгруппа вектора~$k$, т.~е. подгруппа группы Лоренца, оставляющая
вектор~$k$ неизменным. Она будет обозначаться как~$\EuScript L_k$.

Вектор~$k$ называется {\it времениподобным}, если $k^2>0$; {\it
светоподобным\/} или {\it изотропным}, если $k^2=0$; и {\it
пространственноподобным}, если $k^2<0$.

Рассмотрим также группу всех преобразований евклидовой плоскости,
сохраняющих расстояние. Связная компонента этой группы, содержащая
единицу, сама является группой. Мы будем её называть {\it группой
движений евклидовой плоскости\/} и обозначать~$E(2)$.


\myparagraph{Приводящие подпространства.}
 Следуя теореме Жордана-Гёльдера-Нётер, попробуем сразу угадать
композиционный ряд представления~$T$, т.~е. такой монотонный набор
приводящих подпространств представления~$T$, чтобы соответствующие
последовательные факторпредставления оказались неприводимыми:
\begin{equation}
 \{0\}= M^0 \subset M^\parallel \subset M^\perp \subset M^4=M \ .
\label{admset}
\end{equation}
Здесь
 $M^0$~--- нулевое подпространство в~$M$;
 $M^\parallel$~--- подпространство, содержащее все векторы,
     параллельные~$k$;
 $M^\perp$~--- подпространство, содержащее все векторы,
     ортогональные~$k$;
 $M^4$~--- другое обозначение пространства Минковского.

Каждое из этих подпространств, очевидно, инвариантно по отношению
к действию группы~$\EuScript L_k$. Подпредставления в
подпространствах $M^\parallel$ и $M^\perp$ обозначим $T^\parallel$
и $T^\perp$, соответственно.

Далее, набору~(\ref{admset}) соответствует последовательность из
трёх последовательных факторпредставлений в факторпространствах
 $M^\parallel/M^0=M^\parallel$,
 $M^\perp/M^\parallel=M^{\perp/\parallel}$ и
 $M^4/M^\perp=M^{4/\perp}$.
Первое из этих факторпредставлений является подпредставлением. Мы
его уже обозначили как $T^\parallel$. Второе и третье
факторпредставления обозначим $T^{\perp/\parallel}$ и
$T^{4/\perp}$, соответственно.

Неприводимость факторпредставлений $T^\parallel$ и $T^{4/\perp}$
не вызывает сомнений по причине их одномерности. Что касается
факторпредставления~$T^{\perp/\parallel}$, то его неприводимость
(вещественную) мы установим в пункте~\ref{tperppar}.

В пунктах~\ref{unt} и~\ref{tperp} будут доказаны две леммы из
которых следует, что кроме подпространств~(\ref{admset}), других
приводящих подпространств у представления $T$ нет. Тем самым,
композиционный ряд~(\ref{admset}) является единственным. Поэтому в
данном случае утверждение теоремы Жордана-Гёльдера-Нётер о том,
что все композиционные ряды имеют одинаковую длину и
соответствующие факторпредставления у этих рядов эквивалентны,
оказывается тривиальным.


\myparagraph{Гомоморфизм $\EuScript L_k$ в $E(2)$.}
 \label{homom}
 Рассмотрим теперь в пространстве Минковского трёхмерную
гиперплоскость~$N$, задаваемую уравнением:
\begin{equation}
 k\cdot x = const \neq 0  \ .
\label{plane}
\end{equation}
Поскольку группа~$\EuScript L_k$ оставляет вектор~$k$ неизменным и
сохраняет скалярное произведение, гиперплоскость~$N$ инвариантна
относительно действия группы~$\EuScript L_k$.

Введём теперь для векторов из~$N$ отношение
эквивалентности~$\sim$:
\begin{equation}
 a\sim b \ \ \Longleftrightarrow \ \  a-b \,\in\, M^\parallel \ ,
\label{equiv}
\end{equation}
Иначе говоря, два вектора эквивалентны, если их концы лежат на
прямой, параллельной~$k$. Ввиду~(\ref{admset}) и (\ref{plane})
каждая такая прямая целиком содержится в гиперплоскости~$N$.
Факторизуя~$N$ по отношению эквивалентности~(\ref{equiv}),
получаем некоторую двумерную плоскость\footnote{Плоскость
$\widetilde N$ не является {\it вложенной\/} в пространство
Минковского.}~$\widetilde N$. Мы будем обозначать точки
из~$\widetilde N$ и соответствующие им прямые на~$N$ одинаковыми
символами: $\widetilde a$, $\widetilde c$ и т.~п.

Плоскость~$\widetilde N$ обладает естественной евклидовой
структурой. В самом деле, возьмём две произвольные точки
$\widetilde a$ и $\widetilde c$ на~$\widetilde N$. Каждая из
них~--- это класс эквивалентности точек на~$N$. Возьмём по одному
произвольному представителю каждого из этих двух классов: $a$ и
$c$, соответственно. Вычислим величину
\begin{equation}
 \rho\,(a,c)=\sqrt{(a-c)^2} \ .
\label{distN}
\end{equation}
Эта величина не зависит от выбора представителей $a$ и $c$.
Действительно, заменим, скажем, $a$ на $a+\alpha k$.
Учитывая~(\ref{plane}), имеем:
\[
 \rho\,(a+\alpha k,c) = \sqrt{(a-c+\alpha k)^2} =
 \sqrt{(a-c)^2 + 2\alpha(k a-k c) + \alpha^2k^2} =
 \sqrt{(a-c)^2} = \rho\,(a,c) \ .
\]

Кроме того, $(a-c)^2$ всегда больше нуля, и функция
$\rho\,(\,\cdot\,,\,\cdot\,)$ обладает всеми свойствами евклидовой
метрики, поскольку плоскость~$\widetilde N$ можно отождествить с
любой из пространственно-подобных двумерных плоскостей,
содержащихся в~$N$.

Теперь заметим, что и отношение эквивалентности~(\ref{equiv}) и
метрика $\rho\,(\,\cdot\,,\,\cdot\,)$ инвариантны относительно
действия малой группы~$\EuScript L_k$. Таким образом, построено
гомоморфное отображение группы~$\EuScript L_k$ в группу
движений~$E(2)$ евклидовой плоскости~$\widetilde N$.


\myparagraph{Обратное отображение.} \label{inverse}
 В дальнейшем мы будем неоднократно использовать следующие два
предложения, верные как в вещественном, так и в комплексном
случае.

\mystat{}
 {\it Если линейное преобразование действует на некоторый базис
$\{n^i\}_{i=1..4}$ так, что оказываются неизменными скалярные
произведения элементов базиса (включая квадраты базисных
элементов), то это преобразование сохраняет скалярное
произведение.}

\mystat{}
 {\it Если линейное преобразование действует на некоторый базис
$\{n^i\}_{i=1..4}$ так, что оказываются неизменными все величины
$(n^i-n^j)^2$ и $(n^i)^2$, то это преобразование сохраняет
скалярное произведение.}

Первое предложение следует из того, что всякий вектор можно
разложить по векторам базиса. Второе предложение следует из
первого с учётом формулы:
\[
 (n^i-n^j)^2=(n^i)^2+(n^j)^2-2n^in^j   \ .
\]

Зафиксируем теперь произвольное движение плоскости~$\widetilde N$.
Преобразование из малой группы~$\EuScript L_k$, которое при
построенном в пункте~\ref{homom} гомоморфизме переходит в данное
движение на~$\widetilde N$, будем называть {\it искомым}. Покажем,
что искомое преобразование из~$\EuScript L_k$ существует и
единственно. Тем самым будет установлено, что рассматриваемый
гомоморфизм на самом деле является изоморфизмом.

Для этого рассмотрим семейство гиперповерхностей, задаваемых
уравнениями вида
\begin{equation}
 x^2=const \ .
\label{hypsurf}
\end{equation}
Пересечение любой из них с гиперплоскостью~$N$ является
параболоидом. Каждый такой параболоид инвариантен относительно
искомого преобразования. При этом каждая прямая, представляющая
собой точку из~$\widetilde N$, пересекает такой параболоид ровно в
одной точке, и через каждую точку параболоида проходит ровно одна
такая прямая. Следовательно, определено действие искомого
преобразования на любом из параболоидов.

Возьмём теперь на одном из параболоидов четыре точки, не
принадлежащие никакой двумерной плоскости. Четыре вектора
$\{n^i\}_{i=1..4}$, концами которых являются эти четыре точки,
образуют базис в~$M$. Определённое нами на параболоиде
преобразование сохраняет величины $(n^i-n^j)^2$, т.~к. каждая из
них~--- это квадрат соответствующего расстояния в~$\widetilde N$.
Величины~$(n^i)^2$ также сохраняются, т.~к. они постоянны на
параболоиде. Продолжим теперь по линейности построенное
преобразование векторов базиса на всё пространство~$M$. Как
следует из второго предложения, сформулированного в начале этого
пункта, получится преобразование, сохраняющее скалярное
произведение в~$M$. Кроме того, по построению, данное
преобразование сохраняет плоскость~(\ref{plane}), а потому оно
сохраняет неизменным вектор~$k$. Так как исходное движение
плоскости~$\widetilde N$ предполагалось принадлежащим связной
группе $E(2)$, построенное преобразование в~$M$ принадлежит
связной группе~$\EuScript L_k$.

Построенное преобразование является единственным, могущим
претендовать на роль искомого. С другой стороны, построенное
преобразование из~$\EuScript L_k$ при гомоморфизме из
пункта~\ref{homom} переходит в некоторое движение на~$\widetilde
N$. Это движение, по построению, действует на четыре
точки\footnote{Прямые, содержащие концы векторов базиса
$\{n^i\}_{i=1..4}$}~$\widetilde n^i\in \widetilde N$ так же, как и
исходное движение в~$\widetilde N$. Поскольку движение евклидовой
плоскости однозначно определяется своим действием даже на две
несовпадающие точки, построенное преобразование из~$\EuScript L_k$
является искомым. Таким образом, доказана

\mytheorem{} {\it Группы $\EuScript L_k$ и $E(2)$ изоморфны.}


\myparagraph{Факторпредставление $T^{\perp/\parallel}$.}
\label{tperppar}
 Аналогично построенной в пункте~\ref{homom}
факторплоскости~$\widetilde N$, факторпространство
$M^{\perp/\parallel}$ обладает естественной евклидовой
структурой\footnote{Более того, $M^{\perp/\parallel}$ линейно, и в
нём имеется соответствующее скалярное произведение:
 $\widetilde a\cdot \widetilde c = a c$.}, она даётся той же
формулой~(\ref{distN}). Таким образом, малая
 группа~$\EuScript L_k$ действует в $M^{\perp/\parallel}$ как группа
вращений евклидовой плоскости вокруг фиксированной точки. Такая
группа обозначается как $SO(2)$.

Очевидно, в евклидовой плоскости нет такого направления, которое
оставалось бы инвариантным при вращениях. Отсюда вытекает

\mytheorem{}
 {\it Факторпредставление $T^{\perp/\parallel}$ неприводимо
(в вещественном смысле).}


\myparagraph{Неразложимость представления $T$.} \label{unt}
 Предположим, что представление~$T$ разложимо, то есть
пространство~$M$ представимо в виде прямой суммы
 $M=M'\oplus M''$ двух нетривиальных приводящих
подпространств~$M'$ и $M''$. Одно из этих подпространств,
например~$M'$, обязательно содержит какой-нибудь вектор, не
содержащийся в $M^\perp$. При подходящем выборе константы в
уравнении~(\ref{plane}) конец этого вектора будет лежать в
плоскости~$N$. В пункте~\ref{inverse} мы видели, что
преобразованиями из $\EuScript L_k$ этот вектор может быть
переведён в любой другой вектор, конец которого лежит на том же
параболоиде. В том числе он может быть переведён в четыре вектора
$\{n^i\}_{i=1..4}$, составляющие базис в~$M$. Следовательно, $M'$
должно совпадать со всем~$M$. Таким образом, доказана

\mylemma{}
 {\it Всякое приводящее подпространство представления~$T$ либо
совпадает со всем пространством $M$, либо содержится в $M^\perp$.}

Из неё вытекает

\mytheorem{} {\it Представление~$T$ неразложимо.}


\myparagraph{Неразложимость представления $T^\perp$.}
\label{tperp}
 Рассмотрим цилиндр, задаваемый уравнениями:
\begin{equation}
 k\cdot x=0 \ , \qquad x^2 = const \neq 0 \ .
\label{cylinder}
\end{equation}
Докажем, что $\EuScript L_k$ действует на нём транзитивно, т.~е.
всякая точка на цилиндре, может быть переведена во всякую другую.
Обозначим вектор, конец которого указывает в исходную точку, как
$e^1(0)$, а вектор, указывающий в конечную, как $e^1(1)$. Так как
цилиндр связен, то эти две точки можно соединить непрерывной
кривой $e^1(t)$, $t\in[0;1]$.

Введём также вектор $e^2(t)$, непрерывно зависящий от~$t$. Будем
считать, что конец вектора $e^2(t)$ всегда лежит на некотором
одном цилиндре вида~(\ref{cylinder}). Кроме того, потребуем, чтобы
скалярное произведение $e^1(t)\cdot e^2(t)=const$ не зависело от
$t$. И ещё потребуем, чтобы векторы $k$, $e^1(t)$ и $e^2(t)$ были
линейно независимы при $t=0$ и, следовательно, линейно независимы
при произвольном~$t$.

Вектор $e^2(t)$ определён, конечно, неоднозначно. Его всегда можно
заменить на~$e^2(t)+\alpha(t)\,k$, где~$\alpha(t)$~---
произвольная непрерывная функция. Указанную неоднозначность мы
фиксировать никак не будем.

Примем теперь три вектора $k$, $e^1(t)$ и $e^2(t)$ за базис
в~$M^\perp$. Применяя трёхмерный вариант первого предложения из
пункта~\ref{inverse}, получаем непрерывное семейство линейных
преобразований в~$M^\perp$, сохраняющих в~$M^\perp$ скалярное
произведение.

Покажем теперь, что эти преобразования можно достроить до
преобразований из~$\EuScript L_k$. Для этого возьмём произвольный
вектор~$n(0)$, не принадлежащий подпространству~$M^\perp$. Векторы
$k$, $e^1(t)$, $e^2(t)$ и $n(0)$ образуют базис в пространстве
Минковского~$M$. Покажем, что при произвольном~$t$ вектор $n(t)$
можно выбрать так, что следующие четыре величины не будут зависеть
от~$t$:
\begin{equation}
 k\cdot n(t) =const \ , \qquad e^1(t)\cdot n(t) =const \ ,
 \qquad
 e^2(t)\cdot n(t) =const \ , \qquad (n(t))^2 =const \ .
\label{fourth}
\end{equation}

Первые три условия при фиксированном~$t$ выделяют в пространстве
Минковского~$M$ прямую~$\widetilde n$, являющуюся точкой
факторплоскости $\widetilde N$, построенной в пункте~\ref{homom}.
Как было указано в пункте~\ref{inverse}, на этой прямой имеется
ровно одна точка, удовлетворяющая четвёртому из
уравнений~(\ref{fourth}). Это и есть точка, в которую указывает
искомый вектор~$n(t)$.

Применяя теперь первое предложение из пункта~\ref{inverse}, видим,
что построено непрерывное семейство преобразований из
 $\EuScript L_k$, переводящих заданную точку
цилиндра~(\ref{cylinder}) в любую другую заданную точку. Итак,
доказана

\mylemma{}
 {\it Преобразования малой группы $\EuScript L_k$ действуют на
цилиндре}~(\ref{cylinder}) {\it транзитивно.}

Попутно нами было доказано следующее

\mystat{}
 {\it Линейное преобразование пространства $M^\perp$, сохраняющее
в нём скалярное произведение, может быть единственным образом
достроено до линейного преобразования пространства~$M$,
сохраняющего скалярное произведение в~$M$.}

Предположим теперь, что представление~$T^\perp$ разложимо, т.~е.
пространство~$M^\perp$ распадается в прямую сумму
$M^\perp=M'\oplus M''$ двух приводящих подпространств $M'$ и
$M''$. Одно из этих подпространств, например~$M'$, обязательно
содержит какой-нибудь вектор, не содержащийся в~$M^\parallel$.
Конец этого вектора лежит на соответствующем
цилиндре~(\ref{cylinder}). Но тогда, как было только что показано,
и весь цилиндр должен содержаться в $M'$. Но линейная оболочка
цилиндра~(\ref{cylinder}) совпадает с~$M^\perp$, и, следовательно,
$M'=M^\perp$. Таким образом, доказана

\mylemma{}
 {\it Всякое приводящее подпространство представления $T^\perp$
либо совпадает с~$M^\perp$, либо содержится в~$M^\parallel$.}

Из неё вытекает

\mytheorem{}
 {\it Представление $T^\perp$ неразложимо.}

\section*{Комплексное векторное представление}

\myparagraph{Комплексификация.} \label{compl}
 Как было показано в статье~[IV], при исследовании электромагнитного
поля полезно изучить также комплексное векторное представление
малой группы $\EuScript L_k$. Чтобы получить это представление,
нужно просто допустить, что компоненты рассматриваемых векторов
могут быть комплексными числами, и считать, что представление
группы $\EuScript L_k$ действует в нём по тем же формулам, что и в
вещественном случае. Сама группа $\EuScript L_k$ при этом никак не
изменяется.

Комплексифицированное пространство Минковского будет обозначаться
как~$M_{\mathbb C}$. Вообще, далее все пространства и
представления будут помечаться значком соответствующего им поля
скаляров. Если два пространства обозначаются одинаково и
различаются только указанным значком, то предполагается, что
комплексное пространство является комплексификацией вещественного.
Например, вещественное пространство Минковского, будет
обозначаться как~$M_{\mathbb R}$. Мы будем также полагать, что
вещественные пространства являются подмножествами соответствующих
комплексных. Представление группы $\EuScript L_k$, действующее
в~$M_{\mathbb C}$, обозначим $T_{\mathbb C}$.

Произвольный вектор $n\in M_{\mathbb C}$ может быть представлен в
виде суммы своей вещественной и мнимой частей:
$n=n^\mathrm{Re}+in^\mathrm{Im}$. Здесь
$n^\mathrm{Re},n^\mathrm{Im}\in M_{\mathbb R}$. Поскольку
преобразования группы $\EuScript L_k$ являются преобразованиями с
вещественными коэффициентами, можно считать, что преобразованию
комплексного вектора соответствует одновременное преобразование
его вещественной и мнимой частей.

Все найденные в вещественном случае приводящие подпространства и
факторпространства, очевидно, комплексифицируются. Полной
аналогии, однако, между представлениями $T_{\mathbb R}$ и
$T_{\mathbb C}$ нет. Приводимость и разложимость
комплексифицированных представлений нужно исследовать
дополнительно. При доказательстве неразложимости комплексных
представлений важную роль будет играть следующая

\mylemma{}
 {\it Пусть заданы: вектор~$n\in M_{\mathbb C}$; некоторое
двумерное подпространство~$P^2_{\mathbb R}$ в~$M_{\mathbb R}$,
такое, что $n\in P^2_{\mathbb C}$; и некоторое одномерное
подпространство~$P^1_{\mathbb R}$ в~$P^2_{\mathbb R}$, такое, что
$n\notin P^1_{\mathbb C}$. Тогда, умножая~$n$ на подходящий
множитель~$\theta\in\mathbb C$, можно добиться, чтобы
 $(\theta n)^\mathrm{Im}\in P^1_{\mathbb R}$ и
 $(\theta n)^\mathrm{Re}\notin P^1_{\mathbb R}$.}

Чтобы доказать эту лемму, умножим исходный комплексный вектор~$n$
на число вида $e^{i\varphi}$, где переменная~$\varphi$ пробегает
множество вещественных чисел. При изменении переменной~$\varphi$
векторы $n^\mathrm{Re}(t)$ и $n^\mathrm{Im}(t)$ будут двигаться по
некоторому эллипсу в плоскости~$P^2_{\mathbb R}$ (если
$n^\mathrm{Re}$ и $n^\mathrm{Im}$ параллельны, эллипс вырождается
в отрезок). Этот эллипс пересекается с любым одномерным
подпространством~$P^1_{\mathbb R}$, содержащимся
 в~$P^2_{\mathbb R}$. Отсюда очевидно, что при
 подходящем~$\varphi$ мнимая часть вектора окажется в
 $P^1_{\mathbb R}$, а вещественная там содержаться не будет.


\myparagraph{Разложимость $T^{\perp/\parallel}_{\mathbb C}$.}
\label{tperpparcompl}
 Как было показано в пункте~\ref{tperppar}, в пространстве
$M^{\perp/\parallel}_{\mathbb R}$ малая группа $\EuScript L_k$
действует, как группа $SO(2)$. Поскольку группа $SO(2)$ абелева,
то, как следует из леммы Шура, комплексное представление
$T^{\perp/\parallel}_{\mathbb C}$ обязано быть приводимым. Кроме
того, поскольку группа $SO(2)$ компактна, представление
$T^{\perp/\parallel}_{\mathbb C}$ эквивалентно некоторому
унитарному и, следовательно, вполне приводимо. Соответствующие
подпредставления обозначим $T^{(+1)/\parallel}_{\mathbb C}$ и
$T^{(-1)/\parallel}_{\mathbb C}$.

Чтобы сделать этот результат более конкретным, введём в
вещественном пространстве $M^{\perp/\parallel}_{\mathbb R}$
ортонормированный базис~$\{\tilde e^i\}_{i=1,2}$. При этом матрицы
представления $T^{\perp/\parallel}_{\mathbb R}$ примут вид:
\[
 \left(\begin{array}{cc}
          \cos\varphi & -\sin\varphi\\
          \sin\varphi & \cos\varphi
 \end{array}\right)
\]

Если теперь в комплексной плоскости
 $M^{\perp/\parallel}_{\mathbb C}$ перейти к комплексному
\rql спиральному\rqr{}
 базису~$\{\tilde e^{(\lambda)}\}_{\lambda=\pm1}$:
\[
 \tilde e^{(\pm1)} =
   \frac{\tilde e^1 \pm i\,\tilde e^2}{\sqrt{\,2\,}} \ ,
\]
то матрицы представления $T^{\perp/\parallel}_{\mathbb C}$
окажутся диагональными:
\[
\left(\begin{array}{cc}
          e^{-i\varphi} & 0             \\
          0             & e^{+i\varphi}
\end{array}\right)
\]
Таким образом, доказана

\mytheorem{}
 {\it Представление $T^{\perp/\parallel}_{\mathbb C}$ имеет
ровно два приводящих подпространства:
 $M^{(+1)/\parallel}_{\mathbb C}$ и
 $M^{(-1)/\parallel}_{\mathbb C}$.
Соответствующие им подпредставления дают в сумме представление
 $T^{\perp/\parallel}_{\mathbb C}$:
\[
 T^{\perp/\parallel}_{\mathbb C} =
      T^{(+1)/\parallel}_{\mathbb C} \oplus
      T^{(-1)/\parallel}_{\mathbb C} \ .
\]}

В применении к электромагнитному полю (см.~[IV]), приведённая
теорема означает существование плоских монохроматических волн со
спиральностями $+1$ и~$-1$~--- факт общеизвестный.

Заметим ещё вот что:
\begin{itemize}
\item[а.]
Поскольку в пространстве Минковского изначально не вводилось
никакой ориентации, на плоскости $M^{\perp/\parallel}_{\mathbb R}$
её тоже нет. Поэтому невозможно сказать, какое из двух пространств
обозначается $M^{(+1)/\parallel}_{\mathbb C}$, а какое
$M^{(-1)/\parallel}_{\mathbb C}$. Фиксацией этих обозначений
фактически вводится ориентация.
\item[б.]
Если группу Лоренца расширить до {\it полной\/} группы Лоренца,
включив туда пространственные отражения, то малая группа тоже
расширится. При этом группа $SO(2)$, действующая в плоскости
$M^{\perp/\parallel}_{\mathbb R}$ расширится до $O(2)$. При
отражениях комплексные пространства
 $M^{(+1)/\parallel}_{\mathbb C}$ и
 $M^{(-1)/\parallel}_{\mathbb C}$
будут переходить друг в друга, и комплексное представление в
пространстве $M^{\perp/\parallel}_{\mathbb C}$ станет уже
неприводимым.
\end{itemize}


\myparagraph{Подпредставления $T^{(+1)}_{\mathbb C}$ и
$T^{(-1)}_{\mathbb C}$.} \label{tpm}
 В соответствии с установленной в пункте~\ref{tperpparcompl}
разложимостью факторпредставления
 $T^{\perp/\parallel}_{\mathbb C}$, в комплексном пространстве
Минковского $M_{\mathbb C}$ можно выделить два двумерных
приводящих подпространства представления $T_{\mathbb C}$. Эти два
подпространства мы обозначим $M^{(+1)}_{\mathbb C}$ и
$M^{(-1)}_{\mathbb C}$. Соответствующие одномерные
факторпространства
 $M^{(+1)}_{\mathbb C}/M^\parallel_{\mathbb C} =
  M^{(+1)/\parallel}_{\mathbb C}$ и
 $M^{(-1)}_{\mathbb C}/M^\parallel_{\mathbb C} =
  M^{(-1)/\parallel}_{\mathbb C}$ были введены в
пункте~\ref{tperpparcompl}.

Теперь ясно, что в комплексном случае ряд~(\ref{admset}) допускает
уплотнение двумя способами:
\begin{equation}
 \{0\}= M^0_{\mathbb C} \subset
 M^\parallel_{\mathbb C} \subset
 M^{(+1)}_{\mathbb C} \subset
 M^\perp_{\mathbb C} \subset
 M^4_{\mathbb C} = M_{\mathbb C} \ ,
\label{admsetplus}
\end{equation}
\begin{equation}
 \{0\}= M^0_{\mathbb C} \subset
 M^\parallel_{\mathbb C} \subset
 M^{(-1)}_{\mathbb C} \subset
 M^\perp_{\mathbb C} \subset
 M^4_{\mathbb C} = M_{\mathbb C} \ ,
\label{admsetminus}
\end{equation}

В пункте~\ref{tperpcompl} будет показано, что других приводящих
подпространств у комплексного векторного представления нет. Таким
образом, кроме~(\ref{admsetplus}) и~(\ref{admsetminus}), других
композиционных рядов у представления $T_{\mathbb C}$ нет.


\myparagraph{Неразложимость представления $T_{\mathbb C}$.}
\label{tcompl}
 Предположим, что представление $T_{\mathbb C}$ разложимо, т.~е.
пространство $M_{\mathbb C}$ представляется в виде суммы
\begin{equation}
 M_{\mathbb C}=M'_{\mathbb C}\oplus M''_{\mathbb C}
\label{decompt}
\end{equation}
двух приводящих подпространств $M'_{\mathbb C}$ и
 $M''_{\mathbb C}$. Тогда одно из этих подпространств,
например~$M'_{\mathbb C}$, обязательно содержит некоторый
вектор~$n$, не содержащийся в~$M^\perp_{\mathbb C}$. Согласно
лемме из пункта~\ref{compl}, можно без ущерба для общности
считать, что $n^\mathrm{Im}\in M^\perp_{\mathbb R}$ и
$n^\mathrm{Re}\notin M^\perp_{\mathbb R}$.

Введём теперь некоторый вещественный вектор $e^1$. В случае если
вектор $n^\mathrm{Im}$ не содержится в $M^\parallel_{\mathbb R}$,
$e^1$ будет просто другим обозначением для $n^\mathrm{Im}$. Если
же $n^\mathrm{Im}$ содержится в $M^\parallel_{\mathbb R}$, то за
$e^1$ примем произвольный вектор, такой, что
 $e^1\in M^\perp_{\mathbb R}$ и
 $e^1\notin M^\parallel_{\mathbb R}$.

Рассмотрим теперь подгруппу малой группы $\EuScript L_k$,
оставляющую вектор $e^1$ неизменным. Эта подгруппа, согласно
определению вектора $e^1$, в любом случае оставляет неизменным
вектор $n^\mathrm{Im}$. Из результатов пункта~\ref{tperp} следует,
что элементы этой группы могут задаваться некоторым вектором
$e^2$, конец которого всегда лежит на некотором
цилиндре~(\ref{cylinder}), выполняется условие
 $e^1\cdot e^2=const$, и векторы $k$, $e^1$ и $e^2$ линейно
независимы. Конец вектора $e^2$ при этом пробегает некоторую
прямую, параллельную~$k$: $e^2(t)=e^2(0)+t k$.
Уравнения~(\ref{fourth}) для вектора $n^\mathrm{Re}(t)$ принимают
вид:
\[
 k\cdot n^\mathrm{Re}(t) =const        \ , \qquad
 e^1\cdot n^\mathrm{Re}(t) =const      \ , \qquad
 e^2(t)\cdot n^\mathrm{Re}(t) =const   \ , \qquad
 (n^\mathrm{Re}(t))^2 =const           \ .
\]
Третье уравнение можно расписать подробнее:
\begin{equation}
 (e^2(0)+t k)\cdot n^\mathrm{Re}(t) =const \ .
\label{thirdeq}
\end{equation}
Очевидно, если предположить, что
 $(n^\mathrm{Re}(t)-n^\mathrm{Re}(0))\in M^\parallel$, то
уравнению~(\ref{thirdeq}) при $t\neq0$ удовлетворить невозможно.
Следовательно, подпространство~$M'_{\mathbb C}$ должно содержать и
некоторый вещественный вектор, содержащийся в
 $M^\perp_{\mathbb R}$ и не содержащийся в
 $M^\parallel_{\mathbb R}$. Но тогда, согласно второй лемме
пункта~\ref{tperp}, должно
 быть~$M^\perp_{\mathbb R}\subset M'_{\mathbb C}$, и,
следовательно, $M^\perp_{\mathbb C}\subset M'_{\mathbb C}$.

Вернёмся теперь снова к комплексному вектору
 $n\in M'_{\mathbb C}$,
у которого
 $n^\mathrm{Re}\notin M^\perp_{\mathbb R}$ и
 $n^\mathrm{Im}\in M^\perp_{\mathbb R}$.
Поскольку мы уже доказали, что
 $M^\perp_{\mathbb C}\subset M'_{\mathbb C}$,
то отсюда следует, что в $M'_{\mathbb C}$ содержится и такой
вектор, мнимая часть которого равна нулю, а вещественная совпадает
с вещественной частью вектора~$n$. Иначе говоря, доказано, что
$M'_{\mathbb C}$ содержит вещественный вектор, не содержащийся в
$M^\perp_{\mathbb R}$. Но тогда, как следует из леммы
пункта~\ref{unt}, $M_{\mathbb R}\subset M'_{\mathbb C}$, и,
следовательно, пространства $M'_{\mathbb C}$ и $M_{\mathbb C}$
совпадают.

Таким образом, доказана

\mylemma{}
 {\it Всякое приводящее подпространство
представления~$T_{\mathbb C}$ либо совпадает со всем пространством
$M_{\mathbb C}$, либо содержится в $M^\perp_{\mathbb C}$.}

Из неё вытекает

\mytheorem{}
 {\it Представление~$T_{\mathbb C}$ неразложимо.}


\myparagraph{Неразложимость представлений $T^\perp_{\mathbb C}$,
$T^{(+1)}_{\mathbb C}$ и $T^{(-1)}_{\mathbb C}$.}
\label{tperpcompl}
 Докажем сначала следующую лемму:

\mylemma{}
 {\it Всякое подпредставление представления $T^\perp_{\mathbb C}$
содержит подпредставление $T^\parallel_{\mathbb C}$.}

Чтобы доказать эту лемму, обозначим рассматриваемое
подпредставление как $T'_{\mathbb C}$. Приводящее подпространство,
в котором оно действует, обозначим $M'_{\mathbb C}$.

Рассмотрим какой-нибудь ненулевой вектор~$n$ из подпространства
$M'_{\mathbb C}$. Если он лежит в~$M^\parallel_{\mathbb C}$, то
условие леммы выполнено. Предположим теперь, что этот вектор не
содержится в~$M^\parallel_{\mathbb C}$.

Если векторы $k$, $n^\mathrm{Re}$ и $n^\mathrm{Im}$ линейно
зависимы, то, согласно лемме из пункта~\ref{compl}, можно без
ущерба для общности считать, что
 $n^\mathrm{Im}\in M^\parallel_{\mathbb R}$ и
 $n^\mathrm{Re}\notin M^\parallel_{\mathbb R}$.
Далее, согласно первой лемме из пункта~\ref{tperp}, с помощью
преобразований малой группы $\EuScript L_k$ можно изменить вектор
 $n^\mathrm{Re}$ так, что к нему добавится произвольный вектор из
 $M^\parallel_{\mathbb R}$. Вектор $n^\mathrm{Im}$ при этом
останется неизменным. Следовательно,
 $M^\parallel_{\mathbb R}\subset M'_{\mathbb C}$, и,
следовательно, $M^\parallel_{\mathbb C}\subset M'_{\mathbb C}$.

Если же векторы $k$, $n^\mathrm{Re}$ и $n^\mathrm{Im}$ линейно
независимы, то, как следует из предложения пункта~\ref{tperp},
выбирая подходящим образом преобразование малой группы
 $\EuScript L_k$, можно добавить к вектору $n^\mathrm{Re}$
вектор, параллельный~$k$, и оставить при этом неизменным вектор
$n^\mathrm{Im}$. Следовательно, опять
 $M^\parallel_{\mathbb R}\subset M'_{\mathbb C}$, и
 $M^\parallel_{\mathbb C}\subset M'_{\mathbb C}$.
Таким образом, лемма доказана.

Из доказанной леммы вытекают две теоремы:

\mytheorem{}
 {\it Представление $T^\perp_{\mathbb C}$ неразложимо.}

\mytheorem{}
 {\it Представления $T^{(+1)}_{\mathbb C}$ и $T^{(-1)}_{\mathbb C}$
неразложимы.}

Кроме того, из доказанной леммы и теоремы пункта~\ref{tperppar}
следует

\mytheorem{}
 {\it У представления $T^\perp_{\mathbb C}$ кроме подпространств
 $M^\parallel_{\mathbb C}$, $M^{(+1)}_{\mathbb C}$,
 $M^{(-1)}_{\mathbb C}$ и самого $M^\perp_{\mathbb C}$ других
приводящих подпространств нет.}

Таким образом, у представления $T_{\mathbb C}$,
кроме~(\ref{admsetplus}) и~(\ref{admsetminus}), других
композиционных рядов нет.


\myparagraph{Матрицы представлений $T_{\mathbb R}$ и
 $T_{\mathbb C}$.}
Многие из полученных результатов можно наглядно изобразить с
помощью шаблонов для матриц представлений $T_{\mathbb R}$ и
$T_{\mathbb C}$. Если в пространствах $M_{\mathbb R}$ и
$M_{\mathbb C}$ выбрать подходящие базисы, то эти шаблоны примут
блочно-треугольный вид:
\[
 \left(\begin{array}{cccc}      \cline{1-1}
 \multicolumn{1}{|c|}{1} & \ \cdot & \cdot & \cdot \\ \cline{1-3}
  0  & \multicolumn{2}{|c|}{} & \cdot \\
  0  & \multicolumn{2}{|c|}{\smash{\raisebox{1.4ex}{$SO(2)$}}}
     & \cdot \\ \cline{2-4}
  0  &   \ 0   &   0   & \multicolumn{1}{|c|}{1}   \\ \cline{4-4}
 \end{array}\right)
 \qquad
 \left(\begin{array}{cccc}      \cline{1-1}
 \multicolumn{1}{|c|}{1}  & \cdot & \cdot & \cdot \\ \cline{1-2}
  0  & \multicolumn{1}{|c|}{e^{-i\varphi}} & 0 & \cdot \\ \cline{2-3}
  0  &   0   & \multicolumn{1}{|c|}{e^{+i\varphi}} & \cdot \\ \cline{3-4}
  0  &   0   &   0   & \multicolumn{1}{|c|}{1}   \\ \cline{4-4}
 \end{array}\right)
\]
При этом известно, что на месте каждой точки стоит, вообще говоря,
не ноль.

Выяснение того, в каких подпространствах и факторпространствах
действуют различные блоки этих матриц, оставляется читателю.




\newcounter{parnum}
\newlength{\picwidth}
\newcommand*{\myparagraph}[1]{\refstepcounter{parnum}%
\paragraph{\arabic{parnum}. #1}}


\title{О квантовании электромагнитного поля.\\
       VI. Квантование.}
\author{Д.~А.~Арбатский\footnote{http:\dslash d-a-arbatsky.narod.ru/}}
\date{\today}
\maketitle

\begin{abstract}
Описана общая методика квантования линейных полей. Введена
концепция квантования, инвариантного по отношению к действию
некоторой группы. Явно релятивистски-инвариантно строится
пространство квантовых состояний для релятивистских полей.
Устанавливается связь с квантованием полевого осциллятора.
Обосновывается необходимость использования индефинитного
скалярного произведения для электромагнитного поля. Обсуждается
дополнительное условие на \rql физически допустимые\rqr{}
состояния электромагнитного поля. Обсуждаются свойства
пространства состояний электромагнитного поля с точки зрения
функционального анализа. Рассмотрен вопрос о происхождении
антиунитарных преобразований в квантовой теории поля.
\end{abstract}


\myparagraph{Другие подходы к квантованию.}
 Прежде чем приступать к описанию конструкции квантованного
поля, дадим краткое сравнение нашей конструкции с другими
известными подходами к описанию квантовых полей.

\begin{itemize}

\item {\bf Метод формальных манипуляций.}
Этот метод применялся уже в самых ранних работах по квантовой
теории поля. Его суть состоит в том, чтобы вычислять скобки
Пуассона классических величин и считать, что для соответствующих
квантовых величин коммутаторы вычисляются по формуле:
\begin{equation}
 [\,\hat a\,,\,\hat b\,] = i\,\{\,a\,,\,b\,\}\widehat~  \ .
\label{CommRel}
\end{equation}
Приведённая формула, конечно, не может выполняться для всех
величин. Тем не менее, если её применять достаточно осторожно, то
можно выяснить многие свойства квантованных полей, при этом вообще
не уточняя, что они из себя представляют. Этот путь, в
действительности, оказался самым продуктивным с точки зрения
практических приложений.

Что касается того, как вычислялись скобки Пуассона, то делалось
это таким образом, что усмотреть релятивистскую инвариантность
даже в классической теории было очень трудно. Впоследствии,
правда, Пайерлс~\cite{Peierls1952} предложил альтернативный способ
установления коммутационных соотношений, который был явно
релятивистски-инвариантным. Соответствующая классическая скобка,
фигурирующая в выражении (\ref{CommRel}), была даже названа \rql
скобкой Пайерлса\rqr. Довольно сложное \rql физическое\rqr{}
обоснование соотношений (\ref{CommRel}), используемых при
квантовании по методу Пайерлса, имеется в \cite{DeWitt1987}.

Описанный подход нам наиболее близок.

Относительно классического описания релятивистских полей можно
сказать, что инвариантный гамильтонов формализм позволяет
определить скобку Пуассона полностью релятивистски инвариантно.
При таком подходе понятие скобки Пуассона видоизменяется таким
образом, что скобка Пайерлса оказывается её очень частным случаем.
Методика практического вычисления скобок Пуассона в инвариантном
гамильтоновом формализме была нами описана в статье~[I].

Что касается квантования полей, то в данной статье оно будет
произведено {\it конструктивно}: на основе развитого в~[I]
инвариантного гамильтонова формализма, используя обычные
алгебраические методы, мы построим в этой статье квантованные поля
явно.

\item {\bf Метод ящика.}
Этот метод также был популярен в теории поля со времён её
зарождения. Он состоит в том, чтобы рассматривать поле не в
бесконечном пространстве, а в большом ящике с периодическими
граничными условиями. При этом описание поля становится подобно
описанию бесконечного количества осцилляторов.

Введение ящика принципиально разрушает релятивистскую
инвариантность, и уже этой причины достаточно, чтобы не
рассматривать этот метод здесь в подробностях.

Однако, в результате применения этого метода возникло полезное
понятие полевого осциллятора. В этом плане следует заметить, что в
нашем методе имеется аналогичное понятие (введённое в
статье~[IV]). У нас оно вводится несколько более абстрактно, с
помощью понятия индуцированного симплектического представления.
Только таким путём возможно установить правильную
теоретико-групповую природу полевого осциллятора и связать
инвариантные квантования поля с инвариантными квантованиями
осциллятора.

\item {\bf Конструкция Фока и теория Вигнера-Макки.}
В работе~\cite{Fock1932} Фок указал, что пространство состояний
квантованного поля устроено как симметризованная тензорная
экспонента одночастичного подпространства. Впоследствии Вигнером и
Макки был разработан метод поиска всевозможных одночастичных
подпространств, основанный на теории индуцированных унитарных
представлений группы Пуанкаре (см., например,
обзор~\cite{Mackey1968}). Весьма подробное современное изложение
этого подхода имеется в книге~\cite{Weinberg1996}.

Недостатки этого подхода состоят в следующем. Во-первых, операторы
рождения и уничтожения квантов с самого начала вводятся с помощью
довольно сложных формул. Во-вторых, даже когда введены такие
операторы, требуются весьма запутанные рассуждения, чтобы
построить из них локальные операторы {\it поля}. В-третьих, при
таком подходе вообще выпадают из рассмотрения поля с индефинитным
скалярным произведением.

В нашем подходе исходным объектом выступает не одночастичное
квантовое подпространство, а классическое поле, описанное на языке
инвариантного гамильтонова формализма. При этом второй недостаток
автоматически исчезает, т.~к. появление полевых операторов у нас
происходит сразу же при квантовании.

Операторы рождения и уничтожения квантов у нас также могут быть
введены. При этом формулы для них вытекают из самой алгебраической
конструкции квантованного поля, а не постулируются изначально.

Что касается квантованных полей с индефинитным скалярным
произведением, то они у нас возникают столь же естественно, как и
поля с положительно-определённым скалярным произведением. Наша
схема действительно является более общей, как в этом можно
убедиться уже на примере массивного скалярного поля. Благодаря
этой общности, схема совершенно естественно включает в себя
электромагнитное поле как частный случай.

Отметим также, что в нашем подходе имеется также и аналог теории
Вигнера-Макки. Это теория индуцированных симплектических
представлений группы Пуанкаре~[IV]. Теория индуцированных
симплектических представлений является более богатой, чем теория
индуцированных унитарных представлений.

\item {\bf Подход в духе теоремы Стоуна-фон~Неймана.}
В квантовой механике при изучении квантового осциллятора
оказывается полезной теорема Стоуна-фон~Неймана, утверждающая, что
канонические коммутационные соотношения для координат и импульсов
однозначно определяют соответствующее неприводимое унитарное
представление. Естественно, этот подход пытались перенести на
квантовую теорию поля.

Но теорема Стоуна-фон~Неймана на случай квантования систем с
бесконечным числом степеней свободы непосредственно не переносится
(см., например, \cite{Schwarz1975}). Были проведены большие
исследования с целью преодолеть эту трудность. Несмотря на их
несомненную ценность, думается всё же, что они оказались слишком
оторванными от практических приложений.

Здесь стоит также отметить, что если отказаться от положительной
определённости скалярного произведения в пространстве квантовых
состояний, то единственность квантования нарушается уже в случае
одномерного гармонического осциллятора. Поэтому наш путь
отличается от описанного принципиальным образом: мы отказываемся
от требования положительной определённости скалярного
произведения, но взамен вводим концепцию инвариантного
квантования.

Отметим здесь также такую деталь. При формулировке теоремы
Стоуна-фон~Неймана коммутационные соотношения необходимо
записывать в форме соотношений Вейля. В нашем же подходе
фигурируют коммутационные соотношения для самих неограниченных
операторов. При этом сама конструкция квантования строго
определяет, в каком смысле их следует понимать.

\item {\bf Геометрическое квантование.}
Метод геометрического квантования состоит в том, чтобы, опираясь
на геометрию классического фазового пространства, построить
пространство квантовых состояний как некоторое множество функций
на классическом фазовом пространстве (см., например,
\cite{Kirillov1985}).

С точки зрения инвариантного гамильтонова формализма эту схему
естественно применять к инвариантному фазовому пространству $Z$.
Заметим, однако, что у этого подхода сразу же видны значительные
трудности. Во-первых, сделать этот подход математически строгим
для случая систем с бесконечным числом степеней свободы совсем не
просто. Во-вторых, связь с конструкцией Фока в этом случае
осуществляется с помощью не слишком очевидных формул. По этим
причинам такой подход оказывается не очень удобен для практических
приложений. В-третьих, не ясно, каким образом с помощью этого
подхода можно описать случай квантования с индефинитным скалярным
произведением: уже в случае одномерного гармонического осциллятора
возникают большие трудности с определением индефинитного
скалярного произведения.

\item {\bf Метод континуального интеграла.}
Этот метод стал в последнее время очень популярен в связи с тем,
что он будто бы является явно релятивистски-инвариантным. При этом
с континуальным интегралом работают фактически как с символической
формой записи некоторых выражений.

Можно, однако, рассматривать континуальный интеграл как строго
определённый математический объект (см., например,
\cite{Berez1980}). Такой подход наталкивается на такие
существенные трудности, что ни о какой \rql очевидной\rqr{}
релятивистской инвариантности не может быть и речи.

\end{itemize}


\myparagraph{Квантование.} \label{quantization}
 Опишем теперь, как, имея классическое поле, следует строить
поле квантовое.

\begin{enumerate}

\item
Рассмотрим множество (комплексных) функций, постоянных на всём
пространстве $Z$. Обозначим это множество как $\mathcal C$.
Рассмотрим далее прямую сумму $\mathcal C\oplus Z^*_{\mathbb C}$.
Скобка Пуассона превращает эту прямую сумму в комплексную алгебру
Ли.

\item
Отвлекаясь пока от алгебраической структуры множества
 $\mathcal C\oplus Z^*_{\mathbb C}$ и рассматривая все его
элементы как совершенно независимые, будем называть это множество
{\it алфавитом} $\mathcal A$, а его элементы {\it буквами}.
Договоримся также, что элемент $\mathcal C\oplus Z^*_{\mathbb C}$,
рассматриваемый как буква алфавита $\mathcal A$, будет
дополнительно снабжаться \rql шляпкой\rqr{} сверху или сбоку,
например: $\hat a = a\widehat{~}$.

\item
Можно далее ввести формальное умножение элементов алфавита. А
именно, будем считать, что при перемножении (конечного) набора
букв образуется {\it слово}, состоящее из этих букв (порядок
следования букв является существенным):
\[
 \hat a_1 \cdot \hat a_2 \cdot \dots \cdot \hat a_k =
   \hat a_1 \hat a_2 \dots \hat a_k \ ,
 \qquad
 \hat a_1,\hat a_2, \dots,\hat a_k \in \mathcal A \ .
\]
Слова мы будем также обозначать символами со шляпкой, например:
$\hat w = w\widehat{~}$. Множество всех возможных слов обозначим
символом $\mathcal W$; при этом целесообразно включить сюда и
слово нулевой длины, которое можно обозначить как $()\widehat~$.
Естественно считать, что и на множестве слов $\mathcal W$
действует операция умножения, которая \rql склеивает\rqr{} слова.
Таким образом, $\mathcal A$ является {\it системой свободных
образующих}, а $\mathcal W$ является {\it свободной полугруппой с
единицей} (роль единицы в ней играет специально привнесённый
элемент $()\widehat~$).

\item
Подобным же образом определим теперь формальную сумму (конечного)
набора слов, домноженных на произвольные (комплексные)
коэффициенты. А именно, назовём {\it фразой} формальную запись
вида:
\[
 \lambda_1 \hat w_1 + \lambda_2 \hat w_2 +
    \dots + \lambda_m \hat w_m \ ,
 \qquad
 \lambda_1,\lambda_2,\dots,\lambda_m\in\mathbb C \ , \ \
 \hat w_1,\hat w_2,\dots,\hat w_m\in\mathcal W \ .
\]
Будем, кроме того, считать две фразы эквивалентными, если для
любого слова у них совпадает сумма коэффициентов при этом слове.
Множество всех введённых таким образом классов эквивалентности
обозначим символом $\mathcal P$.

На множестве фраз $\mathcal P$ можно теперь естественным образом
ввести операции сложения и умножения на скаляр. А именно, суммой
двух фраз называется фраза, получающаяся соединением исходных фраз
знаком \rql$+$\rqr. Умножение фразы на число определяется как
умножение каждого из коэффициентов на это число. Тем самым
$\mathcal P$ наделяется естественной структурой комплексного
линейного пространства ($\mathcal P$ является {\it свободным
$\mathbb C$-модулем, порождённым $\mathcal W$}).

На множестве фраз $\mathcal P$ можно также ввести естественным
образом операцию умножения двух элементов. Именно, произведение
двух фраз, содержащих по одному слову, определим равенством:
\[
 (\lambda_1 \hat w_1)\cdot(\lambda_2 \hat w_2) =
 (\lambda_1\lambda_2)\,(\hat w_1 \hat w_2)  \ ,
 \qquad
 \lambda_1,\lambda_2\in\mathbb C \ , \ \
 \hat w_1,\hat w_2\in\mathcal W \ .
\]
На произвольные фразы распространим это определение, исходя из
требования дистрибутивности:
\[
 \hat p_1\cdot(\hat p_2+\hat p_3) =
     \hat p_1 \hat p_2 + \hat p_1 \hat p_3 \ ,
 \qquad
 (\hat p_1 + \hat p_2)\cdot \hat p_3 =
     \hat p_1 \hat p_3 + \hat p_2 \hat p_3 \ ,
 \qquad
 \hat p_1,\hat p_2,\hat p_3\in\mathcal P \ .
\]

\item
Таким образом, множество фраз $\mathcal P$ превращается в
(ассоциативную) алгебру ($\mathcal P$ является {\it полугрупповой
алгеброй полугруппы $\mathcal W$} или {\it свободной алгеброй над
полугруппой $\mathcal W$}).

\item
До сих пор при построении алгебры $\mathcal P$ алфавит
 $\mathcal A$ рассматривался как совершенно произвольное
множество. Теперь, используя структуру алгебры Ли в
 $\mathcal C\oplus Z^*_{\mathbb C}$ и соответствие между
подпространством $\mathcal C$ и полем скаляров $\mathbb C$, можно
дополнительно внести в алгебру $\mathcal P$ некоторые {\it
определяющие соотношения}. Это делается с помощью факторизации
алгебры $\mathcal P$ по подходящему идеалу.

Именно, рассмотрим все фразы следующих видов:
\[
 (\lambda\, a)\widehat~ - \lambda\,\hat a \ ,
\]
\[
 (a+b)\widehat~ - (\hat a + \hat b) \ ,
\]
\[
 \{\,a\,,\,b\,\}\widehat~ + i\,(\hat a\,\hat b - \hat b\,\hat a) \ ,
\]
\[
 \hat1 - ()\widehat~ \ .
\]
 Здесь $\lambda\in\mathbb C$,
 $a,b\in\mathcal C\oplus Z^*_{\mathbb C}$,
 $\hat a,\hat b\in\mathcal A$.
 $1$~--- обозначение для функции, равной $1$ на всём инвариантном
фазовом пространстве~$Z$, т.~e.
 $1\in\mathcal C\oplus Z^*_{\mathbb C}$; $\hat1$~---
соответствующий ей элемент алфавита, $\hat1\in\mathcal A$.

Натянем на указанные фразы двусторонний идеал, т.~е. присоединим к
ним те фразы, которые получаются из указанных путём применения
конечного числа операций умножения на скаляр, сложения и умножения
(слева и справа) на произвольные фразы. Факторизуя алгебру
$\mathcal P$ по построенному идеалу получаем алгебру, которую мы
будем называть {\it алгеброй операторов} и обозначать $\mathcal
O$. Элементы этой алгебры называются {\it операторами} и
обозначаются так же, как и соответствующие им фразы.

Проведённая факторизация приводит к тому, что в алгебре операторов
выполняются определяющие соотношения:
\[
 (\lambda\, a)\widehat~ = \lambda\,\hat a \ ,
\]
\[
 (a+b)\widehat~ = \hat a + \hat b \ ,
\]
\[
 \{\,a\,,\,b\,\}\widehat~ = -\,i\,[\,\hat a\,,\,\hat b\,] \ ,
\]
\[
 \hat1 = ()\widehat~ \ .
\]
В третьем соотношении использовано стандартное обозначение для
{\it коммутатора} двух операторов:
 $[\,\hat a\,,\,\hat b\,]=\hat a\,\hat b - \hat b\,\hat a$.

Четвёртое соотношение, в частности, позволяет отказаться от
громоздкого обозначения $()\widehat~$, когда речь идёт об
операторе, и писать всегда $\hat1$.

\item
Введём теперь операцию {\it сопряжения}. Эта операция будет всегда
обозначаться символом ${}^*$, независимо от того, к какого рода
объекту она применяется.

Под сопряжением комплексного числа будем понимать обычное
комплексное сопряжение.

Элементы пространства $\mathcal C\oplus Z^*_{\mathbb C}$ являются
функциями на $Z$. Под сопряжением, применённым к такой функции,
будем понимать переход к функции с комплексно-сопряжёнными
значениями. Очевидно, в результате также получается элемент
пространства $\mathcal C\oplus Z^*_{\mathbb C}$. Более того,
поскольку скобка Пуассона в $\mathcal C\oplus Z^*_{\mathbb C}$
является комплексификацией вещественной скобки Пуассона,
выполняется равенство:
\begin{equation}
 \{\,a^*\,,\,b^*\,\}=\{\,a\,,\,b\,\}^* \ .
\label{poissconj}
\end{equation}

Определение сопряжения естественно переносится на элементы
алфавита $\mathcal A$:
\[
 (\hat a)^*=(a^*)\widehat~ \ .
\]

На слова и на фразы операция сопряжения распространяется по
правилам:
\[
 (\hat a\,\hat b)^* = \hat b^*\,\hat a^*    \ ,
 \qquad
 (\lambda\,\hat a)^* = \lambda^*\,\hat a^*  \ ,
 \qquad
 (\hat a + \hat b)^* = \hat a^* + \hat b^*  \ .
\]
Здесь $\hat a$ и $\hat b$ могут быть буквами, словами и фразами;
$\lambda$~--- комплексное число.

Очевидно, операция сопряжения корректно переносится с фраз на
операторы.

\item
Рассмотрим теперь пространство $Z^*_{\mathbb C}$. Разобьём его в
прямую сумму двух подпространств:
\[
 Z^*_{\mathbb C} = Cr\oplus De \ .
\]
От подпространств $Cr$ и $De$ потребуем, во-первых, чтобы они
переходили друг в друга при сопряжении:
\begin{equation}
 Cr^*=De \ .
\label{cranconj}
\end{equation}
Во-вторых, потребуем, чтобы скобка Пуассона на этих
подпространствах занулялась\footnote{С учётом~(\ref{poissconj}) и
(\ref{cranconj}), достаточно потребовать зануления только на одном
из этих двух подпространств.}:
\[
 \{\,a\,,\,b\,\} = 0 \ , \qquad  a,b\in Cr \ ;
\]
\[
 \{\,a\,,\,b\,\} = 0 \ , \qquad  a,b\in De \ .
\]

Подпространство $Cr$ будем называть {\it рождающим}, а $De$~---
{\it уничтожающим}. В соответствии с приведённой выше
конструкцией, подпространствам $Cr$ и $De$ соответствуют некоторые
подпространства в алгебре операторов; обозначим эти
подпространства $Cr\widehat~$ и $De\widehat~$, соответственно.

Введём также обозначение $\mathcal C\widehat~$ для подпространства
алгебры операторов, отвечающего подпространству $\mathcal C$
алгебры Ли $\mathcal C\oplus Z^*_{\mathbb C}$.

\item
Натянем теперь на подпространство $De\widehat~$ левый идеал, т.~е.
присоединим к имеющимся там операторам те операторы, которые
получаются из указанных в результате конечного числа операций
умножения на скаляр, сложения и домножения слева на произвольные
элементы алгебры $\mathcal O$.

Факторизуем теперь алгебру $\mathcal O$ по построенному левому
идеалу. В результате получится некоторое линейное пространство.
Оно не является алгеброй, но является левым модулем, т.~е. на нём
естественно определено действие операторов слева. Элементы этого
факторпространства называются {\it кет-векторами}. Кет-векторы
обозначаются символами вида $|\,x\,\rangle$, где $x$~--- любой
значок, отличающий данный кет-вектор. Само пространство
кет-векторов обозначается $\mathcal H$ и называется {\it
пространством состояний} квантованного поля.

\item
Можно провести те же самые построения с сопряжёнными объектами,
т.~е. вместо подпространства $De\widehat~$ рассмотреть
подпространство $Cr\widehat~$, натянуть на него правый идеал
вместо левого, и, факторизуя, получить правый модуль вместо
левого. Элементы этого модуля называются {\it бра-векторами}. Сам
же модуль будем, если это не вызывает недоразумений, также
называть {\it пространством состояний} квантованного поля и
обозначать $\mathcal H$.

Бра-векторы обозначаются символами вида $\langle\,x\,|$. Операция
сопряжения естественным образом переносится на кет- и бра-векторы
и устанавливает между ними взаимно-однозначное соответствие.
Говорят, что соответствующие друг другу кет- и бра-вектор
относятся к одному и тому же состоянию квантованного поля.

\item
При факторизации алгебры операторов $\mathcal O$ и переходе к
пространствам кет- и бра-векторов, оператор $\hat1$ переходит в
некоторый кет-вектор $|\,0\,\rangle$ и бра-вектор $\langle\,0\,|$,
соответственно. Эти кет- и бра-векторы описывают состояние,
называемое {\it вакуумом}.

\item
Введём теперь так называемое {\it скалярное произведение} бра- и
кет- векторов. Для произвольных бра-вектора $\langle\,x\,|$ и
кет-вектора $|\,y\,\rangle$ это произведение записывается как
$\langle\,x\,|\cdot|\,y\,\rangle$, или, для краткости,
$\langle\,x\,|\,y\,\rangle$. В результате образуется комплексное
число, т.~е. $\langle\,x\,|\,y\,\rangle \in \mathbb C$.

Потребуем, чтобы операция скалярного произведения обладала
следующими свойствами:
\[
 \langle\,0\,|\,0\,\rangle = 1 \ ,
\]
\[
 \big(\,\lambda\,\langle\,x\,|\,\big)\cdot|\,y\,\rangle =
               \lambda\cdot\langle\,x\,|\,y\,\rangle   \ ,
 \qquad
 \big(\,\langle\,x_1\,|+\langle\,x_2\,|\,\big)\cdot|\,y\,\rangle =
  \langle\,x_1\,|\,y\,\rangle+\langle\,x_2\,|\,y\,\rangle   \ ,
\]
\[
 \langle\,x\,|\cdot\big(\,\lambda\,|\,y\,\rangle\,\big) =
               \lambda\cdot\langle\,x\,|\,y\,\rangle   \ ,
 \qquad
 \langle\,x\,|\cdot\big(\,|\,y_1\,\rangle+|\,y_2\,\rangle\,\big) =
 \langle\,x\,|\,y_1\,\rangle+\langle\,x\,|\,y_2\,\rangle  \ ,
\]
\[
 \big(\,\langle\,x\,|\,\hat o\,\big)\cdot|\,y\,\rangle =
 \langle\,x\,|\,\cdot\big(\,\hat o\,|\,y\,\rangle\,\big)   \ .
\]
В последнем равенстве $\hat o\in\mathcal O$; это свойство
позволяет писать в таких случаях просто:
 $\langle\,x\,|\,\hat o\,|\,y\,\rangle$.

Указанные свойства для скалярного произведения являются
определяющими, т.~е. существует не более одного скалярного
произведения, обладающего указанными свойствами.

\item
Скалярное произведение является эрмитовой формой в $\mathcal H$,
т.~е. всегда
$\langle\,x\,|\,y\,\rangle=(\langle\,y\,|\,x\,\rangle)^*$. Если
скалярное произведение окажется к тому же
положительно-определённым, то оно задаёт в $\mathcal H$ некоторую
топологию. Если $\mathcal H$ относительно этой топологии
пополнить, то оно превращается в обычное гильбертово пространство.
Операторы при этом оказываются заданными на соответствующем
плотном линейном многообразии в $\mathcal H$.

Однако, скалярное произведение может вовсе и не быть
положительно-определённым: это зависит от скобки Пуассона в
$\mathcal C\oplus Z^*_{\mathbb C}$ и выбора подпространств $Cr$ и
$De$. Вопрос об определении топологии в этом случае на примере
электромагнитного поля будет обсуждаться в пункте~\ref{TFIPSEM}.

\end{enumerate}

Итак, в этом пункте было описано, как по алгебре Ли
 $\mathcal C\oplus Z^*_{\mathbb C}$ наблюдаемых классического
поля строятся алгебра $\mathcal O$ операторов квантового поля и
пространство состояний $\mathcal H$. Это построение называется
{\it квантованием} классического поля.


\myparagraph{Градуировка алгебры $\mathcal O$ и пространства
$\mathcal H$. Связь с конструкцией Фока.}
 Рассмотрим теперь в алгебре операторов подпространства
 $\mathcal C\widehat~$, $Cr\widehat~$ и $De\widehat~$. Будем
говорить, что они являются подпространствами {\it степени} $0$,
$+1$ и $-1$, соответственно. Далее, образуя всевозможные
произведения этих операторов, присвоим каждому такому произведению
степень, равную сумме степеней сомножителей. Если два произведения
имеют одну и ту же степень, то припишем ту же степень их сумме.
Нетрудно видеть, что таким образом алгебра операторов,
рассматриваемая как линейное пространство, может быть представлена
в виде суммы своих линейных подпространств, отвечающих разным
степеням:
\[
 \mathcal O = \dots \oplus
   \mathcal O_{-2} \oplus
   \mathcal O_{-1} \oplus
   \mathcal O_0    \oplus
   \mathcal O_{+1} \oplus
   \mathcal O_{+2} \oplus \dots
\]
При этом, если $\hat a\in\mathcal O_i$ и $\hat b\in\mathcal O_j$,
то $\hat a\,\hat b\in\mathcal O_{i+j}$. Короче говоря, алгебра
$\mathcal O$ является градуированной.

Эта градуировка естественным образом переносится на пространство
состояний $\mathcal H$. При этом отрицательным степеням в
пространстве $\mathcal H$ отвечают тривиальные подпространства.
Иначе говоря, пространство состояний $\mathcal H$ является
положительно-градуированным:
\begin{equation}
 \mathcal H =
   \mathcal H_0 \oplus
   \mathcal H_1 \oplus
   \mathcal H_2 \oplus \dots
\label{gradH}
\end{equation}

Если кет-вектор принадлежит одному из пространств $\mathcal H_n$,
то говорят, что в данном состоянии имеется $n$ частиц.

Очевидно также, что всё проведённое построение имеет место и для
сопряжённых объектов, и разложение~(\ref{gradH}) точно так же
выглядит и для бра-векторов.

Нелишне также отметить, что согласно введённому определению, число
частиц всегда неотрицательно, независимо от того, является ли
скалярное произведение в $\mathcal H$ положительно-определённым.

Далее, легко видеть, что подпространства $\mathcal H_i$ и
$\mathcal H_j$ ортогональны относительно скалярного произведения.
Иначе говоря, если $\langle\,a\,|\in\mathcal H_i$ и
$|\,b\,\rangle\in\mathcal H_j$ и $i\neq j$, то
$\langle\,a\,|\,b\,\rangle=0$. Имея это в виду,
разложение~(\ref{gradH}) можно писать в виде:
\begin{equation}
 \mathcal H =
   \mathcal H_0 \dot+
   \mathcal H_1 \dot+
   \mathcal H_2 \dot+ \dots
\label{OrthSum}
\end{equation}
В случае, когда скалярное произведение является
положительно-опре\-де\-лён\-ным, разложение~(\ref{OrthSum})
позволяет установить связь с обычной конструкцией
Фока~\cite{Fock1932}. Поскольку это довольно просто, мы не станем
прослеживать эту связь здесь более подробно.


\myparagraph{Инвариантное квантование.} \label{invquant}
 В пункте~\ref{quantization} квантование классического поля
было описано во \rql вну\-трен\-них\rqr{} терминах инвариантного
гамильтонова формализма. Однако, для выбора подпространств $Cr$ и
$De$ не было предложено никакого конкретного рецепта: были лишь
указаны некоторые необходимые условия выбора этих подпространств.
На практике это приводит к тому, что у одного и того же
классического поля квантований может оказаться слишком много.

Пусть теперь в пространстве $Z^*_{\mathbb C}$ определено линейное
симплектическое представление некоторой группы. Определим тогда
{\it инвариантное квантование} требованием, чтобы подпространства
$Cr$ и $De$ были инвариантными по отношению к действию данной
группы. Иными словами, $Cr$ и $De$~--- приводящие подпространства
указанного представления.

Когда речь идёт о релятивистских полях, в качестве основной группы
инвариантности выступает группа Пуанкаре~$\mathcal P$.
Квантование, инвариантное по отношению к группе Пуанкаре, будем
для краткости называть $\mathcal P$-инвариантным. В статье~[IV] мы
уже видели, каким образом группа Пуанкаре действует в пространстве
$Z^*_{\mathbb C}$, и как приводятся её представления.

Квантование релятивистских полей мы рассмотрим чуть позже, а
сейчас обсудим квантование гармонического осциллятора. Его группа
симметрии~--- это просто однопараметрическая группа временных
сдвигов.


\myparagraph{Квантование гармонического осциллятора.}
\label{harmosc}
 Рассмотрим гармонический осциллятор. Он описывается лагранжианом:
\begin{equation}
 \textstyle
 L = \frac12 \dot\varphi^2 - \frac{m^2}2\varphi^2    \ .
\label{osclag}
\end{equation}
Здесь $\varphi(x)$~--- вещественная функция одного вещественного
аргумента (времени).

Уравнение движения имеет вид:
\[
 (\partial^2 + m^2)\,\varphi = 0   \ .
\]

Инвариантное фазовое пространство $Z$~--- двумерное вещественное
пространство. Симплектическая структура задаётся на нём формулой:
\[
 \omega = \dot\varphi(t) \wedge \varphi(t)    \ .
\]
Здесь формы $\dot\varphi$ и $\varphi$ берутся в один и тот же
произвольный момент времени $t$.

Поскольку лагранжиан~(\ref{osclag}) инвариантен относительно
временных сдвигов, в пространстве $Z$ действует представление
аддитивной группы $\mathbb R$. Полевое представление
 $Z^*_{\mathbb C}$ в данном случае, так же как и у релятивистских
полей, распадается в прямую сумму положительно- и
отрицательно-частотного:
\[
 Z^*_{\mathbb C} = Z^{*\,(+)}_{\mathbb C} \oplus
                   Z^{*\,(-)}_{\mathbb C} \ .
\]
В качестве приводящего базиса возьмём следующие два элемента:
\[
 \textstyle
 a = \frac1{\sqrt{2m}}\,(m\varphi(0) + i\dot\varphi(0)) \ ,
 \qquad
 a^* = \frac1{\sqrt{2m}}\,(m\varphi(0) - i\dot\varphi(0)) \ .
\]
Действительно, поскольку имеет место разложение
 $\varphi(t) =
  \frac1{\sqrt{2m}}(a\,e^{-imt} + a^*e^{+imt})$,
получаем, что $a\in Z^{*\,(+)}_{\mathbb C}$,
 $a^*\in Z^{*\,(-)}_{\mathbb C}$.

Симплектическая структура через формы $a$ и $a^*$ может быть
записана следующим образом:
\[
 \omega = i\,a^*\wedge a  \ .
\]
Скобка Пуассона соответствующих линейных функций равна:
\[
 \{\,a\,,\,a^*\,\} = -i \ .
\]

Таким образом, в соответствии со сказанным в
пункте~\ref{invquant}, имеются ровно два инвариантных квантования:
либо $Cr=Z^{*\,(-)}_{\mathbb C}$ и $De=Z^{*\,(+)}_{\mathbb C}$,
либо $Cr=Z^{*\,(+)}_{\mathbb C}$ и $De=Z^{*\,(-)}_{\mathbb C}$.
Легко видеть, что первое квантование приводит к
положительно-определённому скалярному произведению. При втором
квантовании скалярное произведение оказывается индефинитным.
Причём в разложении~(\ref{OrthSum}) на подпространствах с чётным
числом частиц скалярное произведение определено положительно, а на
подпространствах с нечётным~--- отрицательно.


\myparagraph{Связь инвариантных квантований поля и полевого
осциллятора.} \label{QuantFOsc}
 Полевой осциллятор, как
конечномерную систему, удобнее исследовать алгебраическими
средствами. С другой стороны, между квантованием поля и
квантованием осциллятора имеется тесная связь.

В самом деле, легко видеть, что выбор
 $\mathbb R\times\mathcal L_{k^{(0)}}$-инвариантных рождающих
и уничтожающих подпространств у осциллятора и выбор
 $\mathcal P$-инвариантных рождающих и уничтожающих подпространств
у поля связаны операцией индуцирования. Таким образом, имеется
взаимно-однозначное соответствие между
 $\mathbb R\times\mathcal L_{k^{(0)}}$-инвариантными квантованиями
осциллятора и $\mathcal P$-инвариантными квантованиями поля.

Более того, у соответственных квантований осциллятора и поля
скалярные произведения в пространствах состояний либо одновременно
по\-ло\-жи\-тель\-но-определённы, либо одновременно индефинитны.
Таким образом, возникает удобный критерий знакоопределённости
скалярного произведения.


\myparagraph{Квантование скалярного поля.}
 В статье [IV] было показано, что по отношению к действию
группы Пуанкаре пространство линейных наблюдаемых скалярного поля
$Z^*_{\mathbb C}$ распадается в прямую сумму двух неприводимых
подпространств:
 $Z^*_{\mathbb C}=Z^{*\,(+)}_{\mathbb C}\oplus Z^{*\,(-)}_{\mathbb C}$.
Таким образом, имеется ровно две возможности: либо положить
$Cr=Z^{*\,(-)}_{\mathbb C}$ и $De=Z^{*\,(+)}_{\mathbb C}$, либо
$Cr=Z^{*\,(+)}_{\mathbb C}$ и $De=Z^{*\,(-)}_{\mathbb C}$.
Согласно вычисленным в статье [IV] формулам для скобок Пуассона
соответствующих функций, оба разбиения удовлетворяет всем
требованиям пункта~\ref{quantization}.

Таким образом, скалярное поле допускает ровно два
 $\mathcal P$-инвариантных квантования. Полевой осциллятор в
данном случае~--- обычный одномерный вещественный осциллятор.
Используя результаты пункта~\ref{harmosc} и критерий из
пункта~\ref{QuantFOsc}, получаем, что при первом квантовании
скалярное произведение в пространстве $\mathcal H$ оказывается
по\-ло\-жи\-тель\-но-определённым, а при втором~- индефинитным.

Первое квантование фактически является общепринятым. Не следует
думать, однако, что квантование с индефинитной метрикой заведомо
бессмысленно, т.~к. оно приводит к \rql отрицательным
вероятностям\rqr. В принципе, нельзя исключать полезности такой
теории, в которой состояния рассеяния данного поля будут
составлять подпространство, на котором скалярное произведение
будет положительно-определённым.


\myparagraph{Квантование электромагнитного поля.}
 Рассмотрим теперь квантование электромагнитного поля. Здесь под
электромагнитным полем будет пониматься \rql нефизическое\rqr{}
электромагнитное поле [I].

Как было разъяснено в статье [IV], по отношению к действию группы
Пуанкаре $\mathcal P$ пространство $Z^*_{\mathbb C}$ в этом
случае, как и в случае скалярного поля, распадается в прямую сумму
двух неразложимых подпространств. Поэтому, как и в случае
скалярного поля, есть ровно две возможности: либо
$Cr=Z^{*\,(-)}_{\mathbb C}$ и $De=Z^{*\,(+)}_{\mathbb C}$, либо
$Cr=Z^{*\,(+)}_{\mathbb C}$ и $De=Z^{*\,(-)}_{\mathbb C}$.

Используя критерий из пункта~\ref{QuantFOsc}, легко убеждаемся,
что оба квантования приводят к индефинитной метрике. Далее мы
будем рассматривать только первое из двух указанных квантований,
поскольку именно оно имеет полезную физическую интерпретацию.

Индефинитность скалярного произведения, разумеется, порождает
трудности с вероятностной интерпретацией.

В статье [I] было указано, что состояния рассеяния классического
электромагнитного поля удовлетворяют дополнительному условию~---
условию Лоренца. Естественно предположить, что в квантовой теории
имеется аналог этого условия. При этом хочется особо подчеркнуть,
что это условие должно появляться в квантовой теории не как новый
постулат: оно должно выводиться из динамических законов так же,
как это мы проделали для классического поля. К сожалению,
удовлетворительной теории взаимодействующих полей у нас пока нет.
По этой причине никакого вывода здесь дано не
будет\footnote{Формальная выкладка, конечно, имелась ещё
в~\cite{Bleuler1950}. Кроме того, можно было бы апеллировать к
правилам Фейнмана, но эти правила сами нуждаются в обосновании с
позиций нашей статьи.}. Искомое условие имеет вид:
\begin{equation}
 -\,i\,k_\mu\,\hat a_\mu^{(+)}(k)\,|\,\mathrm{rad}\,\rangle = 0 \ .
\label{Guptacond}
\end{equation}
То есть вектор состояния излучённого поля удовлетворяет указанному
равенству для любого~$k$.

Приведённое условие выглядит точно так же, как и один из вариантов
этого условия в классической теории. Следует, однако, отметить,
что его нельзя записать как
 $-\,i\,k_\mu\,\hat a_\mu^{(-)}(k)\,|\,\mathrm{rad}\,\rangle = 0$ или
как $-\,i\,k_\mu\,\hat a_\mu(k)\,|\,\mathrm{rad}\,\rangle = 0$,
ибо даже вакуум таким условиям не удовлетворяет\footnote{В
координатном представлении условие
 $-\,i\,k_\mu\,\hat a_\mu(k)\,|\,\mathrm{rad}\,\rangle = 0$
записывается как
 $\partial_\mu\hat A_\mu(x)\,|\,\mathrm{rad}\,\rangle = 0$.
В литературе и по сей день нет единства мнений по поводу того,
можно ли дополнительное условие таким образом формулировать. В
таком виде, исходя из аналогии с классической теорией, его
формулировал ещё Ферми. В работах \cite{Ma1949,Belinf1949} было
указано, что такое дополнительное условие ведёт к трудностям с
нормируемостью состояний в квантовом случае. И в работе
\cite{Gupta1950} оно было заменено на
 $\partial_\mu\,\hat A_\mu^{(+)}(x)\,|\,\mathrm{rad}\,\rangle = 0$.
В дальнейшем в литературе одни авторы использовали условие
 $\partial_\mu\,\hat A_\mu(x)\,|\,\mathrm{rad}\,\rangle = 0$
\cite{Thirr1964,Umezawa1958,Dirac1979}. Другие \cite{BogShir1984}
указывали, что такое условие на векторы состояния накладывать
нельзя, т.~к. при этом возникает противоречие с коммутационными
соотношениями. Третьи \cite{Prokh1988} утверждали, что такое
условие можно использовать, но при этом следует проявлять
осторожность, поскольку допустимые векторы состояния оказываются
ненормируемыми. Эти расхождения во мнениях имеют под собой
довольно глубокие причины:

\begin{enumerate}
\item В термин \rql квантование\rqr{} обычно не вкладывался
никакой точный математический смысл. Фактически, квантованное
электромагнитное поле исследовалось путём формальных манипуляций с
алгебраическими символами, но при этом не давалось никакого
конструктивного определения этого объекта. В такой ситуации
дополнительное условие в любом его виде не свободно от критики.
\item Отсутствует удовлетворительная формулировка теории
взаимодействующих полей (хотя бы в рамках теории возмущений).
Правила Фейнмана при этом не имеют достаточно ясной связи с
операторным формализмом, и операторный формализм оказывается
оторванным от практических вычислений (прежде всего от теории
рассеяния).
\end{enumerate}

Что касается первого замечания, то я полагаю, соотношение
 $\partial_\mu\,\hat A_\mu(x)\,|\,\mathrm{rad}\,\rangle = 0$
напрямую не противоречит коммутационным соотношениям. Можно лишь
сказать, что конструкция квантования, изложенная в этой статье, с
таким условием плохо увязывается.

Второе замечание, несомненно, является более существенным. При
рассмотрении динамики классического электромагнитного поля мы
указали, что дополнительное условие на состояния рассеяния
автоматически вытекает из динамики, а вовсе не является
произвольным постулатом. Также была ясно показана роль \rql
нефизических\rqr{} степеней свободы. И хотя мы в данной статье так
и не формулируем теории квантованных взаимодействующих полей, всё
же соображения, основанные на аналогии с классической теорией,
дают основания думать, что построение теории рассеяния с условием
 $\partial_\mu\,A_\mu(x)\,|\,\mathrm{rad}\,\rangle = 0$~--- дело
малореалистическое.}.

Сейчас мы покажем, что условие~(\ref{Guptacond}) обеспечивает
положительность скалярного произведения.


\myparagraph{Положительность скалярного произведения у
электромагнитного поля.}\footnote{Материал данного пункта не имеет
никакой глубокой связи ни с конструктивностью квантования, ни с
алгебраическими, ни с топологическими вопросами, обсуждаемыми в
настоящих статьях. Единственная причина, по которой здесь
обсуждается этот старый (и очень простой) вопрос состоит в том,
что во всех известных мне источниках приводится ошибочное
доказательство (при этом почему-то считают, что из положительности
квадратичной формы на векторах некоторого базиса вытекает и
положительность формы вообще).}
 Рассмотрим полевой осциллятор электромагнитного поля. Будем здесь
писать для краткости $a_\mu$ вместо $\hat a_\mu(+1)$ и $a^*_\mu$
вместо $\hat a_\mu(-1)$. Эти операторы удовлетворяют соотношениям:
\[
 [\,a^*_\mu\,,\,a_\nu\,]\,= g_{\mu\nu}, \qquad
 [\,a^*_\mu\,,\,a^*_\nu\,]=0, \qquad
 [\,a_\mu\,,\,a_\nu\,]=0 \ .
\]
Для интересующего нас квантования также выполняется:
\[
 a_\mu\,|\,0\,\rangle=0 \ .
\]
Дополнительное условие выглядит так:
\begin{equation}
 k^{(0)}_\mu a_\mu\,|\,\mathrm{rad}\,\rangle = 0 \ ,
\label{OscAddCond}
\end{equation}
где $k^{(0)}_\mu$~--- фиксированный вектор на световом конусе.
Подпространство векторов состояния, удовлетворяющих данному
условию, мы обозначим $\mathcal H^\mathrm{rad}$.

Поскольку оператор $k^{(0)}_\mu a_\mu$ имеет определённую
градуировку (его степень равна $-1$), градуировка пространства
состояний $\mathcal H$ переносится на подпространство
 $\mathcal H^\mathrm{rad}$:
\begin{equation}
 \mathcal H^\mathrm{rad}=
   \mathcal H^\mathrm{rad}_0 \dot+
   \mathcal H^\mathrm{rad}_1 \dot+
   \mathcal H^\mathrm{rad}_2 \dot+ \dots
\label{HradSum}
\end{equation}
Иначе говоря, подпространство, выделяемое дополнительным условием
(\ref{OscAddCond}), также распадается в ортогональную сумму
состояний с определённым числом частиц.

Поскольку сумма (\ref{HradSum}) ортогональная, неотрицательность
скалярного произведения достаточно доказать для каждого из
подпространств $\mathcal H^\mathrm{rad}_n$.

Произвольный вектор состояния $|\,n\,\rangle$ из подпространства
 $\mathcal H^\mathrm{rad}_n$ можно представить в виде:
\[
 |\,n\,\rangle = T_{\mu\nu\dots\rho} \,
                 a^*_\mu\,a^*_\nu\dots a^*_\rho \,
                 |\,0\,\rangle    \ ,
\]
где $T_{\mu\nu\dots\rho}$~--- $n$-валентный тензор. Этот тензор
можно без ущерба для общности считать симметричным.

При этом условие (\ref{OscAddCond}) для вектора $|\,n\,\rangle$
оказывается дополнительным условием на тензор
$T_{\mu\nu\dots\rho}$:
\begin{equation}
 k^{(0)}_\mu T_{\mu\nu\dots\rho} = 0 \ .
\label{TOscAddCond}
\end{equation}

Скалярное произведение же вектора $|\,n\,\rangle$ самого на себя
равняется:
\begin{equation}
 \langle\,n\,|\,n\,\rangle =
      (-1)^n\,n!\,\cdot\,
      T^*_{\mu\nu\dots\rho}\,T_{\mu\nu\dots\rho} \ .
\label{Tsum}
\end{equation}

Величина
$(-1)^n\,n!\,\cdot\,T^*_{\mu\nu\dots\rho}\,T_{\mu\nu\dots\rho}$
является суммой положительных вещественных чисел, часть из которых
входит в сумму со знаком \rql плюс\rqr, а часть со знаком \rql
минус\rqr. Этот знак определяется релятивистским правилом
суммирования по повторяющимся индексам и множителем $(-1)^n$.
Покажем, что эта сумма неотрицательна при условии
(\ref{TOscAddCond}).

Поскольку конструкция пространства состояний инвариантна по
отношению к выбору вектора $k^{(0)}_\mu$, этот вектор можно
принять равным:
\begin{equation}
 k^{(0)}_\mu =
 \left(\begin{array}{c|ccc}
       k_0 & 0 & 0 & k_0
       \end{array}\right)_\mu    \ .
\label{kChoice}
\end{equation}

Рассмотрим теперь в сумме (\ref{Tsum}) все слагаемые вида
$(-1)^n\,n!\,\cdot\,T^*_{0\nu\dots\rho}\,T_{0\nu\dots\rho}$. В
силу (\ref{TOscAddCond}) и (\ref{kChoice}), они полностью
сокращаются со слагаемыми вида
$(-1)^n\,n!\,\cdot\,T^*_{3\nu\dots\rho}\,T_{3\nu\dots\rho}$.

Среди оставшихся слагаемых рассмотрим слагаемые вида
$(-1)^n\,n!\,\cdot\,T^*_{\mu0\dots\rho}\,T_{\mu0\dots\rho}$. Они
полностью сократятся с оставшимися слагаемыми вида
$(-1)^n\,n!\,\cdot\,T^*_{\mu3\dots\rho}\,T_{\mu3\dots\rho}$.

И так далее. В результате в сумме (\ref{Tsum}) останутся лишь
слагаемые, индексы которых равны $1$ или $2$. Все эти слагаемые
входят в сумму (\ref{Tsum}) со знаком \rql плюс\rqr.


\myparagraph{Топология и полнота инвариантного фазового
пространства электромагнитного поля.} \label{TFIPSEM}
 При изучении инвариантного гамильтонова формализма
в инвариантное фазовое пространство $Z$ мы вначале включили лишь
решения уравнений движения, гладкие в координатном представлении и
финитные в пространственном направлении. Вопрос же о том, какой в
действительности топологией следует наделять пространство $Z$
вообще не обсуждался. При этом фактически не давалось и точного
определения пространства $Z^*$; не уточнялось, каким же в точности
образом оказываются изоморфными функциональные пространства $Z$ и
$Z^*$; точное определение скобок Пуассона на $Z^*$ также не
давалось. В этом пункте мы покажем, что в пространстве $Z$ можно
ввести такую топологию (и пополнить его относительно этой
топологии), что оно станет очень похожим на гильбертово
пространство. При этом все необходимые для инвариантного
гамильтонова формализма свойства у такой топологии будут
выполняться.

Симплектическая структура сама по себе не задаёт топологии.
Однако, между элементами классического фазового пространства $Z$ и
одночастичными состояниями квантованного поля существует
взаимно-однозначное соответствие:
\begin{equation}
 \underline c \ \longleftrightarrow\
 (I^{-1}\underline c)\widehat~ \ |\,0\,\rangle \ .
\label{ZHisom}
\end{equation}

Рассмотрим теперь для примера скалярное поле. Среди его
инвариантных квантований присутствует квантование с
положительно-определённым скалярным произведением. Соответствие
(\ref{ZHisom}), во-первых, наделяет пространство $Z$ комплексной
структурой, а во-вторых, переносит туда скалярное произведение из
одночастичного квантового пространства $\mathcal H_1$. Таким
образом, $Z$ имеет естественную структуру комплексного гильбертова
пространства\footnote{Полученная топология пространства $Z$
является самой естественной с точки зрения квантования. Бор и
Розенфельд \cite{BohrRos1933} заметили, что при рассмотрении
квантованного электромагнитного поля необходимо производить
усреднение поля по небольшому пространственно-временному объёму:
символ $\hat A_\mu(x)$ при этом сам по себе физической величиной
не является. Проведённое нами рассмотрение показывает, что это
утверждение точно так же справедливо и для классического поля,
если его фазовое пространство наделяется \rql правильной\rqr{}
топологией.}.

Симплектическая структура оказывается непрерывной по паре своих
аргументов относительно полученной топологии; симплектическая
структура задаёт изоморфизм пространств $Z$ и $Z^*$; скобки
Пуассона корректно определяются на всём $Z^*$.

Представим теперь скалярное поле в фурье-представлении в виде:
\[
 \widetilde\varphi(k) = 2\pi\,\delta(k^2-m^2)\cdot a(k)  \ .
\]
Соответствие (\ref{ZHisom}) можно записать более явно так:
\[
 \underline c \ \longleftrightarrow\
 \int d\mu^+_m \cdot i\,a(k)^{\,\underline c} \cdot
               \hat a^*(k) \ |\,0\,\rangle \ ,
 \qquad \mbox{где} \ \
 d\mu^+_m = \frac{d^4\!k}{(2\pi)^4} \cdot
            2\pi\,\delta(k^2-m^2) \cdot
            \theta(k) \ .
\]
И скалярное произведение в пространстве $Z$ принимает вид:
\begin{equation}
 \langle\,\underline c\,,\,\underline d\,\rangle =
 \int d\mu^+_m \cdot
 a^*(k)^{\,\underline c} \cdot
 a(k)^{\,\underline d} \ .
\label{ScalScalProd}
\end{equation}

Буквальное перенесение этой схемы на случай электромагнитного поля
не представляется возможным, поскольку, как мы видели, среди
 $\mathcal P$-инвариантных квантований электромагнитного поля
отсутствуют квантования с положительно-определённым скалярным
произведением.

Здесь имеется, однако, другая возможность. Представим
электромагнитное поле также в фурье-пред\-ста\-вле\-нии как
\[
\widetilde A_\mu(k) = 2\pi\,\delta(k^2) \cdot a_\mu(k)  \ .
\]
По аналогии с формулой (\ref{ScalScalProd}) введём в классическом
фазовом пространстве $Z$ электромагнитного поля скалярное
произведение по формуле:
\begin{equation}
 \langle\,\underline c\,,\,\underline d\,\rangle =
 \int d\mu^+_m \cdot
 M_{\nu\rho} \cdot
 a^*_\nu(k)^{\,\underline c} \cdot
 a_\rho(k)^{\,\underline d} \ .
\label{EMScalProd}
\end{equation}
Здесь $M_{\nu\rho}$~--- произвольная положительно-определённая
эрмитова матрица. В качестве такой матрицы можно взять, например,
$M_{\nu\rho}=\mathrm{diag}\,(+1,+1,+1,+1)_{\nu\rho}$.

Скалярное произведение (\ref{EMScalProd}) является
положительно-определённым. Поэтому оно определяет некоторую
топологию в пространстве $Z$. Важным является то обстоятельство,
что эта топология не зависит от конкретного выбора матрицы
$M_{\nu\rho}$.

Отсюда следует, что указанная топология является
релятивистски-ин\-ва\-ри\-ант\-ной.

Относительно введённой топологии симплектическая структура
оказывается непрерывной по паре своих аргументов; симплектическая
структура задаёт изоморфизм пространств $Z$ и $Z^*$; скобки
Пуассона корректно определяются на всём $Z^*$.

Таким образом, инвариантное фазовое пространство электромагнитного
поля обладает естественной структурой линейного комплексного
топологического пространства, причём топология в нём задаётся
множеством эквивалентных (в смысле топологии) скалярных
произведений. Такие пространства мы назовём {\it пространствами
гильбертова типа}.


\myparagraph{Тензорное произведение пространств гильбертова типа.}
 Сейчас мы покажем, что тензорное произведение пространств
гильбертова типа само является пространством гильбертова типа.

Рассмотрим два таких пространства $X$ и $Y$. Их тензорное
произведение (точнее, алгебраическое тензорное произведение)
определяется следующим образом.

Рассмотрим все формальные произведения вида
 $x\diamond y$, где $x\in X$ и $y\in Y$. Мы будем называть такие
произведения парами. Будем также считать, что пары можно формально
умножать на комплексные числа, образуя при этом выражения вида:
$\lambda\cdot x\diamond y$. И рассмотрим все формальные суммы
вида:
\[
 \lambda_1\cdot x_1\diamond y_1 +
 \lambda_2\cdot x_2\diamond y_2 +
 \dots +
 \lambda_n\cdot x_n\diamond y_n     \ .
\]
Две такие суммы мы будем считать эквивалентными, если у них для
любой пары совпадает сумма коэффициентов при этой паре. Таким
образом, приходим к комплексному векторному пространству, которое
мы обозначим как $X\diamond Y$ (это~--- {\it свободный
 $\mathbb C$-модуль, порождённый декартовым произведением
$X\times Y$}).

В получившемся пространстве рассмотрим элементы вида:
\[
 1\cdot (\lambda x)\diamond y - \lambda \cdot x\diamond y \ ,
\]
\[
 1\cdot x\diamond (\lambda y) - \lambda \cdot x\diamond y \ ,
\]
\[
 1\cdot (x_1+x_2)\diamond y -
   1\cdot x_1\diamond y - 1\cdot x_2\diamond y \ ,
\]
\[
 1\cdot x\diamond (y_1+y_2) -
   1\cdot x\diamond y_1 - 1\cdot x\diamond y_2  \ .
\]
Линейную оболочку указанных элементов обозначим $X\circ Y$.

Факторизуя $X\diamond Y$ по $X\circ Y$, приходим к тензорному
произведению:
\[
 X\otimes Y = (X\diamond Y)/(X\circ Y) \ .
\]

Введём также для краткости обозначение $x\otimes y$. Именно, будем
считать, что при указанной факторизации элемент
 $1\cdot x\diamond y$ пространства $X\diamond Y$ переходит в
элемент $x\otimes y$ пространства $X\otimes Y$.

Если $X$ и $Y$~--- обычные гильбертовы пространства, то, как
известно, их тензорное произведение $X\otimes Y$ обладает
естественной структурой гильбертова пространства. Скалярное
произведение в $X\otimes Y$ при этом определяется вначале для пар
по формуле:
\begin{equation}
 \langle\, x_1\otimes y_1\,,
        \, x_2\otimes y_2\,\rangle_{X\otimes Y}=
 \langle\, x_1\,,\, x_2\,\rangle_X \cdot
 \langle\, y_1\,,\, y_2\,\rangle_Y            \ ,
\label{ScalProdInTensProd}
\end{equation}
а на остальные элементы переносится, исходя из требования
линейности по второму аргументу и антилинейности по первому.
Пополняя $X\otimes Y$ относительно указанного скалярного
произведения, приходим к гильбертову пространству.

В случае пространств гильбертова типа каждое из пространств $X$ и
$Y$ обладает многими эквивалентными скалярными произведениями.
Пространство $X\otimes Y$ при этом, согласно указанной схеме,
наделяется многими скалярными произведениями. Покажем, что все эти
скалярные произведения эквивалентны.

Итак, пусть в одном из двух пространств, например в $X$, мы
переходим от скалярного произведения
$\langle\,\cdot\,,\,\cdot\,\rangle_X$ к эквивалентному скалярному
произведению $(\,\cdot\,,\,\cdot\,)_X$. При этом скалярное
произведение $\langle\,\cdot\,,\,\cdot\,\rangle_{X\otimes Y}$ в
пространстве $X\otimes Y$ заменяется на
$(\,\cdot\,,\,\cdot\,)_{X\otimes Y}$.

Рассмотрим какой-нибудь элемент пространства $X\otimes Y$:
\[
 z=x_1\otimes y_1 + x_2\otimes y_2 + \dots + x_n\otimes y_n \ .
\]
Легко видеть, что этот элемент может быть приведён к такому виду,
что в этой сумме все $y_1,y_2,\dots,y_n$ будут ортогональны друг
другу относительно скалярного произведения в $Y$. Скалярные
произведения этого элемента самого на себя примут тогда особенно
простой вид:
\begin{equation}
 \langle\,z\,,\,z\,\rangle_{X\otimes Y} =
 \langle\,x_1\,,\,x_1\,\rangle_X \cdot
         \langle\,y_1\,,\,y_1\,\rangle_Y +
 \langle\,x_2\,,\,x_2\,\rangle_X \cdot
         \langle\,y_2\,,\,y_2\,\rangle_Y + \dots +
 \langle\,x_n\,,\,x_n\,\rangle_X \cdot
         \langle\,y_n\,,\,y_n\,\rangle_Y  \ ,
\label{ZSqrAng}
\end{equation}
\begin{equation}
 (\,z\,,\,z\,)_{X\otimes Y} =
 (\,x_1\,,\,x_1\,)_X \cdot \langle\,y_1\,,\,y_1\,\rangle_Y +
 (\,x_2\,,\,x_2\,)_X \cdot \langle\,y_2\,,\,y_2\,\rangle_Y +\dots +
 (\,x_n\,,\,x_n\,)_X \cdot \langle\,y_n\,,\,y_n\,\rangle_Y  \ .
\label{ZSqrBra}
\end{equation}

Удобный необходимый и достаточный признак эквивалентности
скалярных произведений
$\langle\,\cdot\,,\,\cdot\,\rangle_{X\otimes Y}$ и \\
$(\,\cdot\,,\,\cdot\,)_{X\otimes Y}$ состоит в следующем:
существует такое $\varepsilon\in\mathbb R$, $\varepsilon>0$, что
для любого $z\in X\otimes Y$ выполняются условия:
\[
 \varepsilon\cdot(\,z\,,\,z\,)_{X\otimes Y} <
 \langle\,z\,,\,z\,\rangle_{X\otimes Y} \ , \qquad
 \varepsilon\cdot\langle\,z\,,\,z\,\rangle_{X\otimes Y} <
 (\,z\,,\,z\,)_{X\otimes Y} \ .
\]

Но если такое $\varepsilon$ существует для скалярных произведений
$\langle\,\cdot\,,\,\cdot\,\rangle_X$ и $(\,\cdot\,,\,\cdot\,)_X$,
то согласно формулам (\ref{ZSqrAng}) и (\ref{ZSqrBra}), оно же
годится и для скалярных произведений
$\langle\,\cdot\,,\,\cdot\,\rangle_{X\otimes Y}$ и
$(\,\cdot\,,\,\cdot\,)_{X\otimes Y}$.

Таким образом, все скалярные произведения в $X\otimes Y$
эквивалентны. Пополняя $X\otimes Y$ относительно топологии,
задаваемой этими скалярными произведениями, приходим к
пространству гильбертова типа.


\myparagraph{Топология пространства состояний квантованного
электромагнитного поля.}
 Как мы видели, у квантованного электромагнитного поля
пространство состояний $\mathcal H$ распадается в ортогональную
сумму подпространств с определённым числом частиц:
\begin{equation}
 \mathcal H =
   \mathcal H_0 \dot+
   \mathcal H_1 \dot+
   \mathcal H_2 \dot+ \dots
\label{PartDecompEMF}
\end{equation}

Подпространство $\mathcal H_0$ одномерно. Поэтому вопрос о его
топологии не встаёт.

Подпространство $\mathcal H_1$, как было показано в
пункте~\ref{TFIPSEM}, фактически отождествляется с фазовым
пространством классического поля $Z$. Оно является пространством
гильбертова типа. Тем самым вопрос о его топологии также решён.

Рассмотрим теперь тензорное произведение пространства $Z$ на себя:
$Z\otimes Z$. В нём естественным образом определяется действие
группы перестановок из двух элементов. Простейшие элементы
преобразуются под действием этой группы как
 $\underline a\otimes\underline b\to
  \underline b\otimes\underline a$, а на остальные элементы это
действие продолжается по линейности.

Оставим среди скалярных произведений в пространстве $Z\otimes Z$
только инвариантные по отношению к действию указанной группы. На
практике удобно просто ограничиться произведениями, для которых в
правой части формулы~(\ref{ScalProdInTensProd}) стоит произведение
одинаковых скалярных произведений.

Тогда группа перестановок действует унитарно\footnote{Под
унитарным преобразованием пространства гильбертова типа мы
понимаем автоморфизм этого пространства, т.~е. взаимно-однозначное
отображение на себя, сохраняющее линейную структуру и каждое из
скалярных произведений.} в $Z\otimes Z$. И $Z\otimes Z$
распадается в ортогональную сумму двух инвариантных
подпространств: симметрического и антисимметрического.

Каждое из этих двух подпространств наследует топологию из
$Z\otimes Z$.

Легко видеть, что двухчастичное квантовое пространство
 $\mathcal H_2$ естественно отождествляется с симметрическим
подпространством в $Z\otimes Z$ и тем самым наделяется
соответствующей топологией.

И так далее. $n$-частичное подпространство $\mathcal H_n$
отождествляется с полностью симметрическим подпространством $n$-ой
тензорной степени $Z$. И наследует оттуда топологию.

Таким образом, каждое из подпространств в ортогональной сумме
(\ref{PartDecompEMF}) является пространством гильбертова типа. С
практической точки зрения это очень удобно, поскольку для
практических приложений гораздо проще определять описанные
пространства с помощью базисов (а не как факторпространства
свободных модулей).

Рассмотрим теперь пространство $\mathcal H$ в целом. На основе
топологий, введённых в его подпространствах, можно вводить разные
топологии во всём пространстве. Например, сходимость
направленности к нулю можно понимать как независимое стремление к
нулю всех проекций этой направленности. Такая топология
представляется в данном случае самой естественной.

Можно же это требование усилить и потребовать дополнительно,
например, чтобы начиная с некоторого момента у направленности лишь
конечное число проекций отличалось от нуля.

Таким образом, в $\mathcal H$ существует много естественных
топологий. При этом само пространство $\mathcal H$ пространством
гильбертова типа не является.

В связи с этим хотелось бы обратить здесь внимание на следующее
обстоятельство. В работе Гупты~\cite{Gupta1950} предлагалось
квантовать электромагнитное поле в обычном гильбертовом
пространстве с положительно-опре\-де\-лён\-ной метрикой. Это
действительно можно сделать, рассматривая квантования,
инвариантные по отношению к более узкой группе, чем группа
Пуанкаре (а именно, по отношению к подгруппе группы Пуанкаре,
оставляющей инвариантным направление оси времени). Подпространства
с фиксированным числом частиц при этом оказываются фактически
совпадающими с построенными выше. Но общее пространство
 $\mathcal H$ наделяется при этом топологией, которая не является
релятивистски-инвариантной. При этом оказывается, что некоторые
состояния поля при квантовании в одной системе отсчёта гильбертову
пространству принадлежат, а в другой нет. Таким образом, старый
формализм Гупты релятивистски-инвариантным не является, даже
неявно.

В своих более поздних работах Гупта пытался отказаться от
насильственного введения знакоопределённой метрики. Однако, его
последняя публикация на эту тему~\cite{Gupta1959} ясно показала,
что никакой явно релятивистски-инвариантной конструкции
квантованного электромагнитного поля у него всё равно нет (не
говоря уже о проблемах с функциональным анализом).


\myparagraph{О происхождении антиунитарных преобразований.}
 В соответствии с введённой процедурой квантования, линейные
преобразования наблюдаемых классического поля, сохраняющие скобки
Пуассона и подпространства $Cr$ и $De$, порождают унитарные
преобразования состояний квантованного поля. Между тем известно,
что в квантовой теории важную роль играют также антиунитарные
преобразования симметрии. Рассмотрим их происхождение на примере
операции обращения времени~--- $T$.

Предположим, что лагранжиан классического поля $T$-инвариантен
(примерами могут служить и скалярное поле и электромагнитное).
Тогда действие оказывается тоже $T$-инвариантным. Тем самым,
полевые функции, удовлетворяющие принципу стационарного действия,
переходят при отражении времени в функции, также удовлетворяющие
принципу стационарного действия.

Таким образом, с точки зрения инвариантного гамильтонова
формализма, операция обращения времени~--- это взаимно-однозначное
отображение инвариантного фазового пространства $Z$ на себя. Как
при этом отображении ведёт себя симплектическая структура? Из
вариационного определения симплектической структуры очевидно, что
она при этом меняет знак. Такие преобразования, меняющие знак
симплектической структуры, естественно называть {\it
антисимплектическими}.

Рассмотрим теперь сопряжённое действие операции обращения времени
на элементы сопряжённого пространства $Z^*_{\mathbb C}$:
\[
 a \stackrel T{\longrightarrow} a_T \ , \qquad
 a,a_T \in Z^*_{\mathbb C} \ .
\]
Скобка Пуассона под действием этого преобразования меняет знак:
\[
 \{\,a_T\,,\,b_T\,\} = - \,\{\,a\,,\,b\,\} \ .
\]
Таким образом, операция обращения времени порождает автоморфизм
алгебры Ли $\mathcal C \oplus Z^*_{\mathbb C}$.

В соответствии с процедурой квантования, изложенной в данной
статье, возникает желание построить соответствующий автоморфизм
алгебры операторов $\mathcal O$ для квантованного поля. Этого,
однако, сделать нельзя. Дело в том, что при построении алгебры
операторов $\mathcal O$ у множества
 $\mathcal C \oplus Z^*_{\mathbb C}$ используется не только
структура алгебры Ли, но и специальное соответствие между
подпространством $\mathcal C$ и множеством скаляров $\mathbb C$
(соотношение $\hat1 = ()\widehat~$). Это означает, что для
перенесения операции обращения времени на квантовый случай
требуются дополнительные соглашения, кроме уже введённой операции
квантования.

Условимся, что слова (т.~е. элементы полугруппы $\mathcal W$) при
обращении времени преобразуются по следующей формуле:
\[
 \hat a\,\hat b \dots \hat c \stackrel T{\longrightarrow}
 \hat c_T \dots \hat b_T\,\hat a_T \ , \qquad
 \hat a,\hat b,\dots,\hat c,\hat a_T,\hat b_T,\dots,\hat c_T
 \in\mathcal A \ ,
\]
т.~е. происходит замена каждой буквы на соответствующую, и буквы
после этого записываются в обратном порядке. При этом очевидно,
что отражение времени на произведение двух слов действует
следующим образом:
\[
 \hat p\,\hat q \stackrel T{\longrightarrow} \hat q_T\,\hat p_T \ ,
 \qquad \hat p,\hat q,\hat p_T,\hat q_T \in \mathcal W \ .
\]
Взаимно-однозначное отображение полугруппы на себя, обладающее
таким свойством можно назвать {\it антиавтоморфизмом}.

Построенный антиавтоморфизм полугруппы слов $\mathcal W$ по
линейности продолжается до антиавтоморфизма алгебры фраз
 $\mathcal P$. Далее, легко проверить, что этот антиавтоморфизм
алгебры фраз порождает антиавтоморфизм алгебры операторов
$\mathcal O$.

Предположим далее, что при обращении времени рождающее
подпространство $Cr$ и уничтожающее $De$ переходят друг в друга.
Это условие обычно выполняется потому, что эти подпространства
являются от\-ри\-ца\-тель\-но- и положительно-частотными,
соответственно. Левый идеал, натянутый на $De\widehat~$, при
обращении времени переходит в правый идеал, натянутый на
$Cr\widehat~$. Получающийся в результате факторизации левый модуль
переходит в соответствующий правый модуль.

Таким образом, операция обращения времени задаёт
взаимно-однозначное линейное соответствие кет- и
бра-век\-то\-ров\footnote{Швингер предлагал~\cite{Schwin1992}
интерпретировать кет-векторы как символы, обозначающие создание
системы в прошлом; бра-векторы как символы, обозначающие
уничтожение системы в будущем; а операторы как действие приборов в
ходе эксперимента. Такая интерпретация оказывается здесь весьма
естественной.}.

Скалярное произведение, как нетрудно видеть, при обращении времени
сохраняется:
\begin{equation}
 \langle\,x\,|\,y\,\rangle=\langle\,y_T\,|\,x_T\,\rangle \ .
\label{TScale}
\end{equation}

Поскольку между кет- и бра-векторами имеется взаимно-однозначное
антилинейное соответствие, то можно проводить все рассуждения в
рамках одного пространства, например, пространства кет-векторов.
Тогда операция обращения времени может рассматриваться как
взаимно-однозначное антилинейное преобразование этого
пространства. Соотношение~(\ref{TScale}) показывает, что при
указанном преобразовании скалярное произведение изменяется на
комплексно-сопряжённое. Преобразование, обладающее такими
свойствами, называют {\it антиунитарным}.


\bigskip
В заключение хочу поблагодарить В.~М.~Шабаева, Л.~Д.~Фаддеева,
В.~А.~Франке, В.~Д.~Ляховского, Л.~В.~Прохорова, В.~В.~Верещагина
и Ю.~М.~Письмака за плодотворные дискуссии.

